\documentclass[a4paper,fleqn,usenatbib,useAMS]{mnras}

\usepackage{times}
\usepackage{amsmath}
\usepackage{amssymb}
\usepackage{float}
\usepackage{lmodern}
\usepackage{graphicx}
\usepackage{epsfig}
\usepackage[a4paper]{geometry}

\usepackage{etoolbox}
\makeatletter
\patchcmd\@combinedblfloats{\box\@outputbox}{\unvbox\@outputbox}{}{%
   \errmessage{\noexpand\@combinedblfloats could not be patched}%
}%
 \makeatother

\def\apj{ApJ}
\def\mnras{MNRAS}

\def\araa{ARA\&A}                
\def\aap{A\&A}                   
\def\aaps{A\&AS}                 
\def\aj{AJ}                      
\def\apjs{ApJS}                  
\def\pasp{PASP}                  
\def\apjl{ApJ}                   

\def\at{The Astronomer$'$s Telegram}
\begin{document}

\title[MUSE observations of a changing-look AGN - I]{MUSE observations of a changing-look AGN\\ I: The re-appearance of the broad emission lines}
\author[S. I. Raimundo]{S. I. Raimundo$^{1}$\thanks{E-mail: s.raimundo$@$dark-cosmology.dk}, M. Vestergaard$^{1,2}$, J. Y. Koay$^{3}$, D. Lawther$^{1}$, V. Casasola$^{4}$, 
\newauthor B. M. Peterson$^{5,6,7}$
\\
$^{1}$ DARK, Niels Bohr Institute, University of Copenhagen, Lyngbyvej 2, 2100 Copenhagen, Denmark\\
$^{2}$ Steward Observatory, University of Arizona, 933 N. Cherry Avenue, Tucson, AZ 85721, USA\\
$^{3}$ Institute of Astronomy and Astrophysics, Academia Sinica, PO Box 23-141, Taipei 10617, Taiwan\\
$^{4}$ INAF-Instituto di Radioastronomia, Via Piero Gobetti 101, 40129, Bologna, Italy\\
$^{5}$ Department of Astronomy, The Ohio State University, 140 West 18th Avenue, Columbus, OH 43210,USA\\ 
$^{6}$ Center for Cosmology and AstroParticle Physics, The Ohio State University, 191 West Woodruff Avenue, Columbus, OH 43210, USA\\
$^{7}$ Space Telescope Science Institute, 3700 San Martin Drive, Baltimore, MD 21218, USA\\
}
\date{Accepted 2019 March 15. Received 2019 March 8; in original form 2019 January 18}

\maketitle

\begin{abstract}
Optical changing-look Active Galactic Nuclei (AGN) are a class of sources that change type within a short timescale of years or decades. This change is characterised by the appearance or disappearance of broad emission lines, often associated with dramatic AGN continuum flux changes that are orders of magnitude larger than those expected from typical AGN variability. 
In this work we study for the first time the host galaxy of a changing-look AGN, Mrk~590, using high spatial resolution optical and near-infrared observations. We discover that after $\sim$ 10 yr absence, the optical broad emission lines of Mrk~590 have reappeared. The AGN optical continuum flux however, is still $\sim$ 10 times lower than that observed during the most luminous state in the 1990s.
The host galaxy shows a 4.5 kpc radius star-forming ring with knots of ionised and cold molecular gas emission. Extended ionised and warm molecular gas emission are detected in the nucleus, indicating that there is a reservoir of gas as close as 60 pc from the black hole. We observe a nuclear gas spiral between radii $r \sim 0.5 - 2$ kpc, which has been suggested as a dynamical mechanism able to drive the necessary gas to fuel AGN. We also discover blue-shifted and high velocity dispersion [O III] emission out to a radius of 1 kpc, tracing a nuclear gas outflow.  The gas dynamics in Mrk~590 suggest a complex balance between gas inflow and outflow in the nucleus of the galaxy.
 
\end{abstract} 


\section{Introduction}
Two of the most easily recognisable features of Active Galactic Nuclei (AGN) are optical-ultraviolet continuum emission from the accretion disc and broad emission lines with a typical velocity broadening of v$_{\rm FWHM}$ $\sim$ 1000 - 10000 km\,s$^{-1}$ \citep{peterson97}. However, broad emission lines are not ubiquitous in AGN. A subset of AGN have narrow emission lines (v$_{\rm FWHM}$ $\lesssim$ 500 km\,s$^{-1}$) but lack broad lines in their spectra. This prompted a historical classification into AGN types: type 1 to identify AGN with broad (and narrow) emission lines and type 2 to identify AGN without broad emission lines \citep{khachikian&weedman74}. The observer's line-of-sight towards the AGN was put forward as an explanation for this dichotomy (\citealt{lawrence&elvis82}, \citealt{antonucci84}, \citealt{antonucci93}). 
Due to the presence of an equatorial dust and gas torus-like structure surrounding the accretion disc and broad line region, broad emission lines would only be observed for viewing angles close to the polar direction. AGN for which the viewing angle is along the equatorial plane would be classified as type 2 due to the broad emission lines being obscured and therefore absent from the spectra.
 
The simple paradigm described above has been challenged by observations of AGN that defy this classification. 
For example the so-called `naked AGN' \citep{hawkins04} or true type 2 AGN (\citealt{panessa09}, \citealt{bianchi17}), AGN in which broad emission lines are not observed even though there is no evidence for obscuration along the line-of-sight. 
It has been suggested that these non-conforming AGN are in a transition phase. Possibly due to low black hole mass accretion rates (e.g. \citealt{bianchi17}) these AGN are not able to produce broad emission lines. Notably, there is another class of AGN that have been observed before and after a transition between types occurred, the so called `changing-look' AGN. These AGN were discovered due to having significantly different spectral features when observed at two or more different epochs. 
`Changing-look' AGN was a term first used to classify AGN that showed an extreme variation in X-ray detected obscuration, transitioning from what would be interpreted as heavily obscured to unobscured and vice-versa, in a short period of time (\citealt{matt03}, \citealt{bianchi05}). This definition has been extended to include optical `changing-look' AGN and quasars, to refer to AGN of low and high luminosity that transition between type 1 and type 2 or vice-versa. There are now a few tens of known changing-look AGN, including: Mrk~590 \citep{denney14}, NGC 2617 \citep{shappee14}, NGC  7603 \citep{tohline&osterbrock76}, Mrk 1018 (\citealt{cohen86}, \citealt{mcelroy16}, \citealt{husemann16}), NGC 1097 \citep{storchi-bergmann93}, NGC 3065 \citep{eracleous&halpern01} and NGC 7582 \citep{aretxaga99}. Transition between intermediate Seyfert types: HE 1136-2304 (\citealt{zetzl18}, \citealt{kollatschny18}) and transitions in quasars e.g.: SDSS J015957.64+003310.5 \citep{LaMassa15}, SDSS J012032.19-005501.9 \citep{li15}, SDSS J155440.25+362952.0 \citep{gezari17} and SDSS J141324.27+530527.0 \citep{wang18}, have also been observed. While most of the first changing-look AGN were discovered serendipitously, systematic searches have recently increased the number of known changing-look AGN. A study by \cite{runco16} of 102 low redshift Seyfert galaxies finds that only 3 Seyferts (3\% of the sample) show a disappearance of broad emission lines, while 38\% show a transition between intermediate types. Systematic searches at the higher luminosity range found a few tens of new changing-look AGN in optical (\citealt{ruan16}, \citealt{macleod18}, \citealt{yang18}) and infrared (\citealt{ross18}, \citealt{stern18}) quasar catalogues. To identify AGN that transitioned between types, two or more epochs of observations are needed. The transitions in the literature have been observed to occur within a timescale of years up to a few decades, but the current timescale upper limit is likely due to the limited time span covered by historical monitoring data.

The transition in optical changing-look AGN is still not fully understood. The favoured scenario in most of the studies mentioned above is that this transition is due to a change in mass accretion rate as opposed to changes in obscuration (e.g. \citealt{sheng17}, \citealt{stern18}). In several of these type transitions a large change in the AGN photometry or AGN continuum flux is also observed, with changes of a factor of a few up to a hundred (e.g. \citealt{denney14}, \citealt{husemann16}, \citealt{macleod18}) in addition to the changes in the broad lines. These are dramatic flux changes that differ from the typical 10\% - 20\% flux variations  in the timescale of months to years observed in AGN (\citealt{collier&peterson01}, \citealt{kelly09}, \citealt{macleod10}). 

Continuum variations have been suggested as the cause for the type transition. \cite{elitzur14} suggest a mechanism to explain the broad line appearance/disappearance based on a disc-wind scenario for the AGN broad line region. They suggest that as the AGN continuum luminosity decreases, the distance from the BLR clouds to the accretion disc will decrease, reducing the fraction of continuum photons intercepted by the BLR clouds. In case of a decrease in AGN continuum luminosity, the model of \cite{elitzur14} predicts a decrease in the broad line strength. This transition is controlled by changes in $L_{\rm bol}$ with the variable $L_{\rm bol}/M^{2/3}_{\rm BH}$ controlling the transition between spectral classes (such as type 1, type 2 and intermediate types).

If a change in black hole mass accretion rate is responsible for the change in AGN continuum flux, such change needs to occur within a timescale of years or decades. This timescale is much shorter than the characteristic timescale for significant changes in the surface mass density of a standard thin accretion disc \citep{shakura&sunyaev73}, i.e. the viscous timescale $\sim$ 10$^{5}$ yr (e.g. \citealt{noda&done18}). Variations on shorter timescales have recently been put forward to explain the mass accretion rate variation timescales observed in changing-look AGN. These include for example, the propagation of a cold front in an inflated accretion disc (\citealt{ross18}, \citealt{stern18}) or the propagation of a cold front and a state transition (as seen in black hole binaries) in an accretion disc dominated by radiation pressure \citep{noda&done18}.

Additionally, for AGN where a single transition from type 1 to type 2 and a dramatic decrease in AGN continuum flux are observed, it is possible that the AGN will not return to a type 1. Such a transition could be associated with the turn-off of the AGN, as it is not guaranteed that the AGN will turn on again in the future. While it is not clear what controls the black hole mass accretion rate, a shortage of gas supply from the galaxy or lack of gas transport mechanisms may limit the gas available for AGN fuelling. 

In this work we study in detail and for the first time the host galaxy of Mrk~590, a changing-look AGN, using optical and near-infrared integral field spectroscopy. 
We study the past history of the host galaxy, determine its gas content and distribution and search for dynamical mechanisms that may drive gas to and from the black hole. 
The goal of the study presented here is to obtain clues on the host nuclear environment and physics of a changing-look AGN.

\subsection{Mrk 590}

The target of our study is the changing-look AGN in Mrk~590. Mrk~590 is an Sa galaxy at z $=$  0.026385 (493 pc/$''$) with more than 40 years of multi-wavelength data available (\citealt{denney14} and references therein). This galaxy has an active nucleus whose AGN activity type has varied in between observations. From the 1970's to the 1990's Mrk~590's flux increased by a factor of $\sim$10. In the 1990's this galaxy was a typical Seyfert type 1 galaxy, accreting at a high mass accretion rate corresponding to an Eddington ratio of $\sim$0.1 (e.g., \citealt{peterson04}; see \citealt{koay16a}). It showed a bright AGN continuum over-imposed on the host galaxy starlight emission and the presence of broad Balmer emission lines in the optical \citep{peterson98}. The broad emission lines were the target of a reverberation mapping campaign that determined Mrk~590's black hole mass: M$_{\rm BH} = 4.75 \pm 0.74 \times 10^{7}$ M$_{\odot}$ \citep{peterson04}. However, since then, the AGN continuum flux has decreased by a factor of more than 100 across all wavebands and sometime between 2006 and 2012 the broad H$\beta$ emission line completely disappeared \citep{denney14}.

\cite{denney14} showed that the type transition in Mrk~590 did not result from line of sight obscuration. This conclusion was based on two arguments: 1) there was no evidence for intrinsic obscuration in the X-ray spectrum of the source; 2) the [O III] 5007 \AA\, flux also showed some flux variations on longer timescales. As the [O III] emission is produced in the narrow line region which is located much further out (out to kpc scales) than the broad line region, the flux changes would require an impossibly large obscuring system to hide both the broad line and narrow line regions. \cite{denney14} suggest that the change in type was due to a variation of the AGN mass accretion rate. Further support for this hypothesis came from radio observations that show that the radio flux variations of Mrk590 over the last few decades are consistent with the trends in optical and X-rays, ruling out obscuration in Mrk590 \citep{koay16b}.

Ultraviolet emission observed in November 2014 with the \emph{Hubble Space Telescope} (\emph{HST}) shows interesting features \citep{mathur18}. \cite{mathur18} find a broad Mg II $\lambda$2800 \AA\, line and an increase of more than a factor of 2 in the UV continuum compared to earlier 2014 observations. Earlier that year (in January 2014) the optical broad H$\beta$ line was not present in the spectra \citep{denney14}. The H$\beta$ and Mg II emitting gas are expected to be co-spatial in the broad line region. However, \cite{mathur18} suggest that due to the lower responsivity of the Mg II line with respect to H$\beta$ (\citealt{korista&goad00}, \citealt{cackett15}), the broad Mg II component may never have disappeared completely when the AGN transitioned to a lower state flux.

Due to its recently observed spectral changes, Mrk~590 is the target of an on-going multi-wavelength monitoring campaign to observe the AGN as it transitions between states. 
In this paper we present the results of MUSE optical observations and SINFONI near-infrared observations of Mrk~590 that probe the stars and gas in the vicinity of the AGN and out to kiloparsec scales. We will show that by 2017 October 28 the AGN optical broad emission lines have reappeared.

This is the first of two papers analysing the host galaxy properties of Mrk~590. In this first paper we will describe the data analysis, present the maps of ionised and molecular gas dynamics and discuss the gas excitation mechanisms and stellar population of Mrk~590.
In the second paper (Raimundo et al. in prep, hereafter Paper II) we will discuss the dynamical structures observed in light of the fuelling and feedback of the AGN.

\section{Data reduction}
\subsection{MUSE observations}
\label{sec:MUSEreduction}
We obtained observations of Mrk~590 with the MUSE optical integral-field spectrograph on the Very Large Telescope (VLT). The observations were divided into four nights (2017 October 28, November 12, 14 and 16) for a total on-source exposure time of 2h27min. The field of view was rotated by 90 degrees between each exposure. The data were processed using the command line tool \textsc{esorex} (version 3.12.3) and the ESO recipes for the MUSE pipeline (version 2.2.0) \citep{weilbacher14}. The data from each night were processed separately using the standard calibration procedure. Each night had two science exposures and one sky exposure in the sequence object-sky-object. The necessary calibration files were created with the recipes \textsc{muse\_bias}, \textsc{muse\_flat}, \textsc{muse\_wavecal} and \textsc{muse\_twilight}. A representation of the MUSE wavelength-dependent line spread function, needed for an accurate sky subtraction, was calculated using \textsc{muse\_lsf}. The line spread function profiles we calculate are very similar to the ones provided with the MUSE pipeline package, something that was also observed during the commissioning of MUSE. 
Using \textsc{muse\_scibasic} several calibrations and corrections were applied to each exposure to remove the instrument signature: the exposures were bias subtracted and flat-field corrected. Wavelength and geometrical calibrations were applied and a correction for relative illumination was carried out. For this task we used the geometry tables provided in the MUSE pipeline package. As a final product \textsc{muse\_scibasic} creates the pixel tables, an intermediate MUSE data product mapping the properties of each CCD pixel. In the post-processing analysis, the pixel tables were flux calibrated using observations of a spectrophotometric standard star (\textsc{muse\_standard}) and a model of the sky spectrum was obtained based on the off-target sky exposure (\textsc{muse\_create\_sky}). The final step with the recipe \textsc{muse\_scipost} does a flux calibration of the pixtables of each exposure, corrects for sky emission and telluric features and creates a data cube. The spatial offset between each exposure is calculated using \textsc{muse\_exp\_align}. We fit a Moffat 2D spatial profile to the white light image of the M2V star (SDSS J021432.74-004600.8), a foreground star in the field of view of the observations, and calculate the star's total flux for each exposure. We then rescale all exposures to the stellar flux level measured in the night with the best (photometric) conditions to account for flux variations due to sky transmission. 
To obtain the final data cube and increase the signal-to-noise ratio we align, flux correct and combine the pixtables from all the exposures using \textsc{muse\_exp\_combine}.

The final cube has a spatial and spectral dimension of 321 $\times$ 312 $\times$ 3681 pixels, covering a field of view of 1 $\times$ 1 arcmin$^{2}$ with a pixel spatial sampling size of 0.2 $\times$ 0.2 arcsec$^{2}$ and a spectral sampling of 1.25 \AA. The wavelength range is 4750 - 9350 \AA.

\subsubsection{Spatial and spectral resolution}
In addition to the data cube we create a sky cube from the off-source sky exposures to determine the spectral resolution of the data. The spectral resolution is determined by fitting a Gaussian function to unblended sky emission lines.
We fit three sky emission lines, at 5577\,\AA, 6923\,\AA\ and 8399\,\AA\, across the field of view. The median spectral resolution across the field of view is $\sigma = 60\pm 6$ km\,s$^{-1}$ (FWHM = 2.6 \AA) at 5577 \AA, $\sigma = 44\pm 8$ km\,s$^{-1}$ (FWHM = 2.37 \AA) at 6923 \AA\ and $\sigma = 37\pm 4$ km\,s$^{-1}$ (FWHM = 2.42 \AA) at 8399 \AA.
The spatial resolution is calculated by fitting the spatial profile of the star in the field of view using a Moffat function. The spatial resolution varies slightly from night to night but the final value measured in the combined white light image is FWHM $= 0.68 \times 0.66$ arcsec.

\subsection{SINFONI observations}
\label{sec:SINFONIreduction}
We have processed and analysed archival SINFONI/VLT observations of Mrk~590. The observations were carried out in the K-band on two nights in October 2007 (observing programme: 080.B-0239) using a spatial scale of 0.1 $\times$ 0.05 arcsec/pixel and a field-of-view of 3 $\times$ 3 arcsec$^{2}$. The data reduction was done using \textsc{esorex} (version 3.12.3) and SINFONI pipeline recipes version 3.0.0. The method for the data reduction followed the standard SINFONI data reduction cascade with two extra \textsc{idl} routines: one for the sky emission subtraction \citep{davies07_oh} and a bad pixel removal routine for 3D
data cubes (\cite{davies10} - a 3D version of LA cosmic \citep{vandokkum01}) as outlined by \cite{raimundo13} and \cite{raimundo17}. The cubes for the different exposures and nights were shifted using the header values for the spatial offsets, combined and re-sampled to a spatial scale of 0.05 $\times$ 0.05 arcsec$^{2}$. The spectral wavelength range covered is 1.95 - 2.45 \micron.

\section{Data analysis}
The two data cubes obtained from MUSE and SINFONI can be used to extract information on the presence, distribution and dynamics of several ionised and molecular gas tracers and on the stellar population properties and kinematics. In this section we describe the methods used to extract such information in each of the data cubes.
\subsection{MUSE optical observations}
\label{sec:muse_analysis}

\subsubsection{Spatial binning}
\label{sec:binning}
The final data cube has an average signal-to-noise ratio in the continuum of S/N $\sim$ 17 per spatial pixel within the central 40 $\times$ 40 arcsec, increasing to S/N $\sim$ 150 at the AGN position. The noise is determined from the variance at $\sim 5750$ \AA, which is calculated as a by-product of the MUSE pipeline data reduction. To allow for a robust extraction of the stellar kinematics and the gas dynamics we produce several spatially binned cubes using the Voronoi binning technique described by \cite{cappellari&copin03}. The Voronoi binning technique consists of adaptive binning, where the size of the bins is not fixed but adapted to the signal-to-noise measured locally. Spatial regions with insufficient signal-to-noise ratio are therefore grouped and the data within those bins is averaged. This results in a loss of spatial resolution but an increase in the signal-to-noise ratio. Pixels with virtually no signal were masked out before the binning. We also masked out a circular region around the foreground star present in the MUSE field-of-view.

\subsubsection{Stellar kinematics}
\label{sec:kinematics}
We derive the gas and stellar kinematics by identifying and modelling the gas emission lines and the stellar absorption lines. 
To determine the stellar kinematics we use the Penalized Pixel-Fitting (pPXF) method (\citealt{vandermarel&franx93}, \citealt{cappellari&emsellem04}). This method uses a set of spectral templates and the observed galaxy spectra to determine the stellar line of sight velocity distribution (LOSVD) at each spatial position of the data cube. The LOSVD is parameterised in the form of a Gauss-Hermite series expansion \citep{vandermarel&franx93}.
 
Our spectral templates consist of a polynomial curve to account for the AGN emission contribution and a set of stellar templates from the MILES single stellar population model library (FWHM $= 2.5$ \AA, \citealt{sanchez-blazquez06}, \citealt{falcon-barroso11}) broadened to the spectral resolution of our spectra. For an accurate extraction of the stellar kinematics, pPXF is used on a data cube that has been binned to reach a high signal-to-noise of S/N $\sim$ 100 per pixel in the continuum, following the method described in Section~\ref{sec:binning}. We fit the observed wavelength range 4750 - 7450 \AA\, after masking wavelength regions of strong emission lines.
pPXF iterates to find the best-weighted template and line of sight velocity distribution parameters to fit the observed spectra at every spatial bin.  We recover the first four moments of the LOSVD ($V$, $\sigma$, $h3$, $h4$). We estimate the errors with Monte Carlo simulations. We perturb the input data by adding random noise to the spectra, with similar properties to the noise of our data, and by changing the wavelength limits for the fit and the initial guesses for the velocity and velocity dispersion. We do 100 realisations of the stellar kinematics fit and determine the errors from the standard deviation of the results. The errors in the mean velocity ($V$) and velocity dispersion ($\sigma$) are typically 5 km\,s$^{-1}$ in the nucleus and increase to $\sim$ 10 - 15 km\,s$^{-1}$ at a distance of 5 - 10 arcsec from the nucleus due to the lower S/N in the stellar absorption lines. The typical errors are $<0.03$ for $h3$ and $h4$.

We calculate the galaxy systemic velocity using the method of \cite{krajnovic06} implemented in the routine \textsc{fit\_kinematic\_pa} as used by \cite{cappellari07}. This method uses the stellar velocity field to determine the global kinematic position angle of the galaxy: PA = 118.0$\pm 1.2$ degrees and the systemic velocity: V$_{sys} = 7833$ km\,s$^{-1}$. The gas and stellar velocities quoted henceforth are relative to this systemic velocity.

\subsubsection{Gas dynamics}
\label{sec:gas_analysis}

To determine the gas dynamics, we use \textsc{gandalf}, a Gas AND Absorption Line Fitting algorithm (\citealt{falcon-barroso06}, \citealt{sarzi06}). The algorithm fits the stellar continuum and emission lines simultaneously to ensure an accurate determination of the kinematics, as some of the emission lines overlap in wavelength with stellar absorption lines. This algorithm has been used successfully in several studies of emission line kinematics (e.g. \citealt{serra08}, \citealt{krajnovic15}) obtaining similar results to other emission line fitting techniques (e.g. \citealt{humire18}).
\textsc{gandalf} uses as input the stellar templates and stellar line of sight velocity distribution determined from p\textsc{pxf} and fits the emission lines using Gaussian profiles while adjusting the amplitude and optimal combination of the stellar templates. The Gaussian profiles determined allow us to constrain the central wavelength, dispersion and flux of each emission line.

Using the method of Section~\ref{sec:binning} we bin the datacube to a minimum S/N that is high enough to obtain the stellar kinematics accurately, but still preserving the high spatial resolution in the nucleus of the galaxy. To extract the emission line properties it is not necessary to use a binned cube with S/N = 100 as in the previous section. There is nevertheless a minimum S/N required for a robust extraction of the line of sight stellar velocity distribution. S/N $\sim$ 40/\AA\ is typically required at stellar velocity dispersion values similar to those of Mrk~590 \citep{kuijken&merrifield93}. We use a data cube binned to a minimum S/N $\sim 50$ per pixel which corresponds to S/N $\sim 40$/\AA, to extract the gas dynamics. We correct the spectra for Galactic reddening, using the reddening curve parameterisation of \cite*{cardelli89} and the V-band extinction A$_{V} = 0.101$ mag, obtained using the extinction calculator of the NASA/IPAC Extragalactic Database.

Several emission lines are observed in the spectra. We identify and fit the flux, velocity and velocity dispersion of the following transitions: H$\beta$ $\lambda$4861, [OIII] $\lambda$$\lambda$4959, 5007, [NI] $\lambda$5198, [NI] $\lambda$5200, He I $\lambda$5876, [OI] $\lambda$6300,6364, H$\alpha$ $\lambda$6563, [NII] $\lambda$$\lambda$6548,6583 and [SII] $\lambda$$\lambda$6716,6731. 
Interestingly, in the nucleus of the galaxy, H$\alpha$ and H$\beta$ (and to a lower significance, He I) show a broad component. These broad lines are associated with the AGN and will be discussed in detail in Section~\ref{sec:agn_spec}.
Due to the small size of the broad line region where the broad lines are produced (typically a few light days in size), this emission is spatially unresolved in our observations. 
The broad emission flux peaks at the AGN position and has a spatial extent that is consistent with the PSF of the observations. We use the peak of this emission to identify the AGN position.

\begin{figure}
\centering
\includegraphics[width=0.45\textwidth]{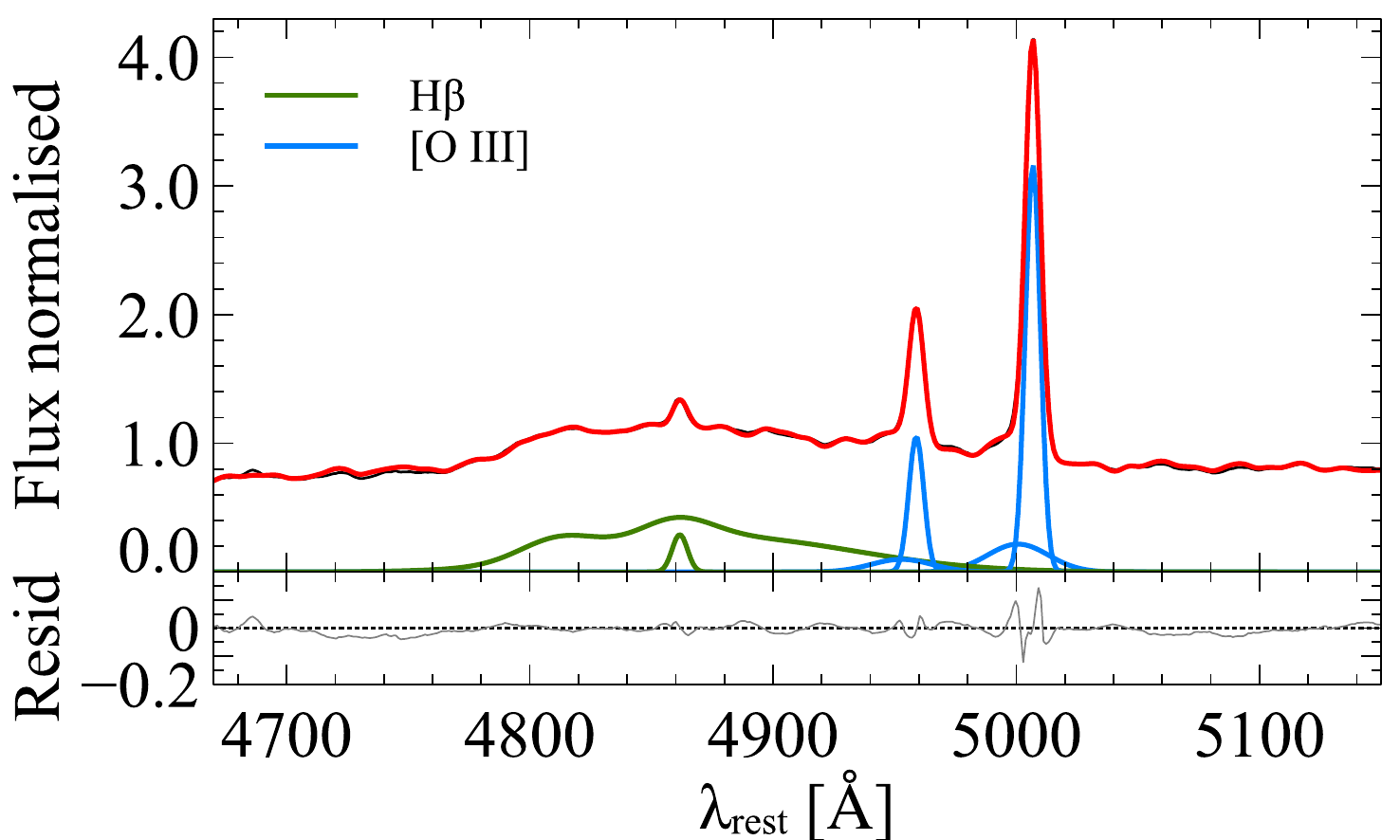}\\
\includegraphics[width=0.45\textwidth]{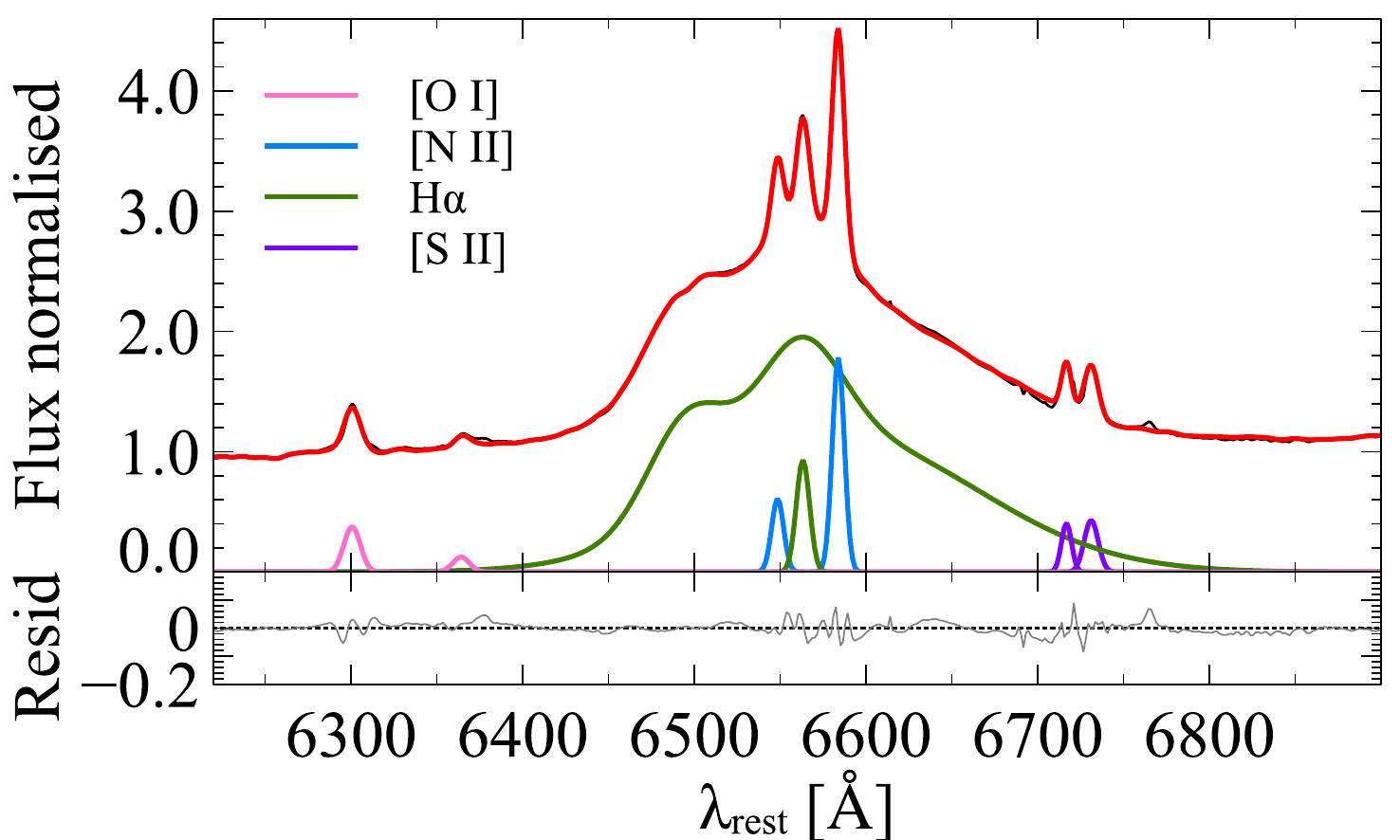}
\caption{Example of the emission line fitting procedure in the nucleus of Mrk~590. The nuclear spectrum was extracted from a nuclear region of size $0.8 \times 0.8$ arcsec$^{2}$. The top panel shows the \textsc{gandalf} fit for the H$\beta$ spectral region while the bottom panel shows the H$\alpha$ spectral region. In each panel the galaxy spectrum is shown at the top and the residuals at the bottom. The galaxy spectrum in shown in black with the best fit model shown in red. The flux has been normalised. The individual line components are illustrated by the coloured lines. In the top panel the green lines show the broad and narrow H$\beta$ emission line components while the blue lines show the broad and narrow [O III] emission line components. In the bottom panel the broad and narrow H$\alpha$ emission line components are shown in green and the [O I], [N II] and [S II] emission lines in pink, blue and purple, respectively.}
\label{halpha_hbeta}
\end{figure}

To constrain the broad line properties we first fit a nuclear spectrum by integrating the central 0.8 $\times$ 0.8 arcsec$^{2}$ region. This region is dominated by the emission from the AGN, and includes a strong contribution from the H$\alpha$ and H$\beta$ broad emission lines. As the AGN emission is unresolved, we expect the velocity and velocity dispersion parameters of the broad lines to be the same everywhere in the field-of-view and the line flux to have a spatial distribution that reflects the PSF of the observations. We use several Gaussian components for H$\alpha$ and H$\beta$, as the line profiles are complex, with a `blue shoulder' and a red side tail. We use three Gaussian components to fit the broad emission line: a broad component centred at the same velocity as the narrow component and two extra broad components, one redshifted and one blue-shifted to account for the asymmetry observed in the broad emission line profiles. The central wavelengths and widths of the Gaussian components are constrained to be the same for both the H$\alpha$ and H$\beta$ emission lines. The fit to the H$\alpha$ and H$\beta$ wavelength region in the nucleus of the galaxy is shown in Fig~\ref{halpha_hbeta}. The nuclear spectrum shown in the figure was extracted from a region of size $0.8 \times 0.8$ arcsec$^{2}$. The top panel of Fig~\ref{halpha_hbeta} shows the fit to the H$\beta$ line wavelength region, with the broad and narrow H$\beta$ emission line components in green and the broad and narrow [O III] emission line components in blue. The broad H$\beta$ emission line shown includes the three Gaussian broad components referred to above. The bottom panel of Fig~\ref{halpha_hbeta} shows the fit to the H$\alpha$ emission line wavelength region, with the broad and narrow H$\alpha$ emission line components in green and the [O I], [N II] and [S II] lines in pink, blue and purple, respectively. As for H$\beta$ the broad H$\alpha$ emission line shown includes the three Gaussian broad components referred to above.

We determine the best fit central wavelength and velocity dispersion for all the broad line components in the nuclear spectrum. We hold these parameters fixed and adjust the broad line intensity when doing a pixel by pixel emission line fitting with \textsc{gandalf} throughout the field-of-view.

The Gaussian parameters that characterise the narrow emission lines are all fit independently, except for: the doublets, H$\alpha$ and H$\beta$ and the [N I], [O I] and [S II] lines. For doublets we require each component to share the same kinematics and have relative strengths fixed by the theoretical ratio of the transition probabilities. We assume [OIII] 4959/ [O III] 5007 = 0.33 \citep{storey&zeippen00} and [N II] 6583/ [N II] 6548 = 2.95 (e.g. \citealt{acker89}).
The central velocity of H$\alpha$ and H$\beta$ are tied. The central velocities of the [N I], [O I] and [S II] lines are tied to the central velocity of H$\alpha$, due to their lower S/N.
We set a conservative threshold for the detection of lines, and mask out all the pixels for which the ratio between the amplitude of the emission line (A) and the noise in the spectrum (N) is lower than 5: A/N $<$ 5 \citep{sarzi06}. The velocity dispersion values are corrected in quadrature for the instrumental broadening. We estimate the errors in the velocities with Monte Carlo simulations as in Section~\ref{sec:kinematics}. The typical errors in the line of sight velocity and velocity dispersion are 3 km\,s$^{-1}$ and 5$-$10 km\,s$^{-1}$, respectively. 

\subsection{SINFONI near-infrared observations}

\begin{figure}
\centering
\hspace{-0.1cm}\includegraphics[width=0.45\textwidth]{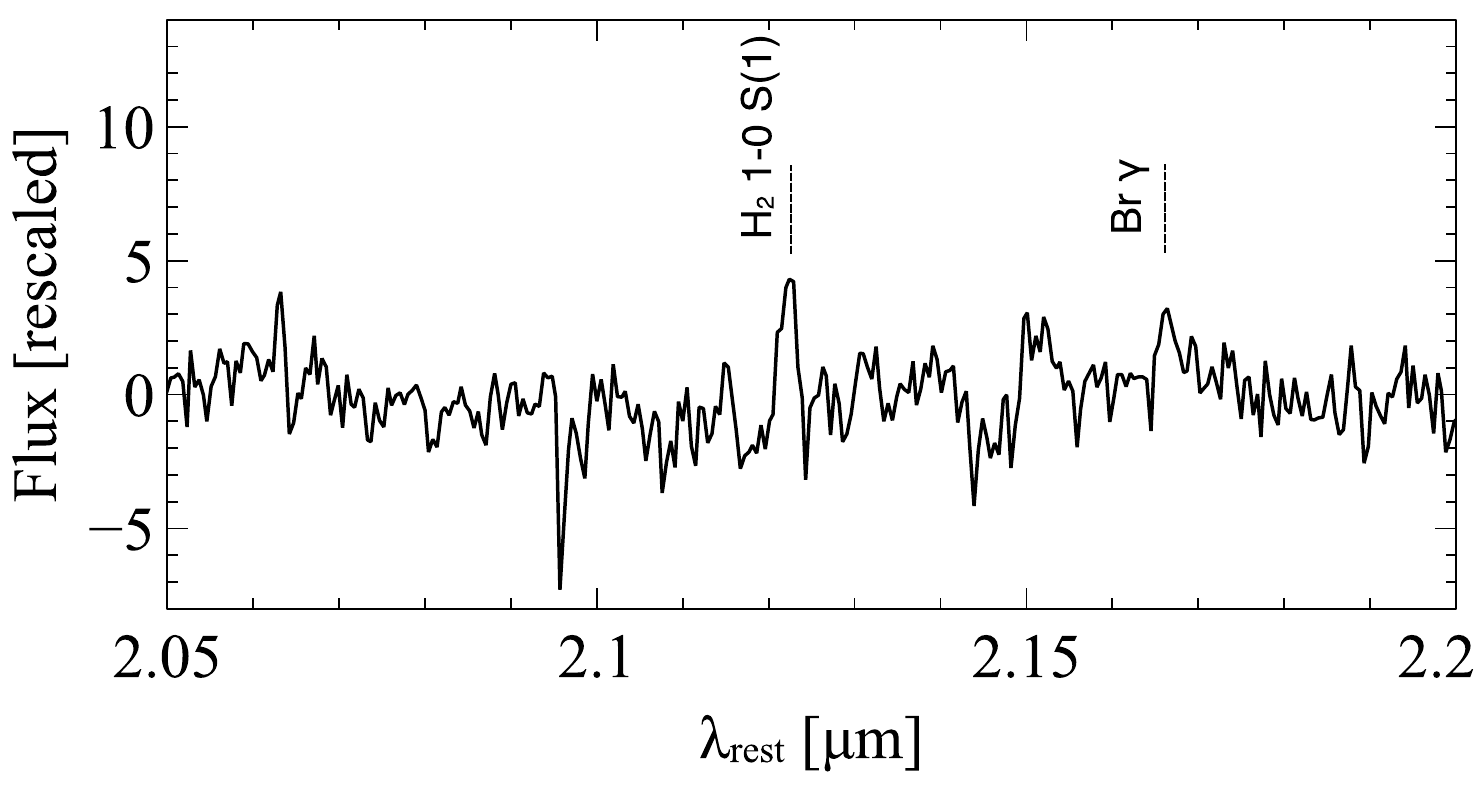}
\caption{Nuclear K-band spectrum obtained from the mean of the spectra within the central 0.5 $\times$ 0.5 arcsec$^{2}$. The spectrum has been spectrally binned by a factor of 2 and rescaled for visualisation purposes by subtracting the mean flux value. The H$_{2}$ 1-0 S(1) 2.12 $\micron$ emission that traces warm molecular gas is observed in the spectrum.}
\label{mean_spec_sinfoni}
\end{figure}

The S/N per pixel in the SINFONI data cube is relatively low which requires us to bin the data cube spatially prior to carrying out the analysis. We start by examining the integrated nuclear spectrum. In Fig.~\ref{mean_spec_sinfoni} we show the mean nuclear spectrum in the central 0.5 $\times$ 0.5 arcsec$^{2}$. The spectrum has been spectrally binned by a factor of 2 and rescaled by subtracting the mean flux value. The ro-vibrational transition H$_{2}$ 1-0 S(1) at 2.12 $\mu{m}$, which traces the warm molecular gas distribution, is clearly detected albeit at a low signal-to-noise.

\begin{figure}
\hspace{-1cm}\includegraphics[width=0.6\textwidth]{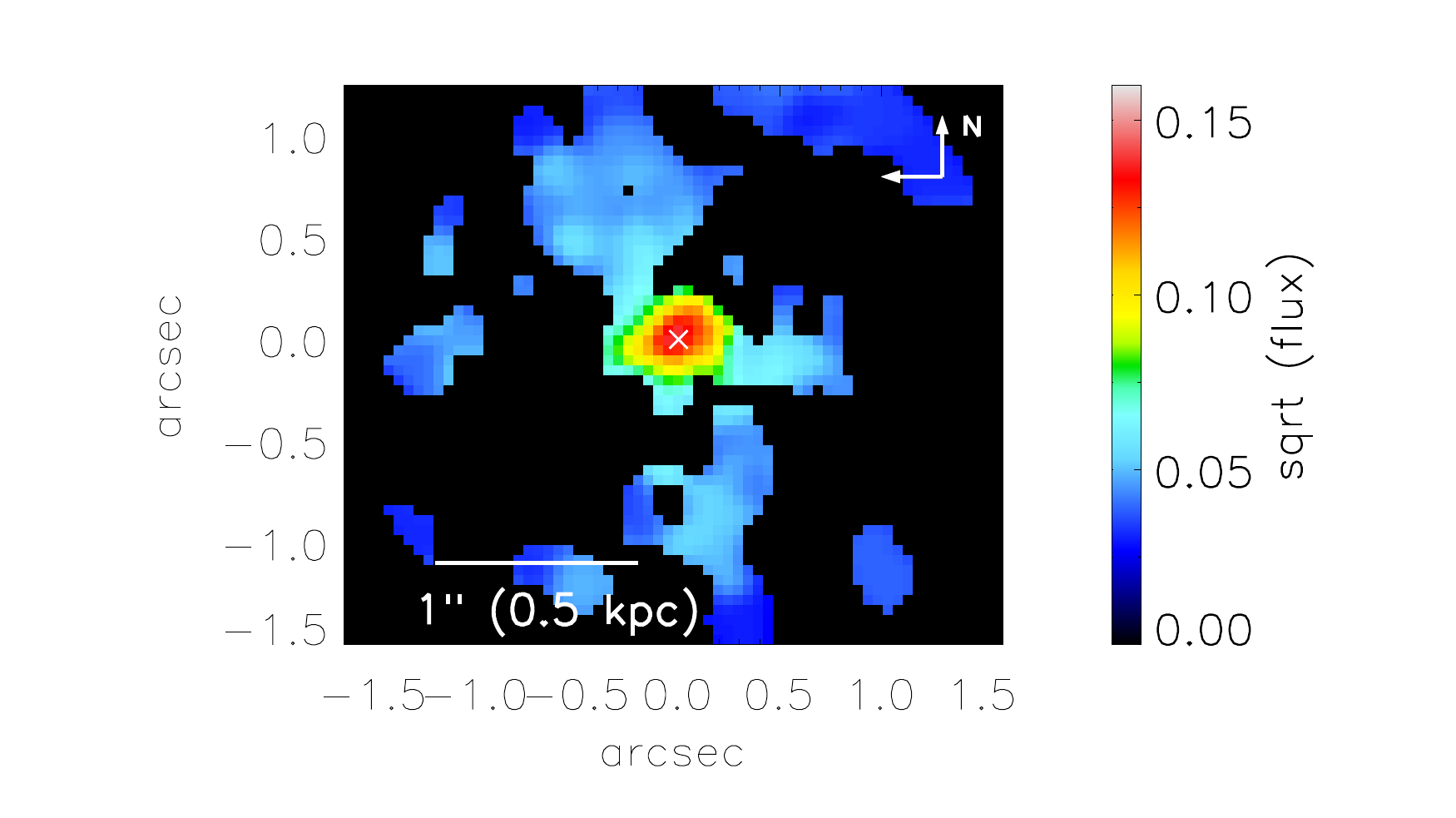}
\caption{H$_{2}$ 1-0 S(1) 2.12 $\micron$ line flux map. The white cross indicates the likely position of the AGN, identified as the peak in the Br$\gamma$ narrow emission.}
\label{h2_flux}
\end{figure}

We bin the cube to a minimum S/N $\sim 20$ in the continuum using the Voronoi binning technique described in Section \ref{sec:binning}, before fitting the H$_{2}\,2.12$ $\micron$ line. The H$_{2}\,2.12$ $\micron$ line is detected at a S/N $\sim$ 5 per bin in the nucleus and at S/N $\sim$ 2 per bin in an extended region towards the North-South direction and to the West.
In Fig.~\ref{h2_flux} we show the H$_{2}$ 2.12 $\micron$ line flux map. We do not determine the H$_{2}$ kinematics due to the low signal-to-noise of the data. 
Regions where the emission line peak is less than 2$\sigma$ above the continuum emission noise were masked out of the map.
The central point, marked with the cross corresponds to the peak of the narrow Br$\gamma$ emission which, in the absence of other signatures, we assume indicates the AGN position. 

\section{Results and Discussion}

\begin{figure*}
\centering
\includegraphics[width=0.6\textwidth]{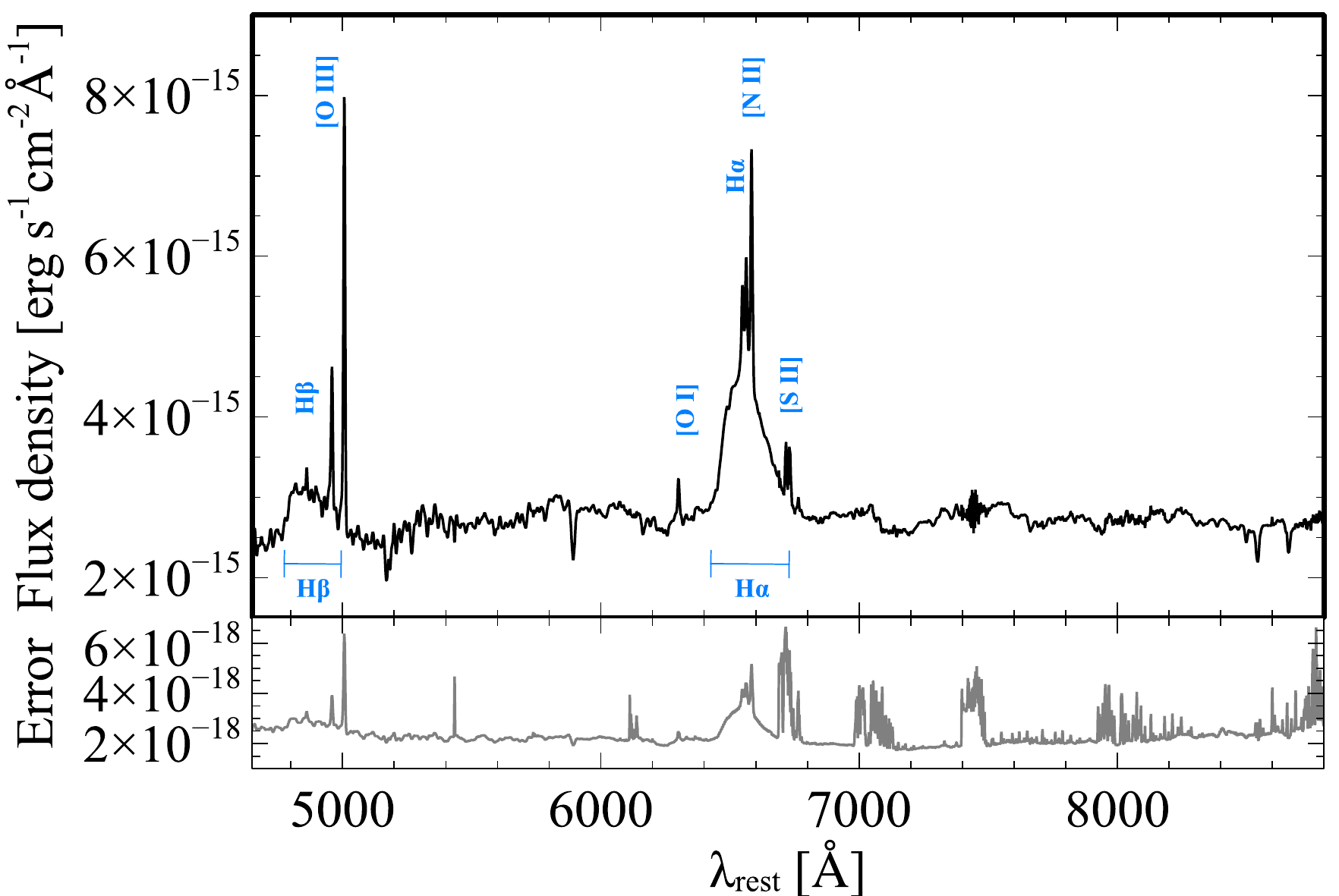}
\caption{Nuclear integrated spectrum of Mrk~590 extracted using a circular aperture of 3 arcsec diameter centred at the AGN position. The top panel shows the spectrum and the bottom panel the associated flux density uncertainties.}
\label{slit_spec_2003}
\end{figure*}

\begin{figure*}
\centering
\includegraphics[width=0.6\textwidth]{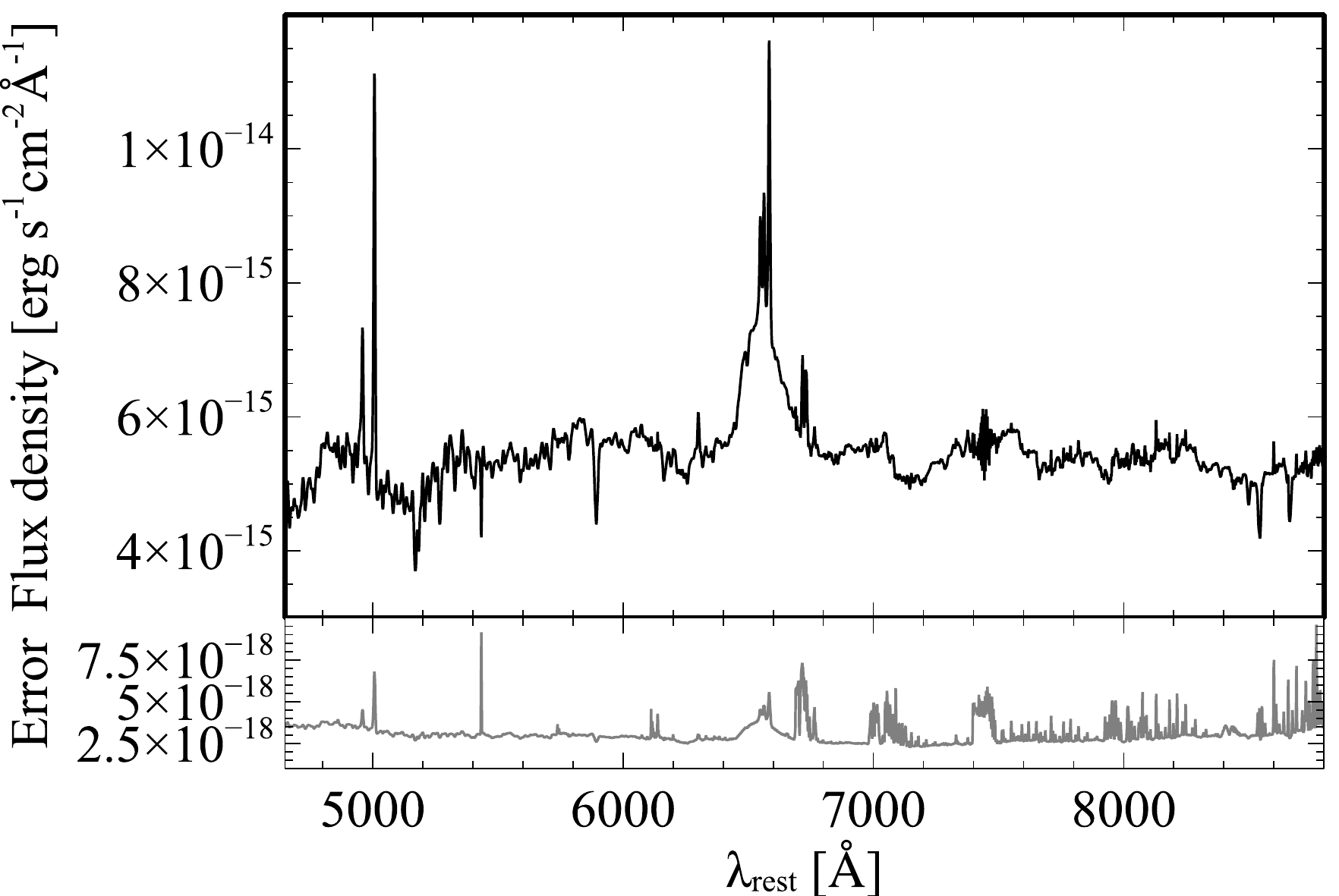}
\caption{Nuclear integrated spectrum of Mrk~590 extracted using a pseudo-slit of 5 arcsec by 7.6 arcsec centred at the AGN position and along PA = 0 deg to match the observations carried out in 1996 and presented by \citealt{peterson98}. The top panel shows the spectrum and the bottom panel the associated flux density uncertainties.}
\label{slit_spec_nuc}
\end{figure*}

\subsection{AGN spectrum}
\label{sec:agn_spec}

Previous to our MUSE observations, optical spectra of Mrk~590 indicated that the optical broad emission lines (H$\alpha$ and H$\beta$) had disappeared \citep{denney14}. This was a dramatic change from the higher AGN luminosity state in the 1990s when the broad emission lines and AGN continuum were clearly observed. Our MUSE observations show that by 2017 October 28 the optical broad emission lines had reappeared, which indicates a second AGN type change after a period of more than 10 years without the presence of optical broad emission lines. Fig.~\ref{slit_spec_2003} shows the integrated nuclear spectrum extracted from a 3 arcsec diameter region. The broad H$\beta$ and broad H$\alpha$ lines can be clearly identified. Narrow components of H$\beta$ and H$\alpha$ are also detected, in addition to strong [O III] and [N II] lines. To directly compare our measurements with those carried out during previous flux states, we extract nuclear spectra using pseudo slits that match previous observations. We extract a nuclear spectrum using a pseudo slit of 5 arcsec width and 7.6 arcsec aperture, chosen to match the slit used to obtain Mrk~590's spectrum in 1996 \citep{peterson98}. We also determine the integrated spectrum to match the aperture used in the 2014 measurement with the 1.3m McGraw-Hill telescope at the MDM Observatory \citep{denney14}. According to \cite{denney14}, out of the epochs analysed in their paper, the 1996 and 2014 spectra have two of the best spectrophotometric calibrations. 
In Fig.~\ref{slit_spec_nuc} we show the spectrum using the pseudo slit of 5 arcsec width and 7.6 arcsec aperture. 

With our observations in 2017 we measure a total 5100 \AA\ flux of F = 4.5$\times 10^{-15}$ erg s$^{-1}$ cm$^{-2}$ \AA$^{-1}$ in the aperture corresponding to the 1996 measurement. This flux is almost two times lower than what was measured in 1996 ($\sim 8$$\times$ $10^{-15}$ erg s$^{-1}$ cm$^{-2}$ \AA$^{-1}$, \citealt{denney14}). With an aperture similar to that used in 2014, the flux we measure is just slightly higher (F = 5.5$\times 10^{-15}$ erg s$^{-1}$ cm$^{-2}$ \AA$^{-1}$) than the flux level measured in 2014 (F$\sim$5$\times 10^{-15}$ erg s$^{-1}$ cm$^{-2}$ \AA$^{-1}$, \citealt{denney14}). This indicates that although the optical broad emission lines are now observed in Mrk~590, the AGN has not returned to the luminous state observed in the 1990s. To a first order, the increase in flux is due to the AGN flux increase, as the host galaxy flux is expected to be constant. Comparing the 5100 \AA\ optical flux we observe, with the historical values in \cite{denney14}, it appears that the AGN is at a flux level close to that observed in 2003 and still $\sim$ 10 times lower than that observed in the 1990s. It also shows that the production of broad optical emission lines is not necessarily accompanied by a significant increase in the AGN optical continuum flux. The broad lines re-appeared between 2014 and 2017 while there was only a slight increase in optical continuum flux. 
Our continued monitoring campaign shows that the optical broad emission lines are present for at least 3 months (the full duration of the spectral monitoring campaign) while the optical continuum flux is at a similar low level. Since the measured time delay between continuum and broad line region variations is 19 days \citep{peterson93a}, which is significantly shorter than the duration of our spectral monitoring, we can conclude that the optical continuum flux is indeed at a low level while the broad lines are being produced. A full analysis of the monitoring data will be presented in an upcoming publication.

Standard AGN accretion disc emission would produce a characteristic optical-ultraviolet `blue-bump' spectral feature, with ultraviolet emission being produced in the inner region of the accretion disc and optical emission being produced at larger radii in the accretion disc (e.g. \citealt{raimundo12}). The observations of optical broad emission lines in this work and the ultraviolet emission line and continuum measurements of \cite{mathur18} indicate that ultraviolet continuum photons are being produced by the AGN. However, standard accretion disc emission is not seen in Mrk~590, since we observe that the optical continuum has not increased significantly. This indicates that there may be an alternative mechanism to generate ultraviolet continuum emission or that if an accretion disc is present or being built up, only the ultraviolet emission from the inner radii is currently being produced or able to reach the observer.

The near-infrared data we analyse here are from October 2007. As can be seen in Fig.~\ref{mean_spec_sinfoni}, there is no evidence of strong broad Br$\gamma$ emission. 
Slightly earlier and later spectroscopy from 2006 and 2009 \citep{muller-sanchez18} also do not detect broad Br$\gamma$ emission. 
Optical spectroscopy from 2006 shows only a weak broad H$\alpha$ component \citep{denney14} which by 2013 had disappeared. The near-IR data are likely sampling the transition period during which the broad emission lines disappeared.

\subsubsection{The re-appearance of the broad emission lines}
With our recent observations we have determined that Mrk~590 is one of the few AGN that had broad emission lines disappear and reappear within a timescale of years to decades.
A similar behaviour has also been observed in Mrk 1018 (\citealt{mcelroy16}, \citealt{husemann16}). We compare our nuclear spectrum with the compilation shown by \cite{denney14}. In the spectra taken in or after 2013 there are no significant signatures of the H$\beta$ or H$\alpha$ broad components. In our data from 2017 October the broad components are clearly seen and have an asymmetric line shape similar to what was observed in 2003. This implies that the optical broad lines disappeared within a time span of 7 years between 2006 and 2013, and have reappeared within a time span of 4 years, sometime between 2013 and 2017. We note that narrow H$\beta$ is also clearly visible in the nuclear spectrum we obtain from the MUSE data (Fig.~\ref{slit_spec_2003}), in contrast with the spectrum taken between 2006 and 2013 shown in \cite{denney14}, where, at most, a weak signature of narrow $H\beta$ is visible. 
This is likely due to a combination of factors: the aperture used to extract the spectra in those observations and the fact that the AGN flux has increased in 2017 and the $H\beta$ line is now more clearly seen in contrast with the galaxy emission, for a given sensitivity. Narrow H$\beta$ may have been present in the 2006 - 2013 spectra but not significantly detected due to the factors mentioned above.

Comparing the spectra from 2003 and earlier (seen in \citealt{denney14}) with our nuclear spectrum (Fig.~\ref{slit_spec_2003}), we can see a variation in the shape of the broad emission lines as a function of epoch, which is associated with a change in physical properties or dynamics in the broad line region. In a future publication we will compare in detail the line shapes and AGN fluxes for different epochs based on our monitoring of Mrk~590.

\subsection{Stellar kinematics}
\label{sec:stellar_kinematics}
\begin{figure*}
\centering
\includegraphics[width=0.48\textwidth]{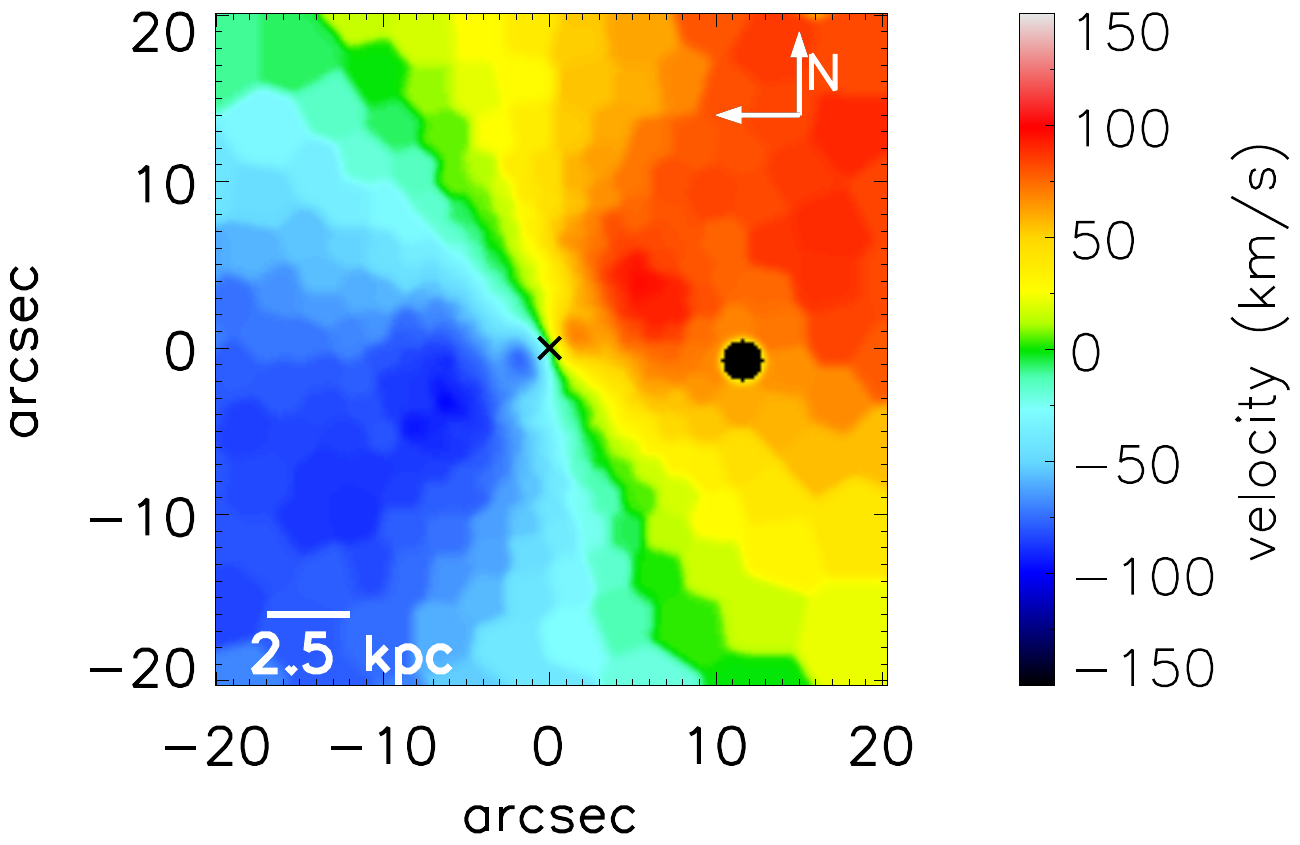}\hspace{0.2cm}
\includegraphics[width=0.48\textwidth]{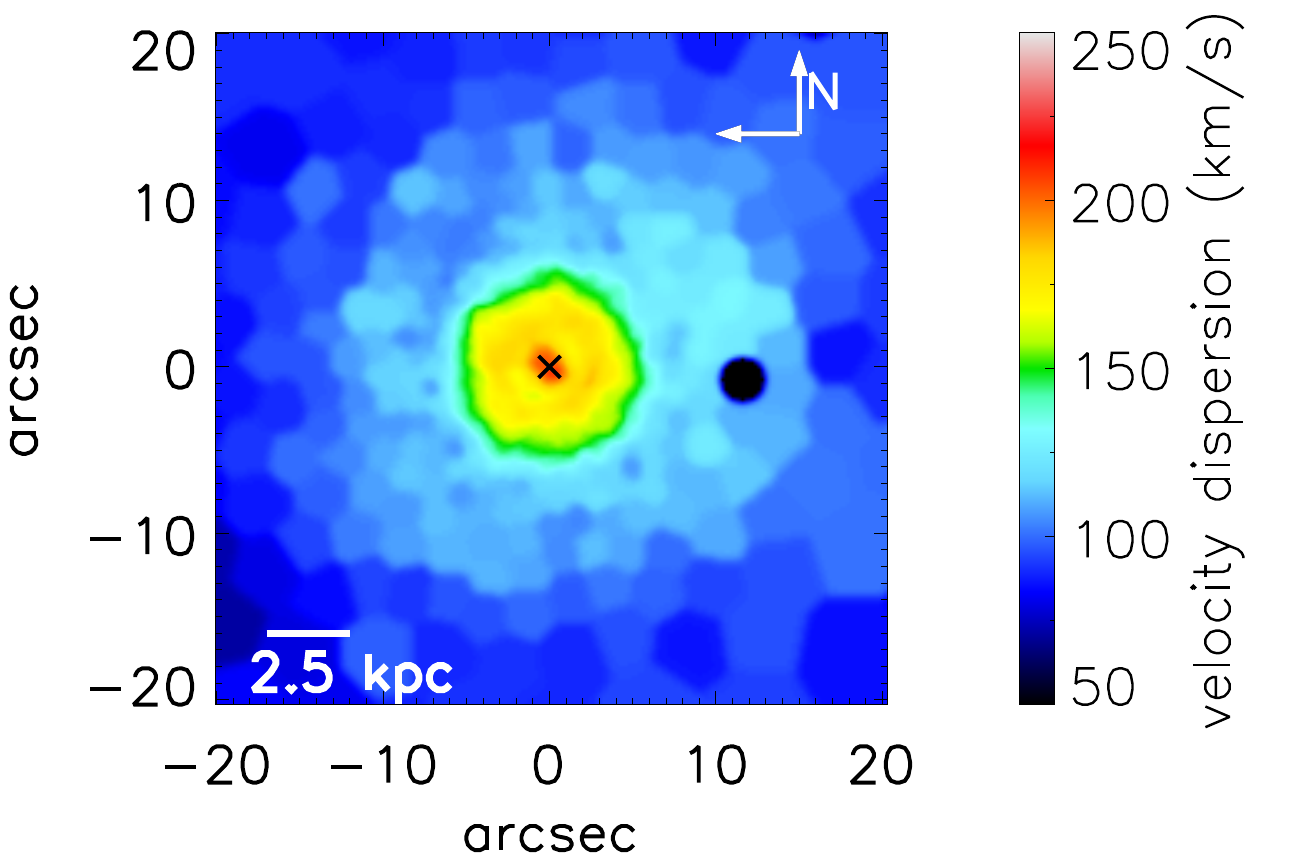}\\
\includegraphics[width=0.48\textwidth]{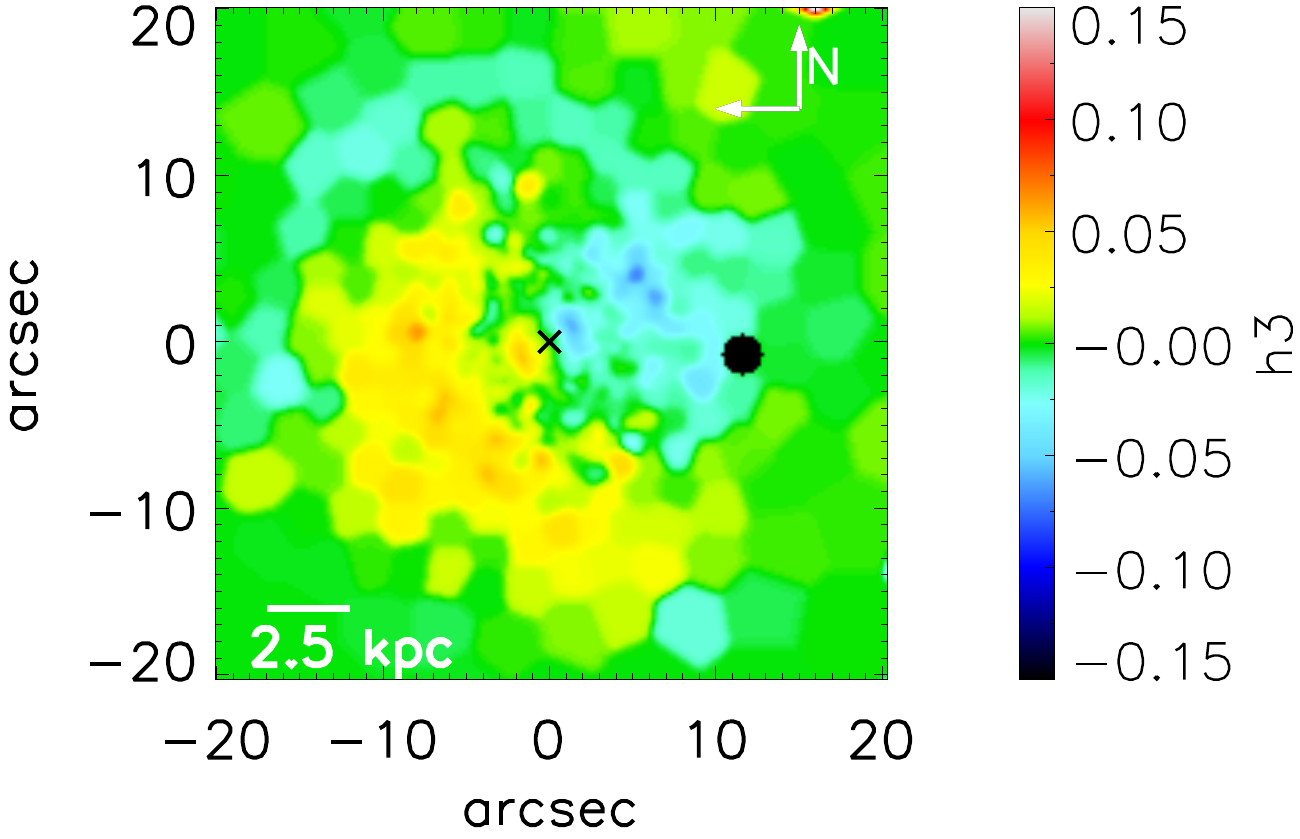}\hspace{0.2cm}
\includegraphics[width=0.48\textwidth]{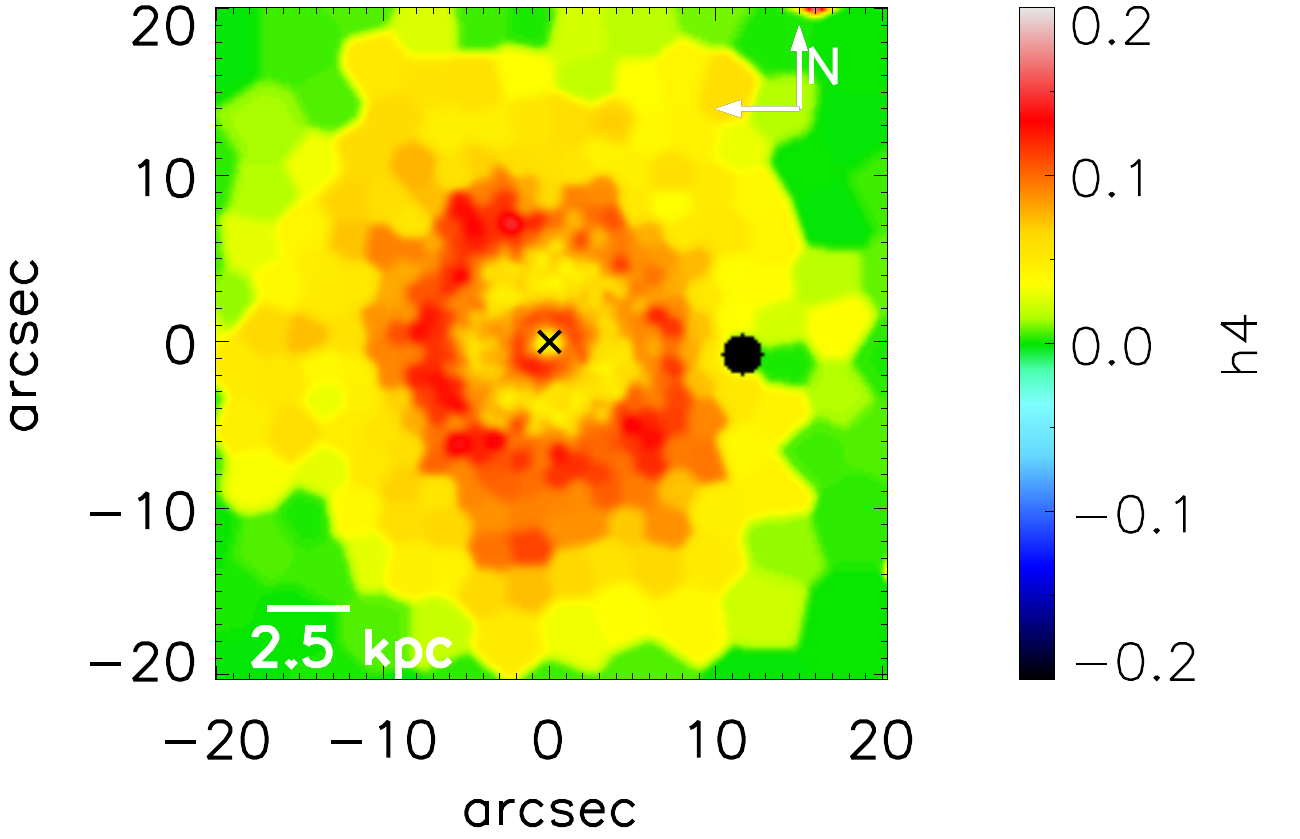}
\caption{Moments of the stellar line of sight velocity distribution. Top row: mean velocity $V$ and velocity dispersion $\sigma$ in km\,s$^{-1}$. Bottom row: Gauss-Hermite coefficients $h3$ and $h4$ indicating respectively asymmetric and symmetric deviations from a Gaussian line shape. We observe a velocity field (top left panel) characteristic of ordered rotation with increased velocity dispersion (top right panel) in the central 10 arcsec of the galaxy. The mean velocity has been corrected for the systemic velocity V$_{sys}$ = 7833 km\,s$^{-1}$ and the velocity dispersion has been corrected for the instrumental broadening. The measurement errors are discussed in the text. The black cross indicates the position of the AGN. The black circle to the West of the nucleus corresponds to the position of a foreground star that has been masked out of the image.}
\label{stellar_losvd}
\end{figure*}

\begin{figure*}
\centering
\includegraphics[width=0.48\textwidth]{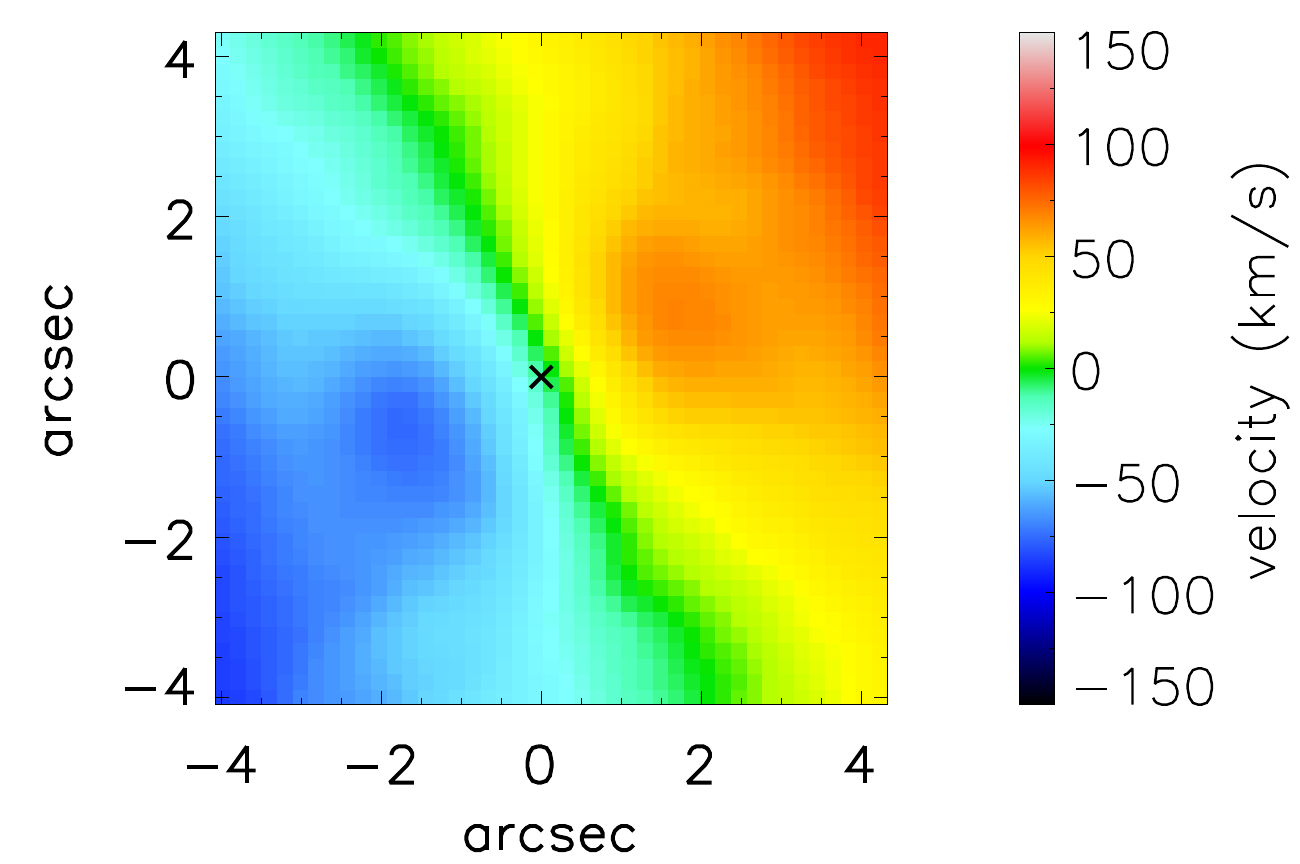}\hspace{0.2cm}
\includegraphics[width=0.48\textwidth]{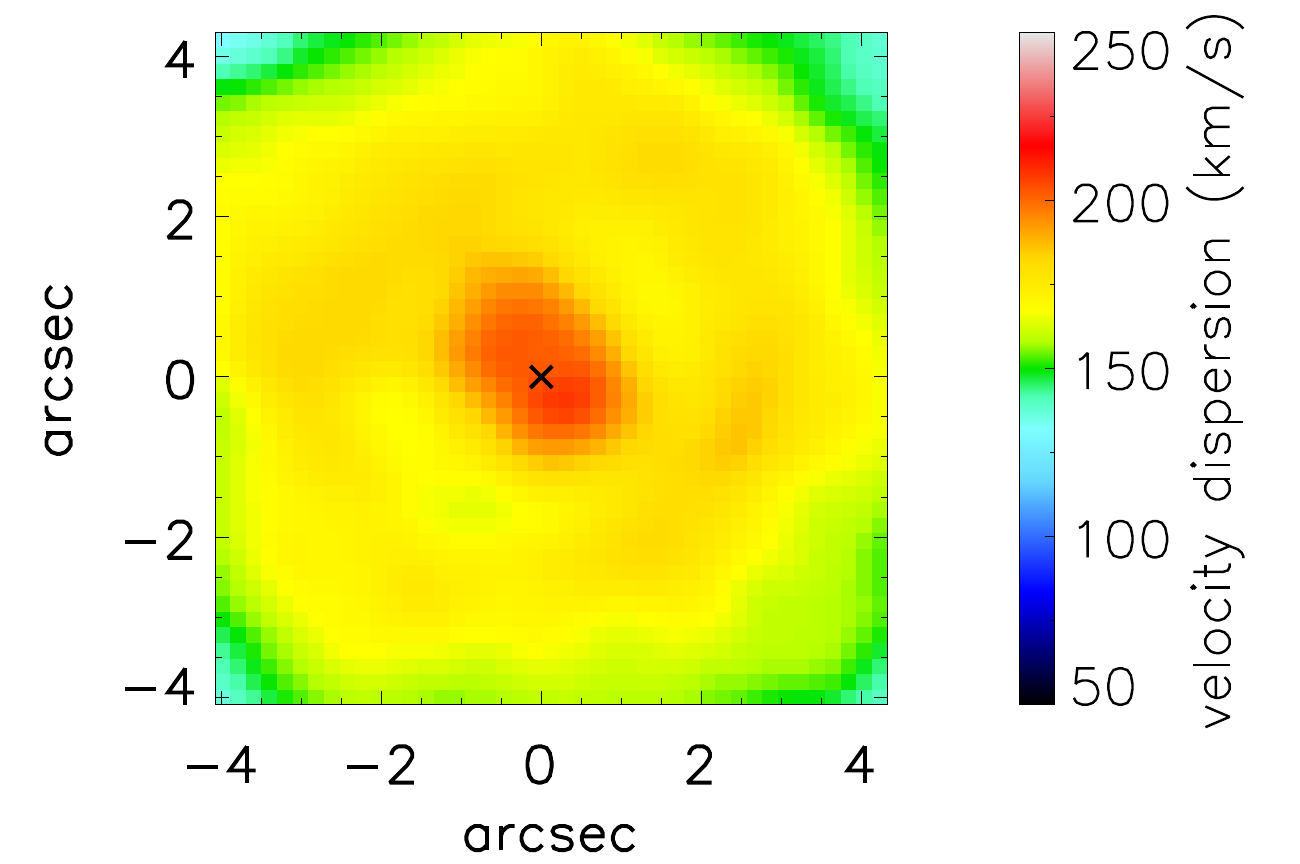}\\
\includegraphics[width=0.48\textwidth]{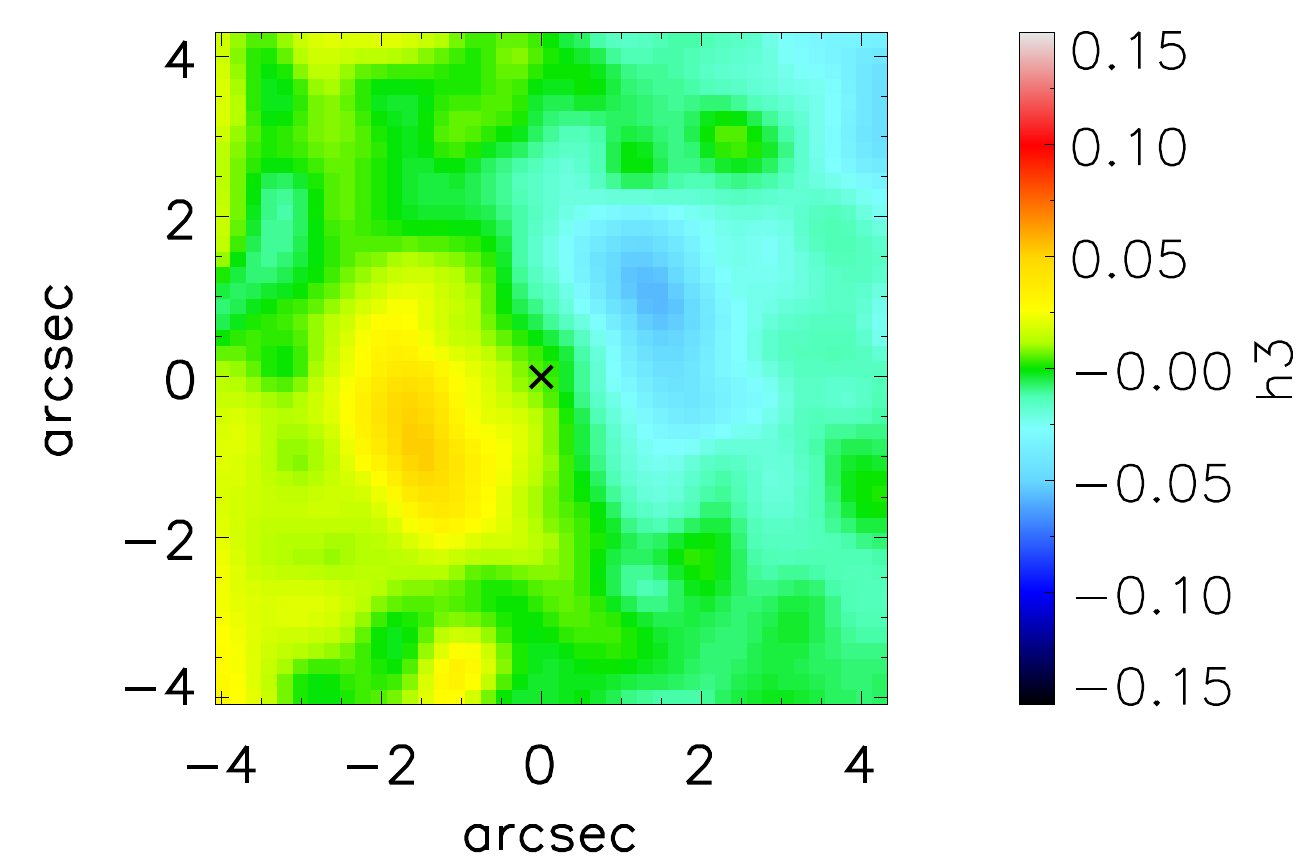}\hspace{0.2cm}
\includegraphics[width=0.48\textwidth]{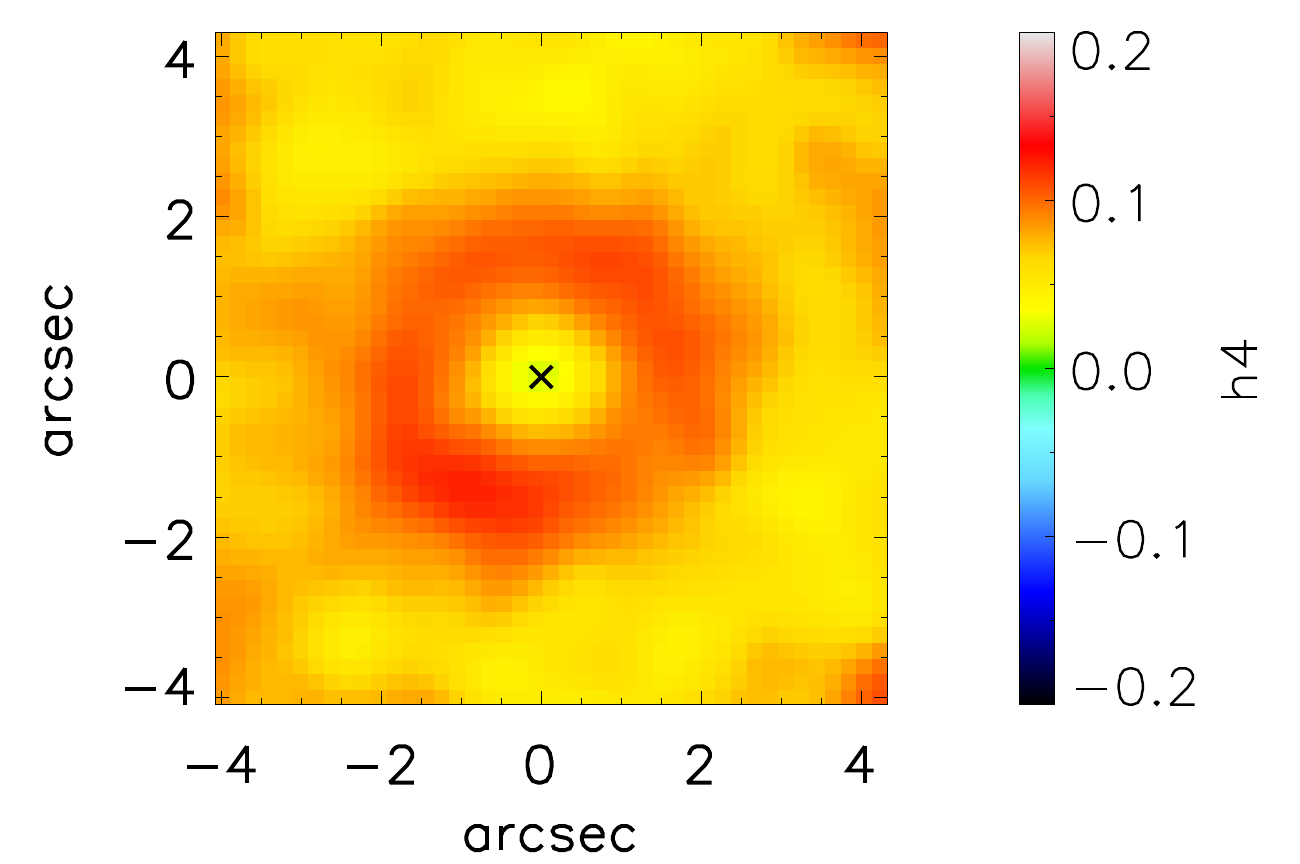}
\caption{A close-up view of the moments of the stellar line of sight velocity distribution within the central $\sim$8 arcsec. Top row: mean velocity $V$ and velocity dispersion $\sigma$. Bottom row: Gauss-Hermite coefficients $h3$ and $h4$. A drop in the velocity and increased velocity dispersion is observed at $\sim$2$-$3 arcsec from the nucleus. The $h4$ moment shows a ring of increased values at $\sim$ 1 - 2 arcsec from the nucleus corresponding to a region of lower velocity dispersion. The mean velocity has been corrected for the systemic velocity V$_{sys}$ = 7833 km\,s$^{-1}$ and the velocity dispersion has been corrected for the instrumental broadening. The black cross indicates the position of the AGN.}
\label{stellar_losvd_zoom}
\end{figure*}

We determined the first four moments of the stellar LOSVD ($V$, $\sigma$, $h3$, $h4$) using the method described in Section~\ref{sec:kinematics}. We show the results in Fig.~\ref{stellar_losvd}. A zoom-in of the nuclear stellar kinematics is shown in Fig.~\ref{stellar_losvd_zoom}.

The first two moments are the mean velocity ($V$) and velocity dispersion ($\sigma$), and the higher moments of the LOSVD ($h3$ and $h4$) are used to measure deviations of the LOSVD from a Gaussian profile \citep{vandermarel&franx93}. The moment $h3$ indicates an asymmetry in the line profile: a negative $h3$ is associated with a distribution with a tail towards velocities higher than the systemic velocity (towards longer wavelengths) while a positive $h3$ is associated with a distribution with a tail towards velocities lower than the systemic velocity (towards shorter wavelengths). The moment $h4$ indicates a symmetric deviation from a Gaussian line profile. Positive values of $h4$ indicate that the LOSVD is more peaked than a Gaussian while negative values of $h4$ indicate a more flat-topped LOSVD profile.
From Fig.~\ref{stellar_losvd} we can see that the stellar velocity shows an overall rotation pattern, with the stellar velocity dispersion increasing towards the centre of the galaxy. There are however some perturbations, as the velocity and velocity dispersion are not monotonic but show a wavy pattern as a function of radius. There is a region at 3 arcsec from the centre where the mean velocity has a dip accompanied by an increase in the stellar velocity dispersion (Fig.~\ref{stellar_losvd_zoom}). There is also a region at $\sim$ 10 arcsec from the centre, coinciding with a region of low ionised gas flux, where the mean stellar velocity has a local minimum. 
A wavy pattern in the LOSVD could be caused by the contribution of retrograde orbits, created by the presence of a bar (\citealt{bettoni89}, \mbox{\citealt{wozniak&pfenniger97}}). The bar will also create dynamical resonances that can affect the stellar and gas dynamics.
There is evidence from \emph{HST} data (\citealt{pogge&martini02}, \citealt{bentz06}) and from our MUSE imaging analysis, of a bar within the central arcseconds of Mrk~590.
A detailed imaging and dynamics analysis will be done in paper II.

The value for $h3$ is relatively low, between $[-0.07, +0.07]$, meaning that the LOSVD is not strongly asymmetric. An anti-correlation between $V$ and $h3$ as observed in our data, indicates a tail of lower velocity material and can be explained by a superposition of different stellar components, for example a superposition of stars in a rotating disc and stars with lower rotation velocity from the bulge (\citealt{halliday01}, \citealt{krajnovic06}, \citealt{riffel17}). 
The high stellar velocity dispersion in the centre is likely tracing the bulge component while the lower velocity dispersion is tracing the disc dominated regions. The regions of lower velocity dispersion, outside the central 10 arcsecs and in a ring of $r \sim 3$ arcsec could be associated with kinematically colder regions that still retain the kinematics of the gas from which they formed. 
These low velocity dispersion regions also tend to show higher $h4$ ($\sim$ 0.1 - 0.15). This could be due to younger stellar populations at these locations, since recently formed stars that are still in the disc structure are expected to have more peaky LOSVD (e.g. \citealt{riffel17}). 

\subsection{Ionised gas}
\label{sec:ionised_gas}
\begin{figure*}
\begin{center}
\hspace{0.1cm}
\includegraphics[width=0.2\textwidth]{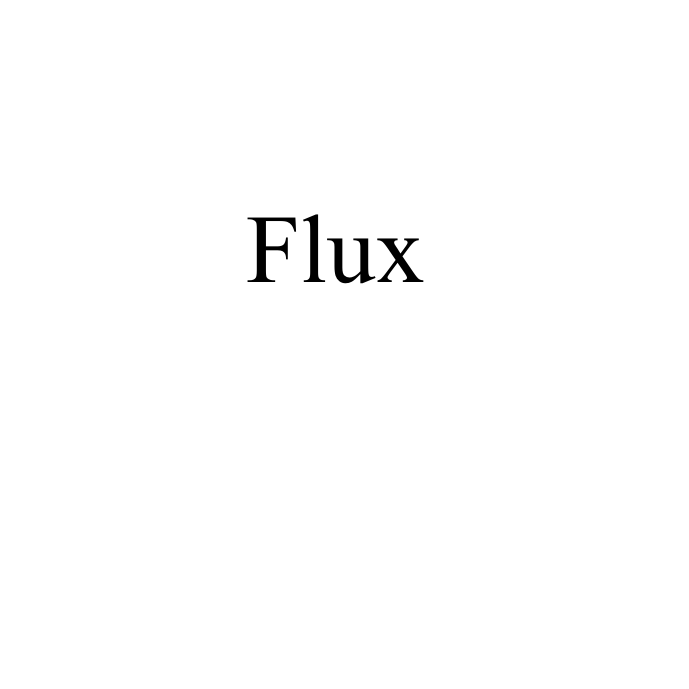}\hspace{1.0cm}
\includegraphics[width=0.2\textwidth]{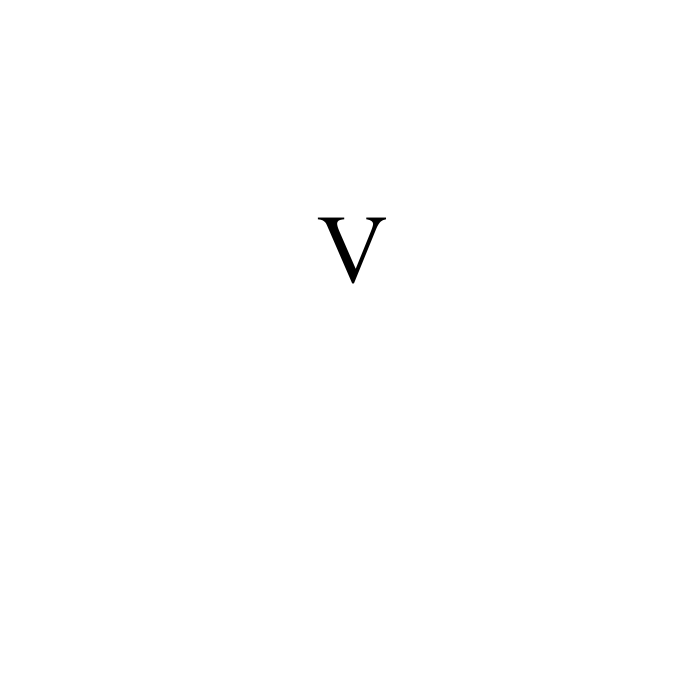}\hspace{1.7cm}
\includegraphics[width=0.2\textwidth]{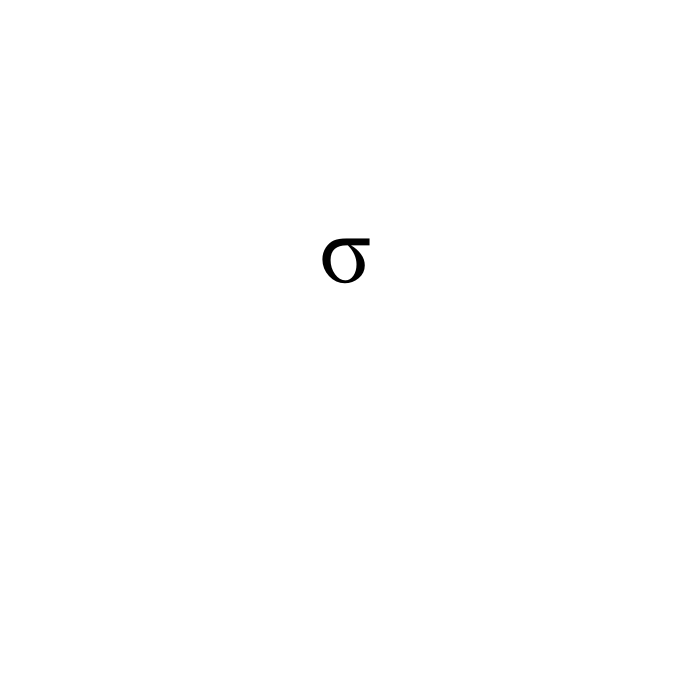}\\[-2.2cm]
\hspace{-1.0cm}
\includegraphics[width=0.2\textwidth]{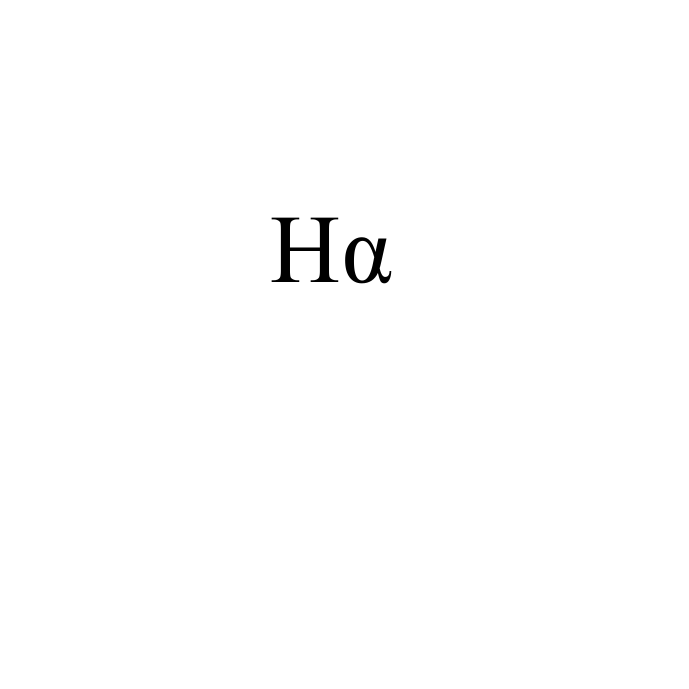}\hspace{-2.0cm}
\includegraphics[width=0.38\textwidth]{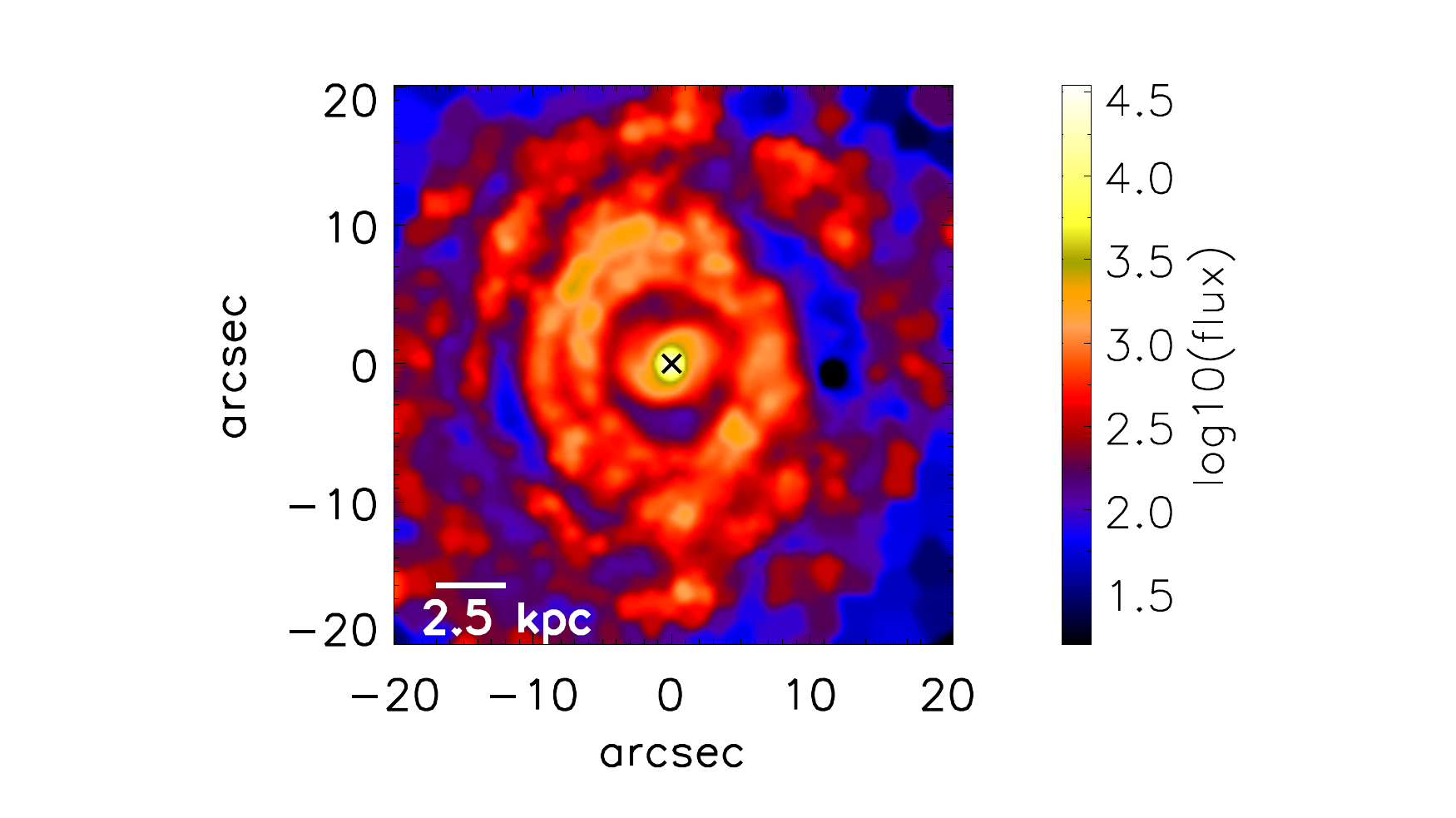}\hspace{-2.0cm}
\includegraphics[width=0.38\textwidth]{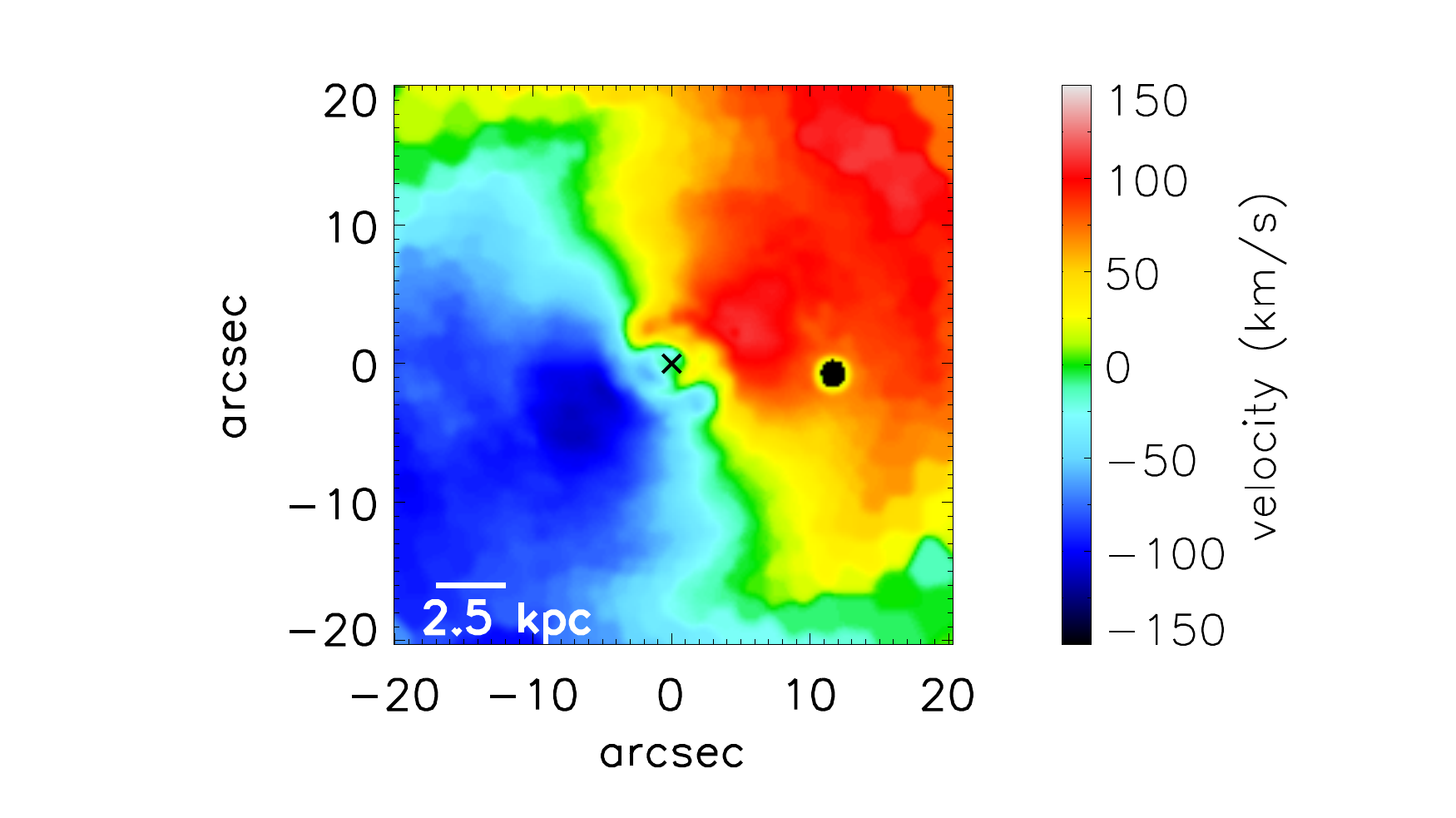}\hspace{-1.5cm}
\includegraphics[width=0.38\textwidth]{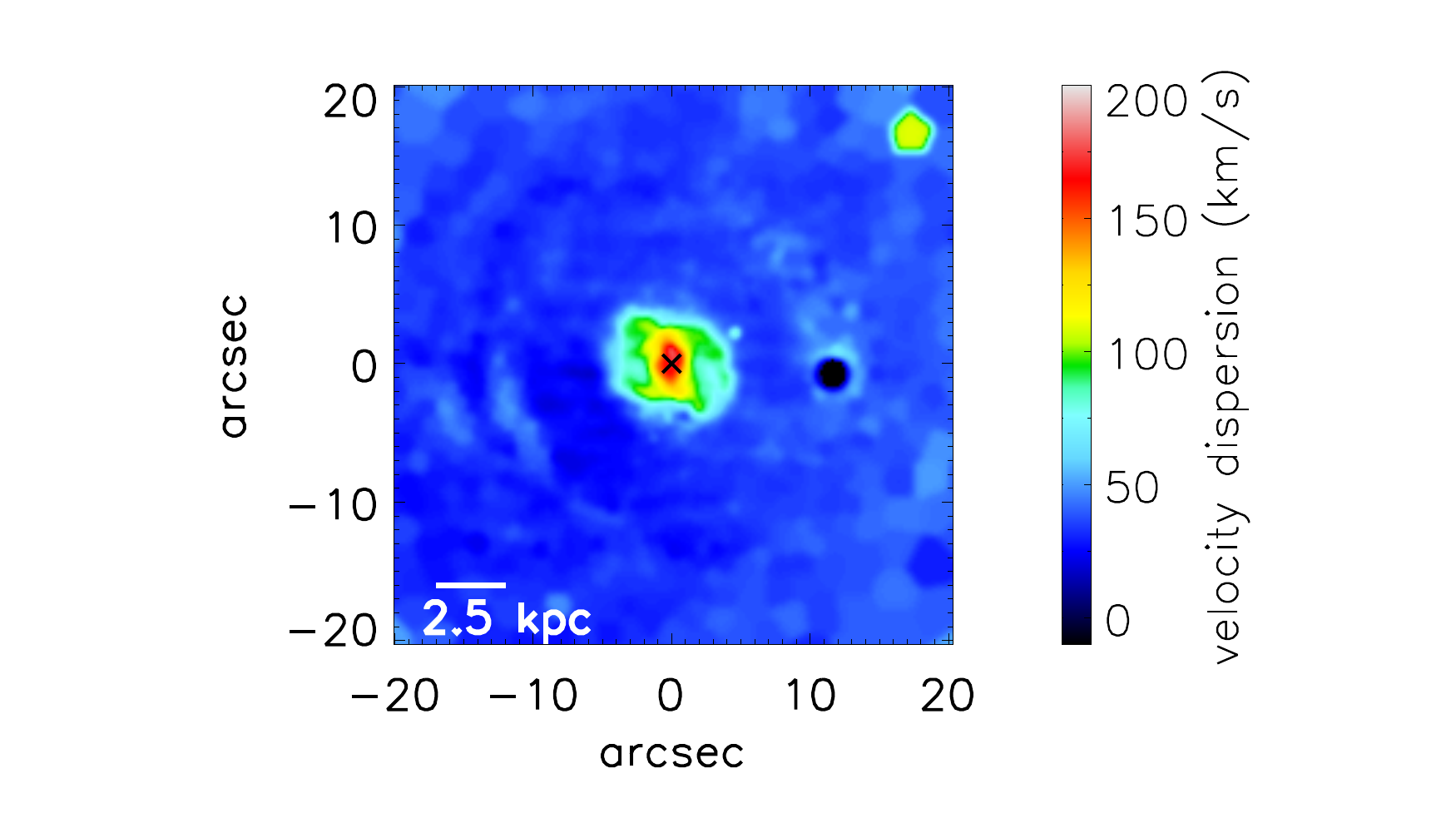}\\[-0.4cm]
\hspace{-1.0cm}
\includegraphics[width=0.2\textwidth]{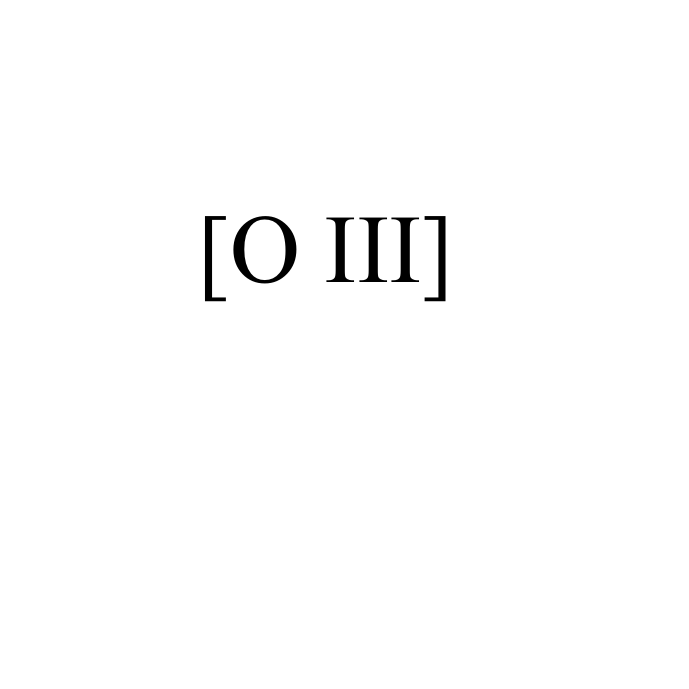}\hspace{-2.0cm}
\includegraphics[width=0.38\textwidth]{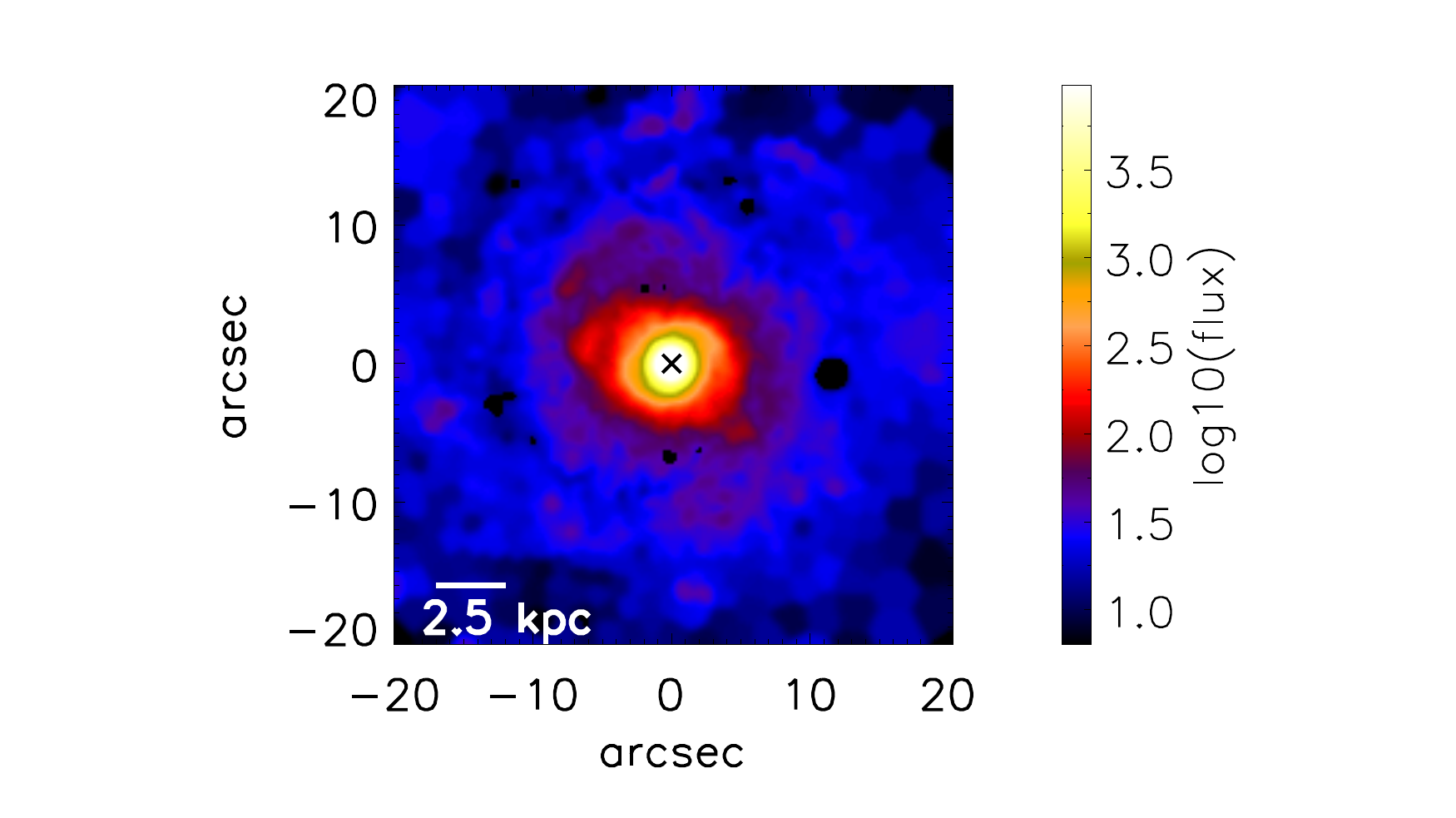}\hspace{-2cm}
\includegraphics[width=0.38\textwidth]{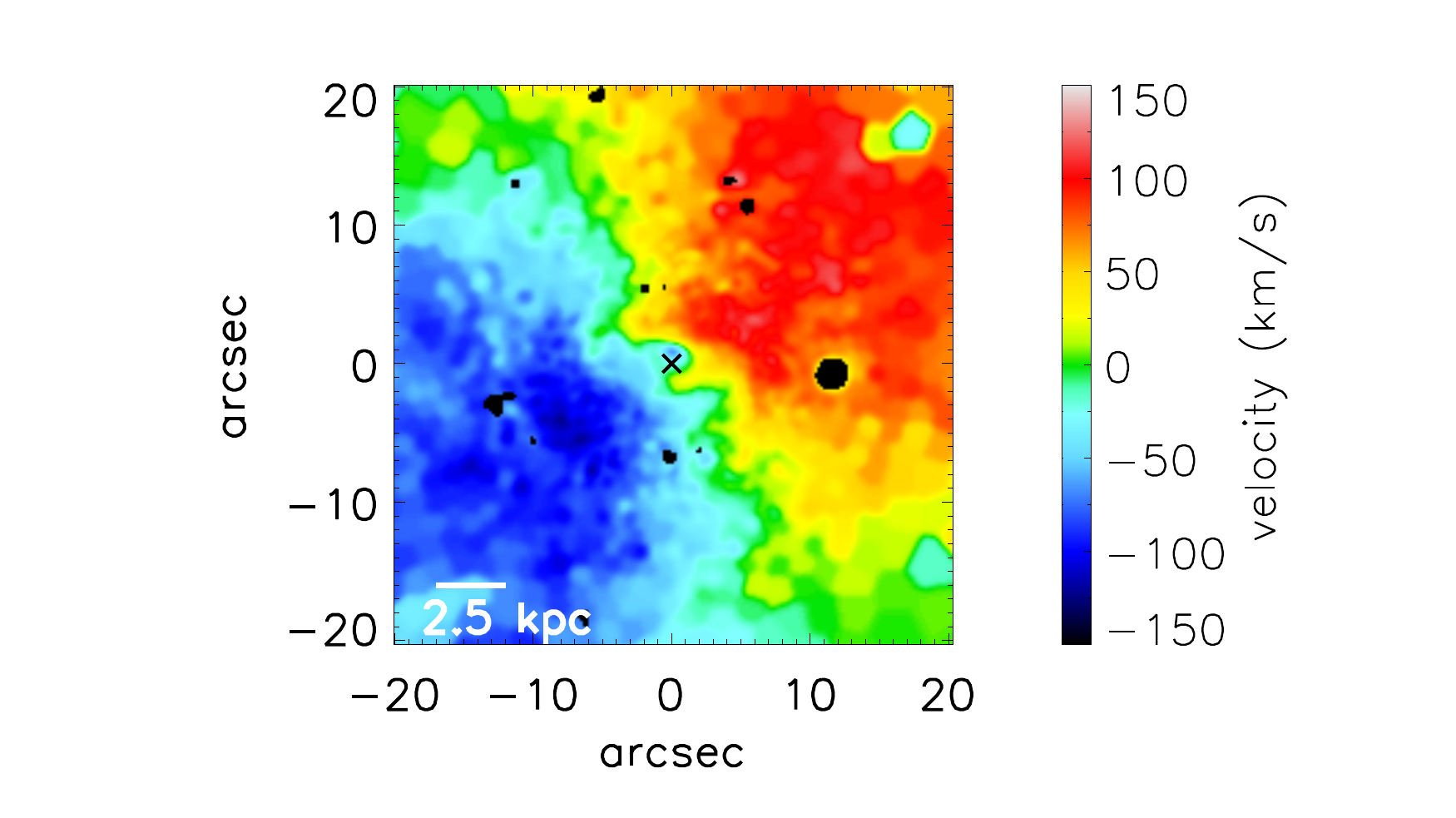}\hspace{-1.5cm}
\includegraphics[width=0.38\textwidth]{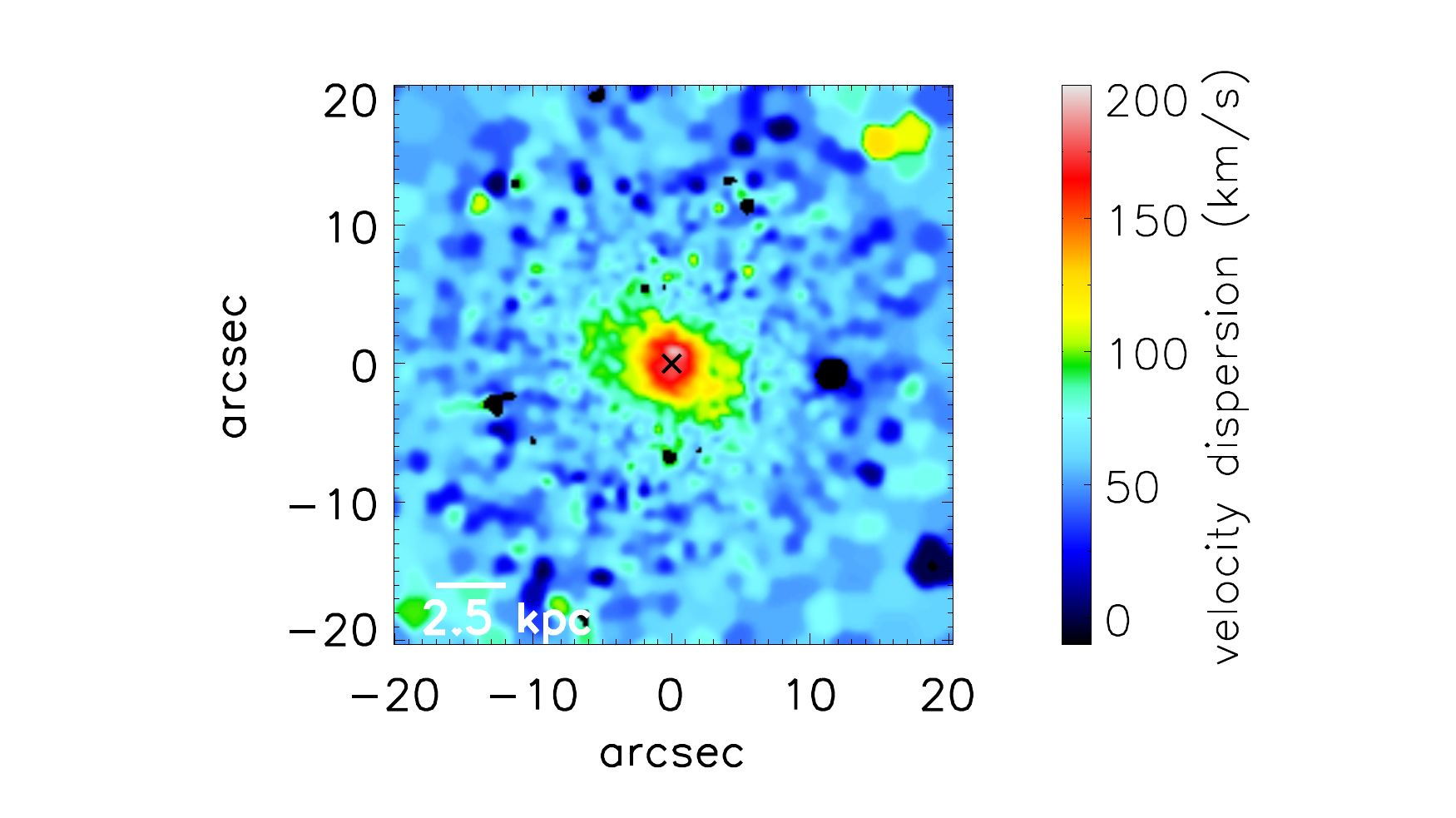}\hspace{-1.5cm}\\[-0.4cm]
\hspace{-1.0cm}
\includegraphics[width=0.2\textwidth]{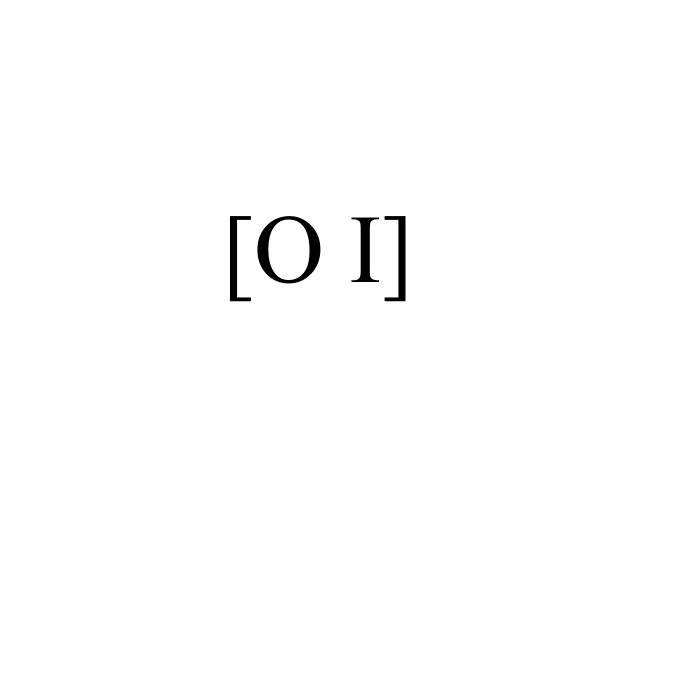}\hspace{-2.0cm}
\includegraphics[width=0.38\textwidth]{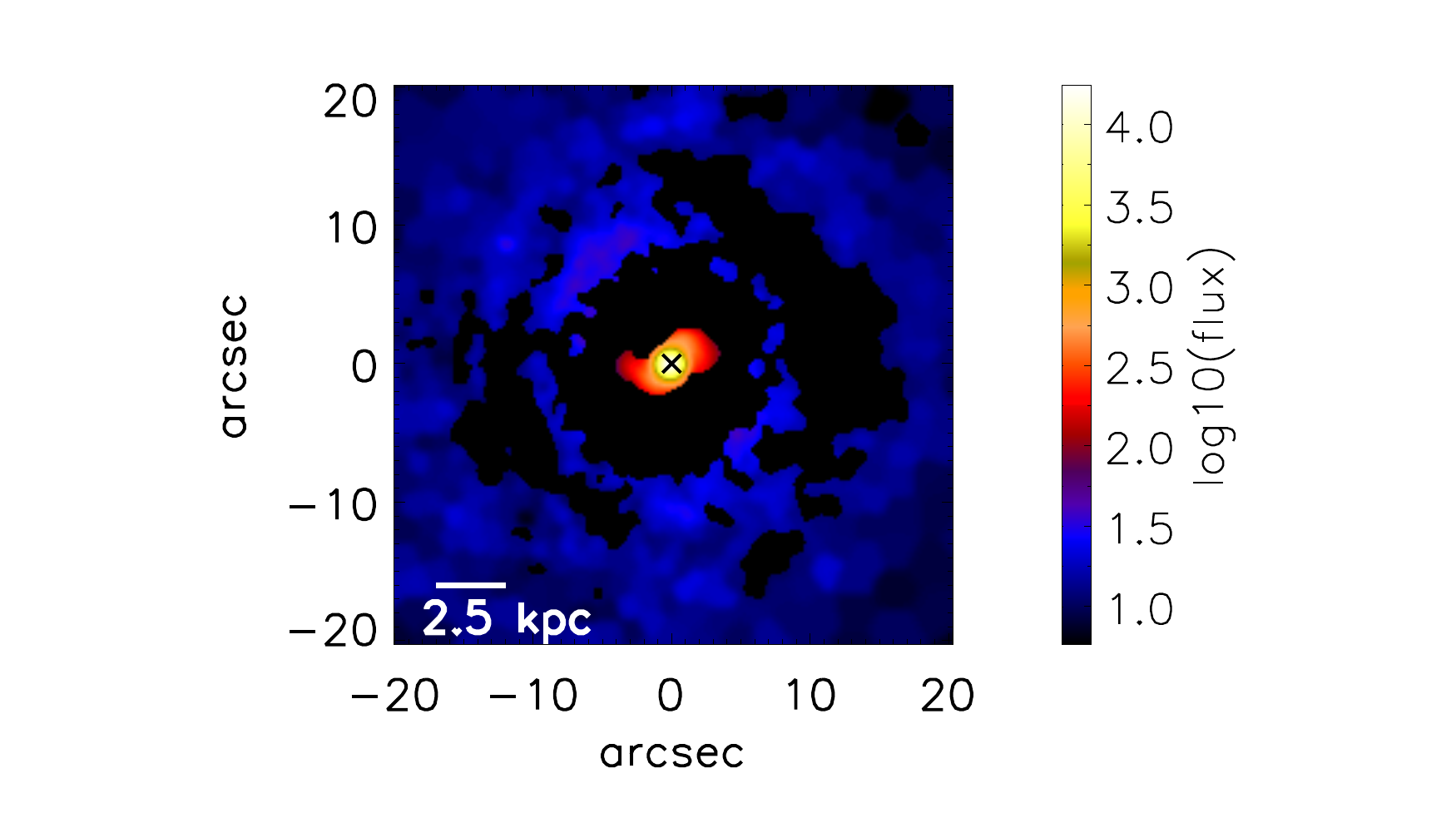}\hspace{-2cm}
\includegraphics[width=0.38\textwidth]{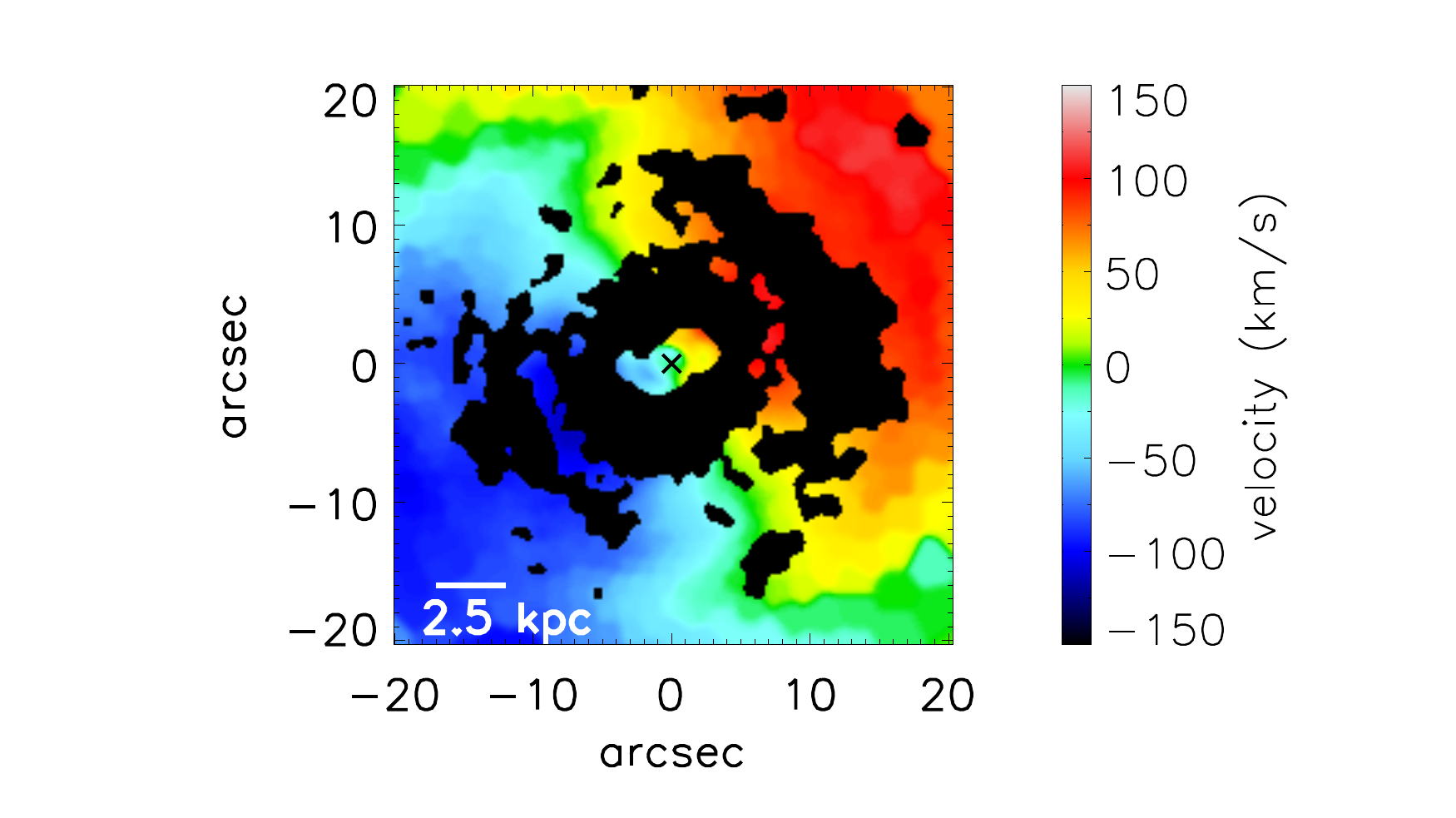}\hspace{-1.5cm}
\includegraphics[width=0.38\textwidth]{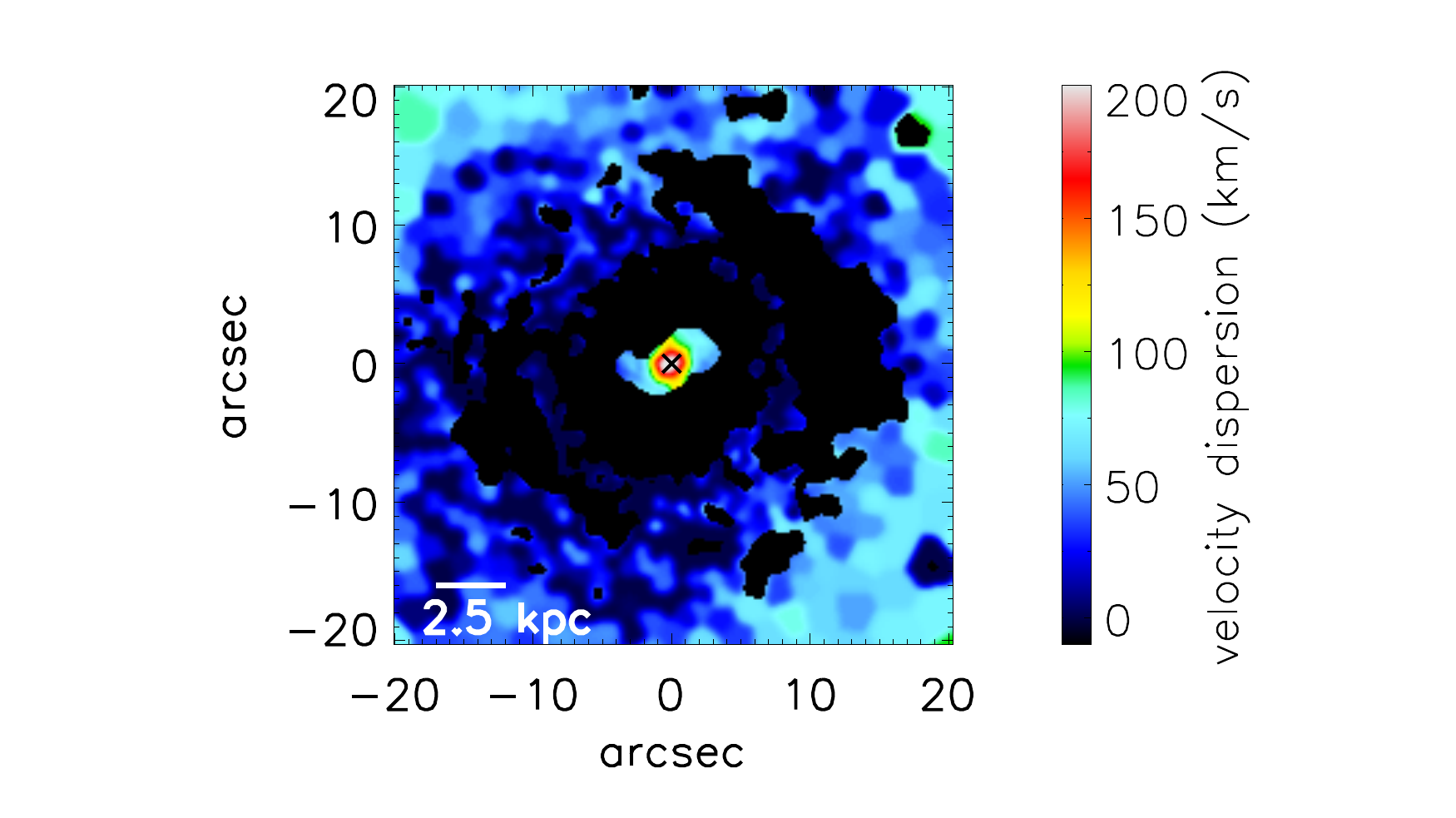}\hspace{-1.5cm}\\[-0.4cm]
\hspace{-1.0cm}
\includegraphics[width=0.2\textwidth]{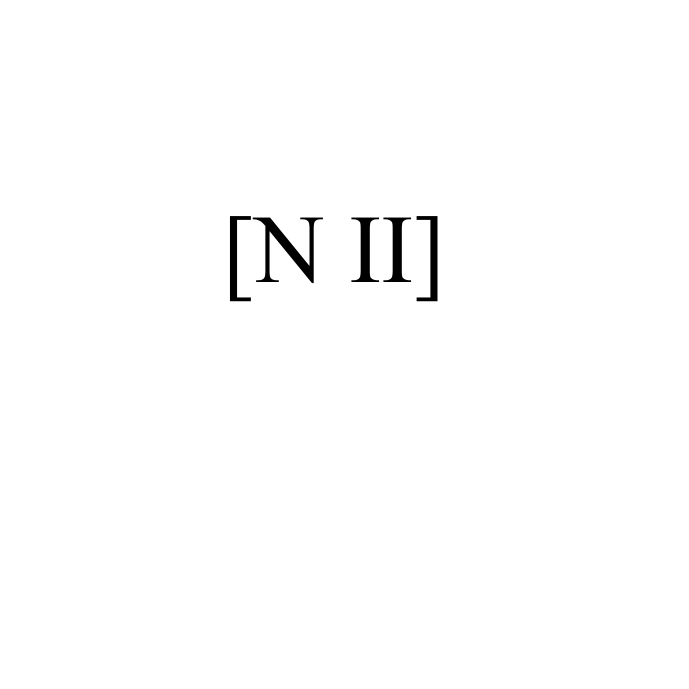}\hspace{-2.0cm}
\includegraphics[width=0.38\textwidth]{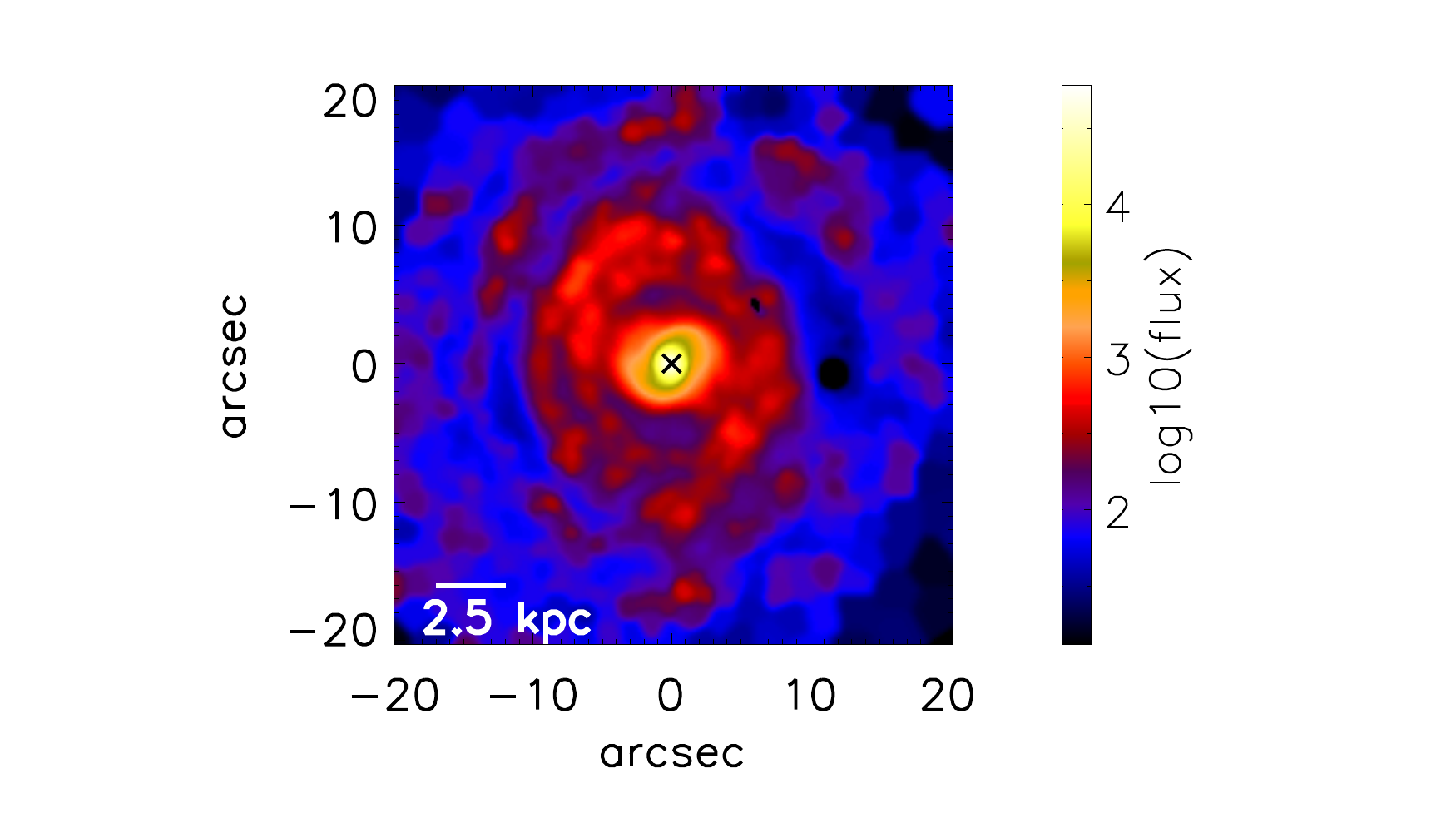}\hspace{-2cm}
\includegraphics[width=0.38\textwidth]{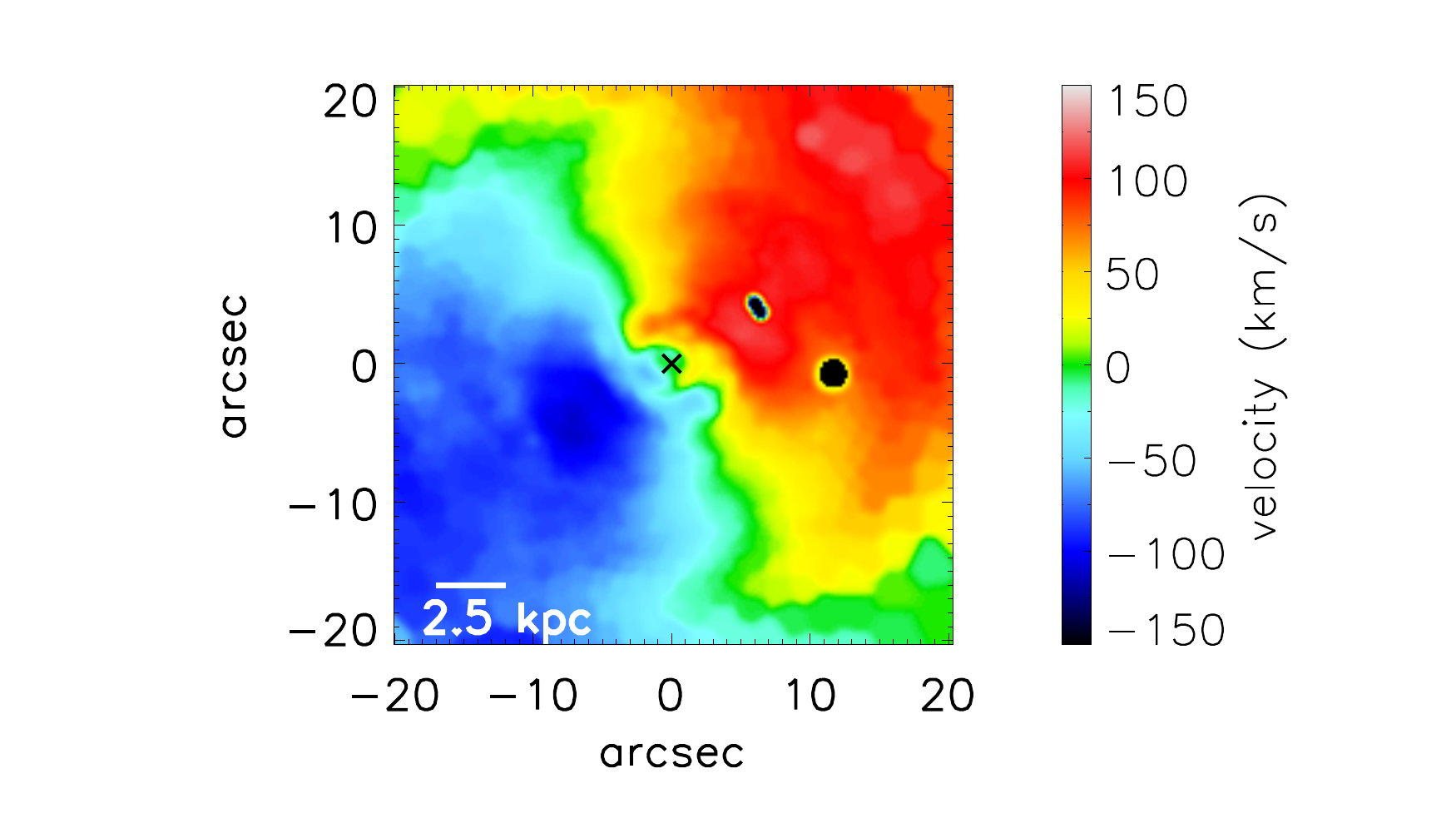}\hspace{-1.5cm}
\includegraphics[width=0.38\textwidth]{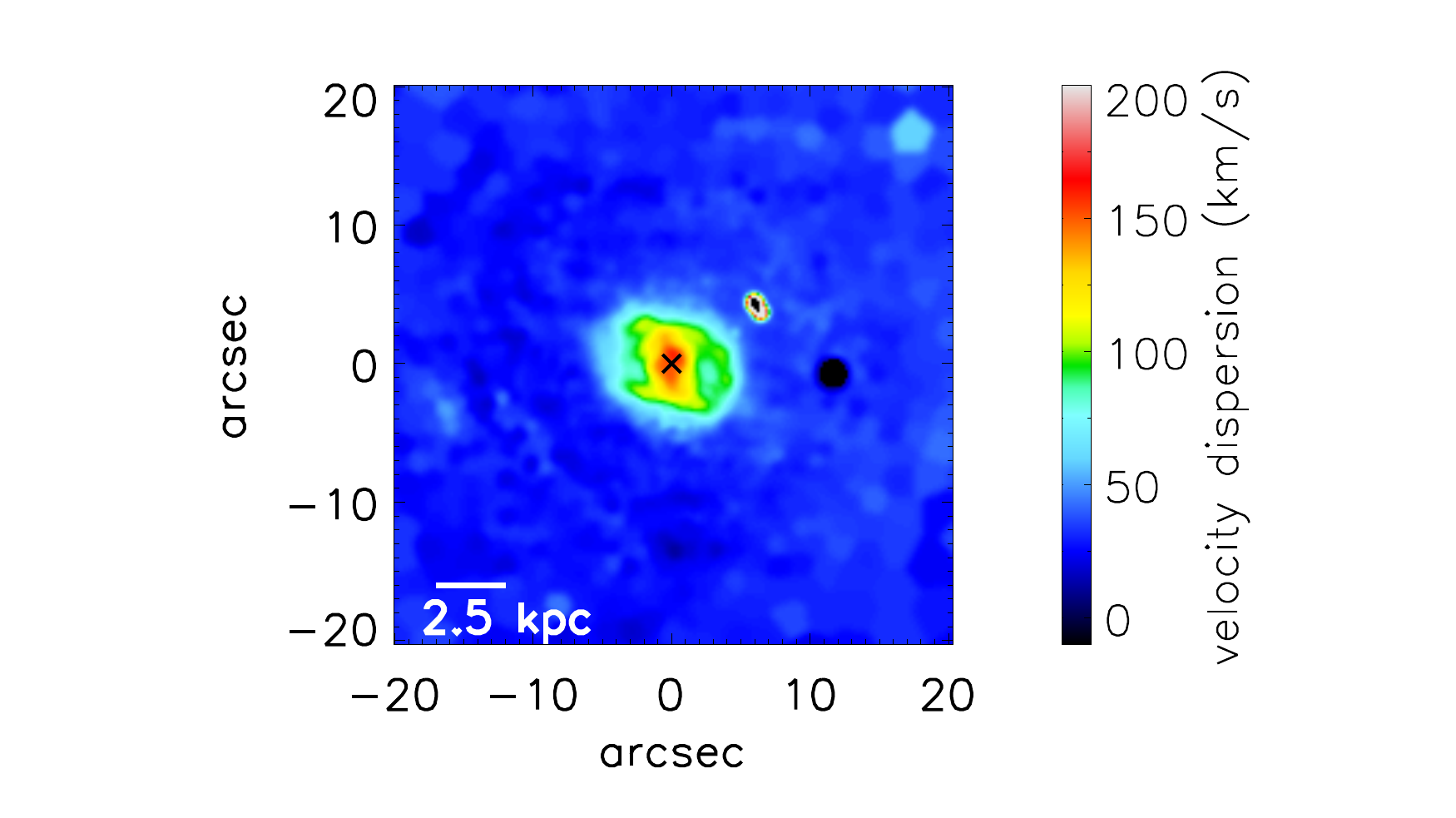}\\[-0.4cm]
\hspace{-1.0cm}
\includegraphics[width=0.2\textwidth]{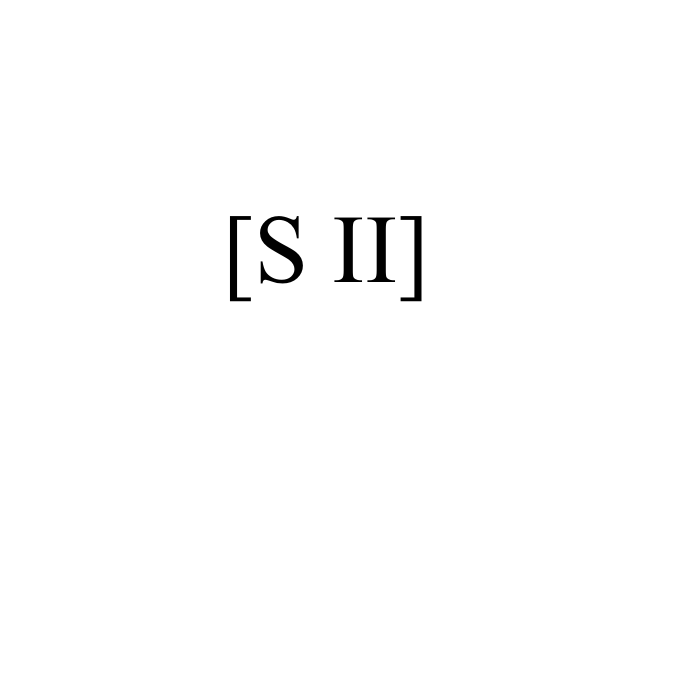}\hspace{-2.0cm}
\includegraphics[width=0.38\textwidth]{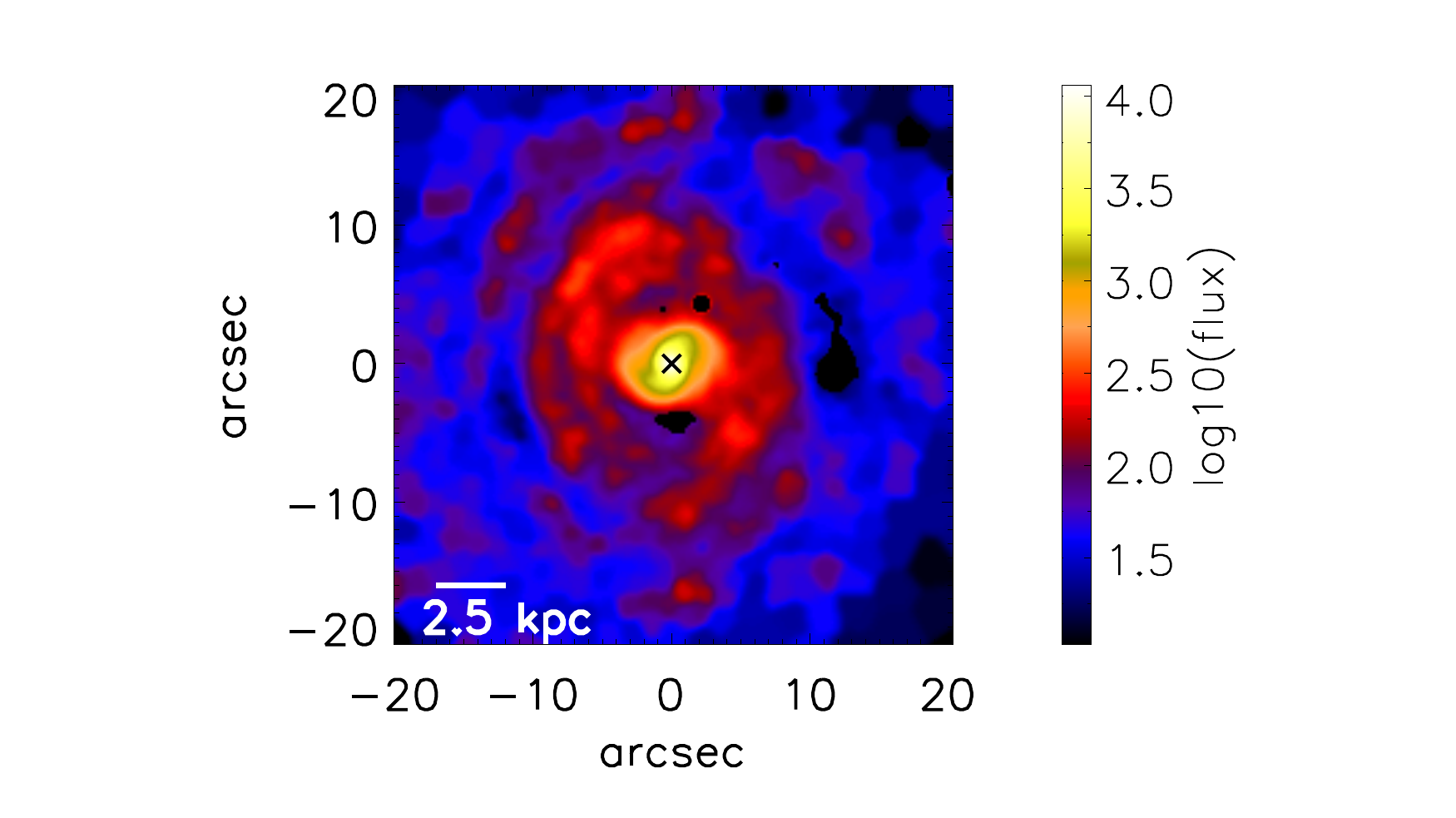}\hspace{-2cm}
\includegraphics[width=0.38\textwidth]{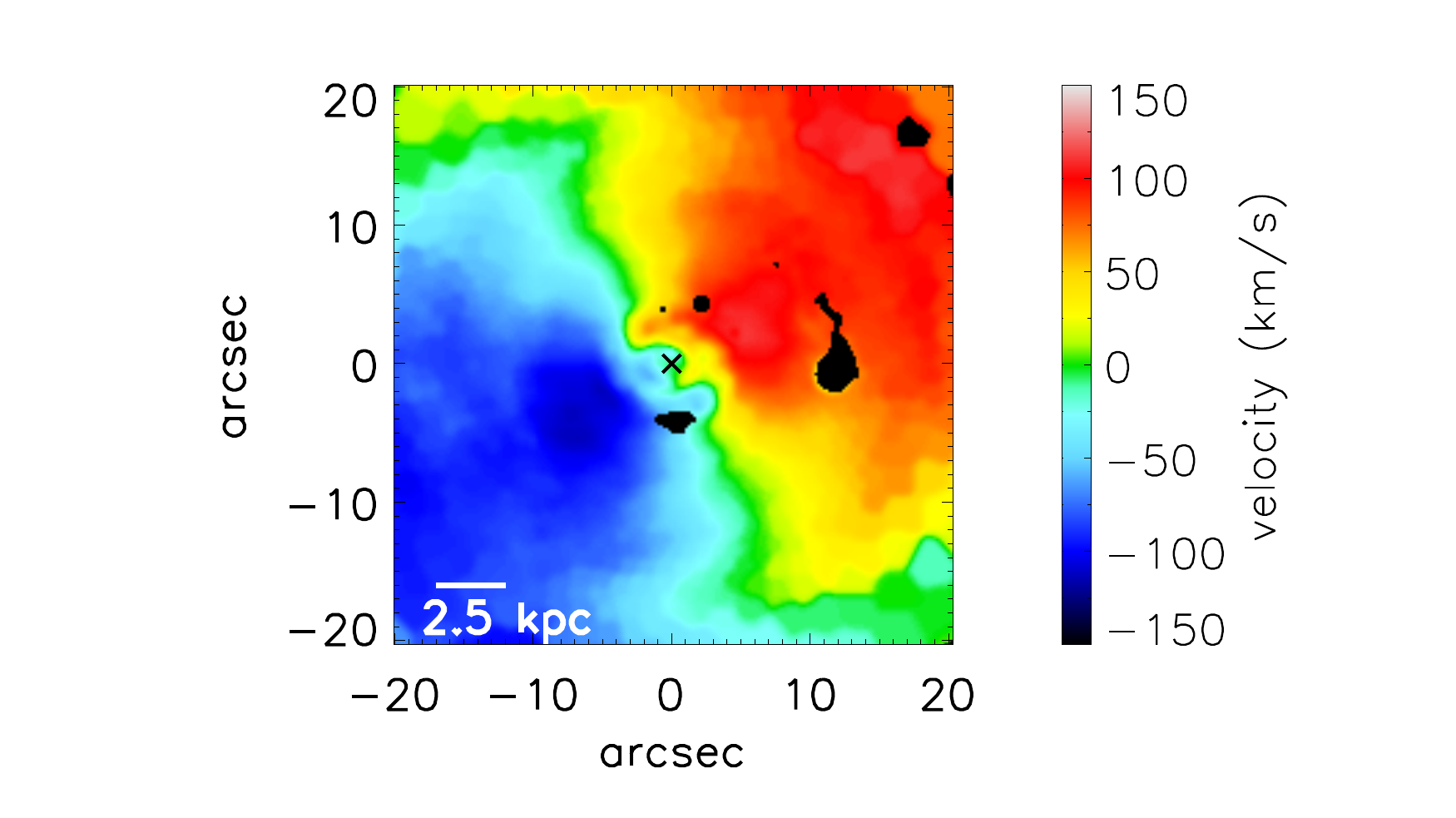}\hspace{-1.5cm}
\includegraphics[width=0.38\textwidth]{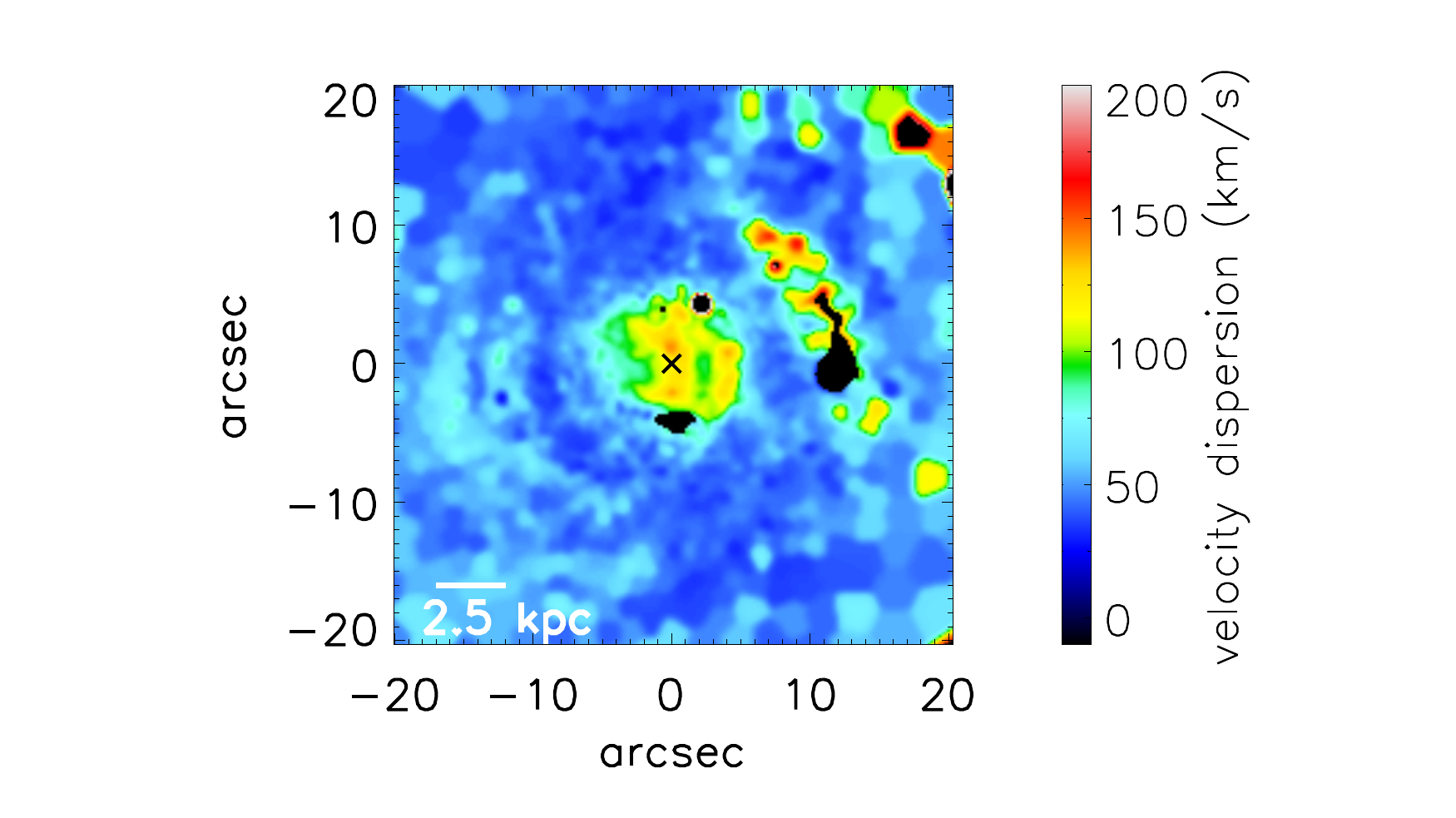}
\end{center}
\caption{Maps of ionised gas tracers. Each row corresponds to a different tracer, from top to bottom: H$\alpha$, [O III], [O I], [N II] and [S II]. Only the narrow components of the lines are shown in the panels. Columns from left to right show the line flux (in a logarithmic scale for visualisation purposes), velocity and velocity dispersion maps, respectively. Black cross indicates the AGN position as determined from the H$\alpha$ broad emission line peak flux. North is up, East is left. Regions in black have been masked out due to low signal to noise (see text for details).}
\label{gas_dynamics}
\end{figure*}

\begin{figure*}
\centering
\hspace{0.1cm}
\includegraphics[width=0.2\textwidth]{plots/labels_flux.pdf}\hspace{1.0cm}
\includegraphics[width=0.2\textwidth]{plots/labels_V.pdf}\hspace{1.7cm}
\includegraphics[width=0.2\textwidth]{plots/labels_sigma.pdf}\\[-2.2cm]
\hspace{-1.0cm}
\includegraphics[width=0.2\textwidth]{plots/labels.pdf}\hspace{-2.0cm}
\includegraphics[width=0.38\textwidth]{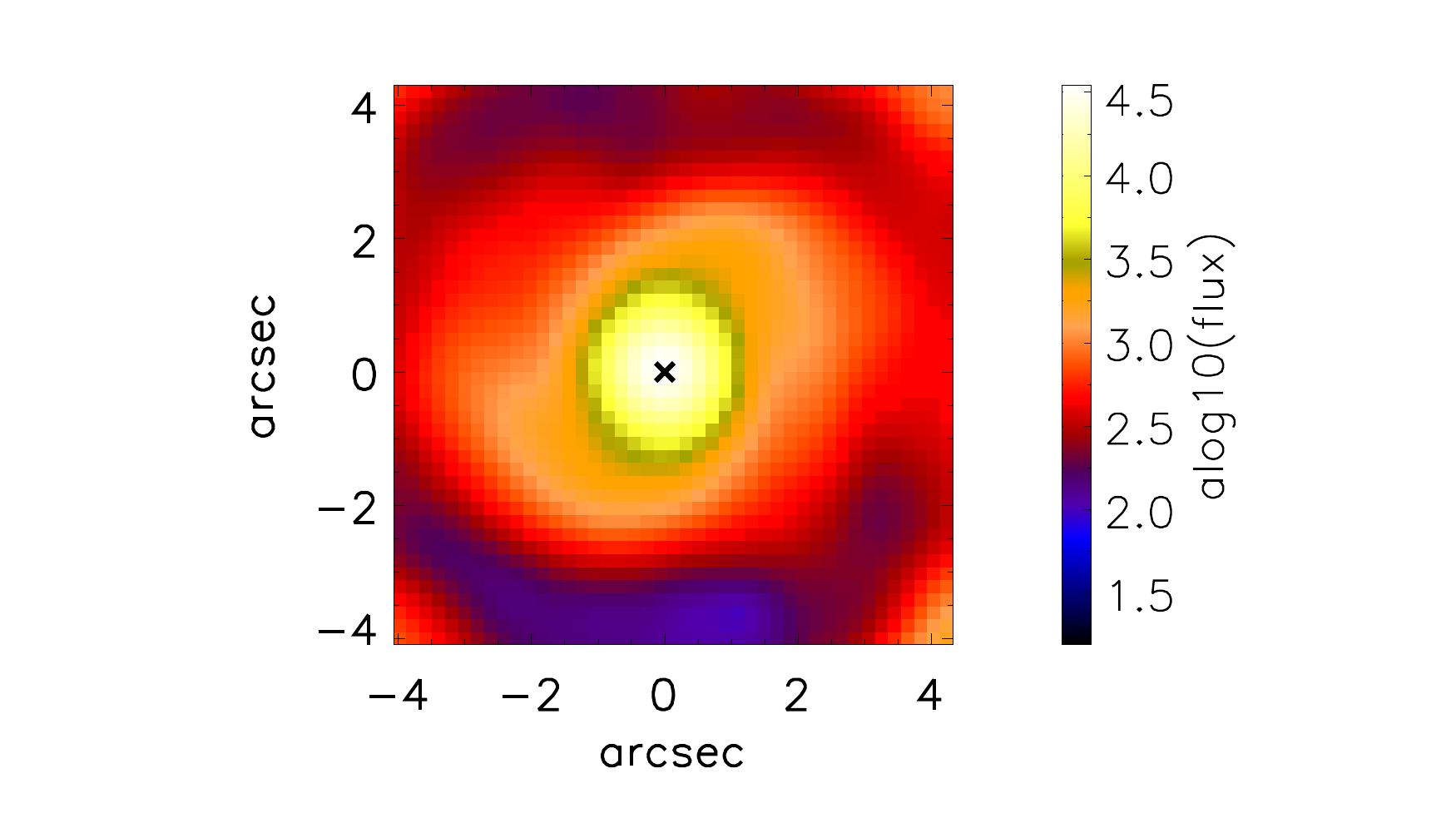}\hspace{-2cm}
\includegraphics[width=0.38\textwidth]{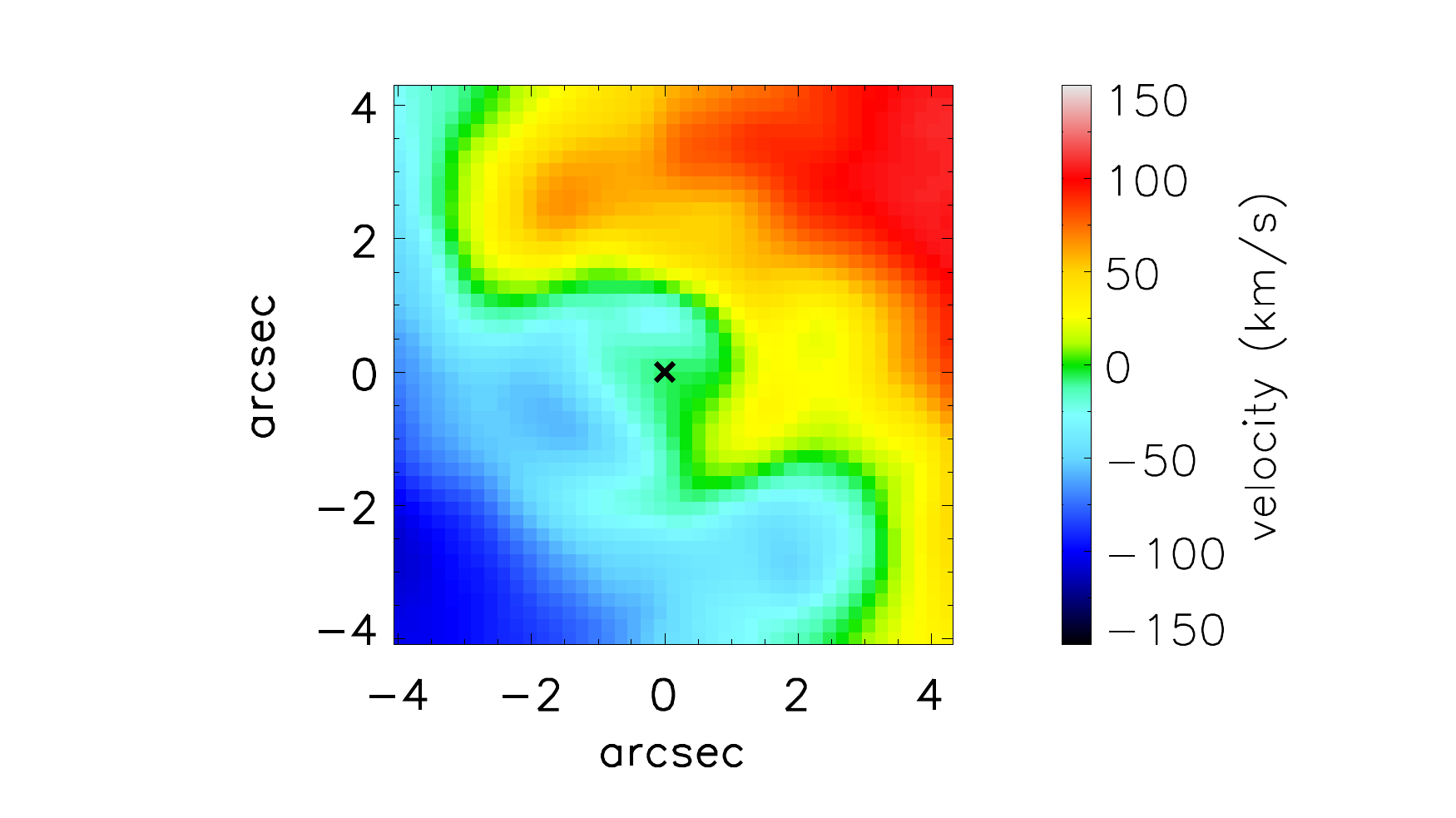}\hspace{-1.5cm}
\includegraphics[width=0.38\textwidth]{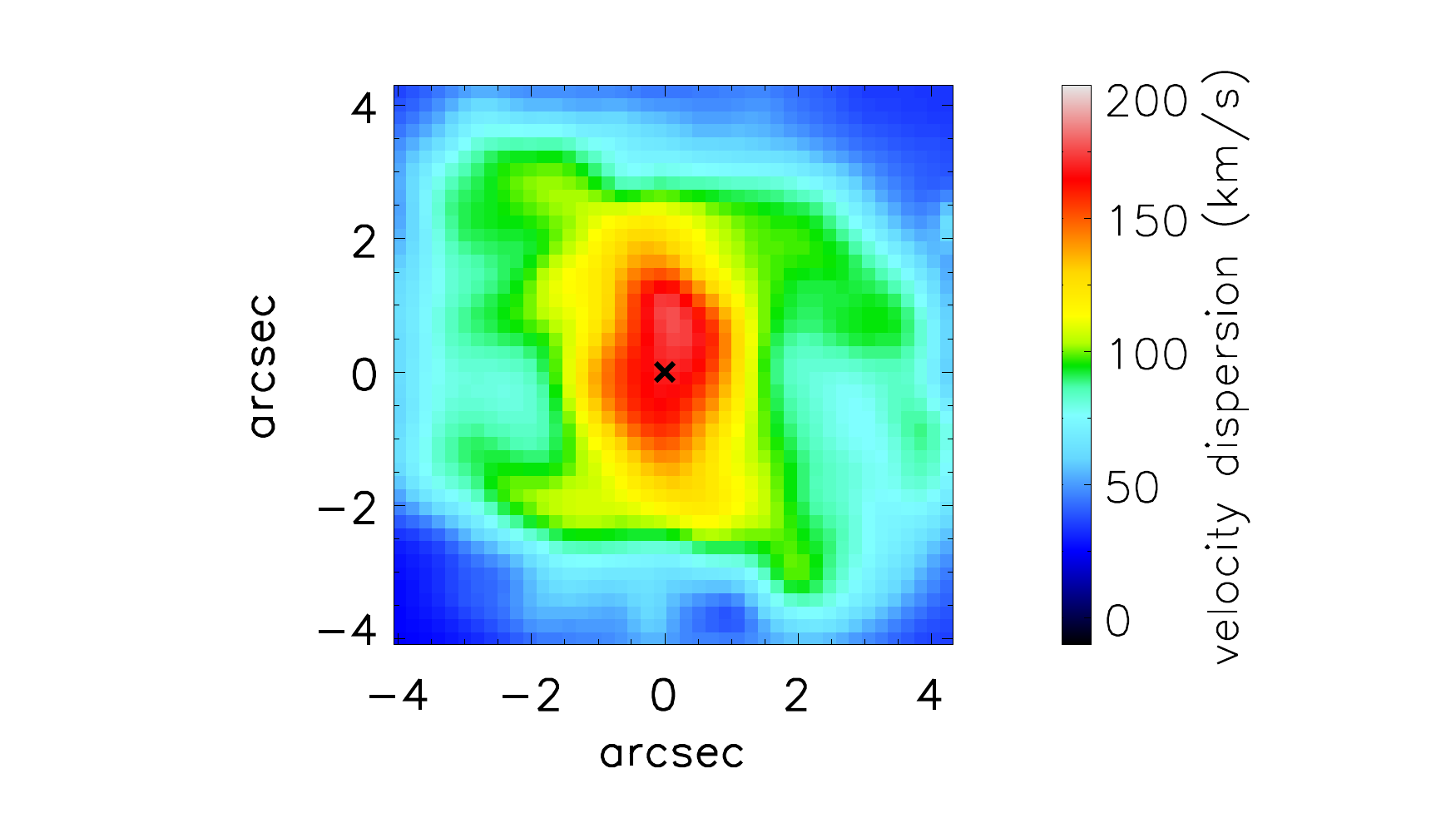}\\[-0.4cm]
\hspace{-1.0cm}
\includegraphics[width=0.2\textwidth]{plots/labels_OIII.pdf}\hspace{-2.0cm}
\includegraphics[width=0.38\textwidth]{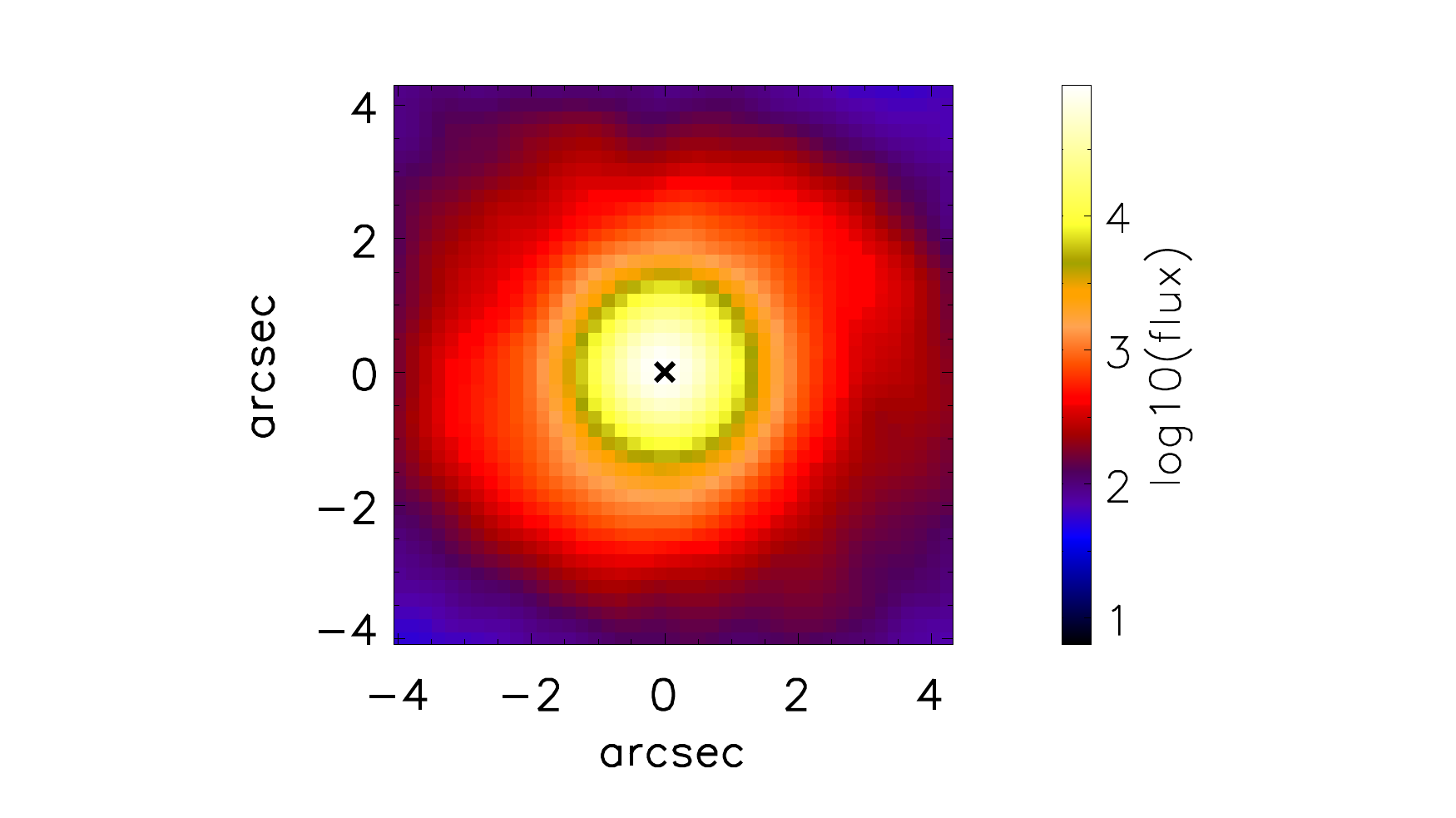}\hspace{-2cm}
\includegraphics[width=0.38\textwidth]{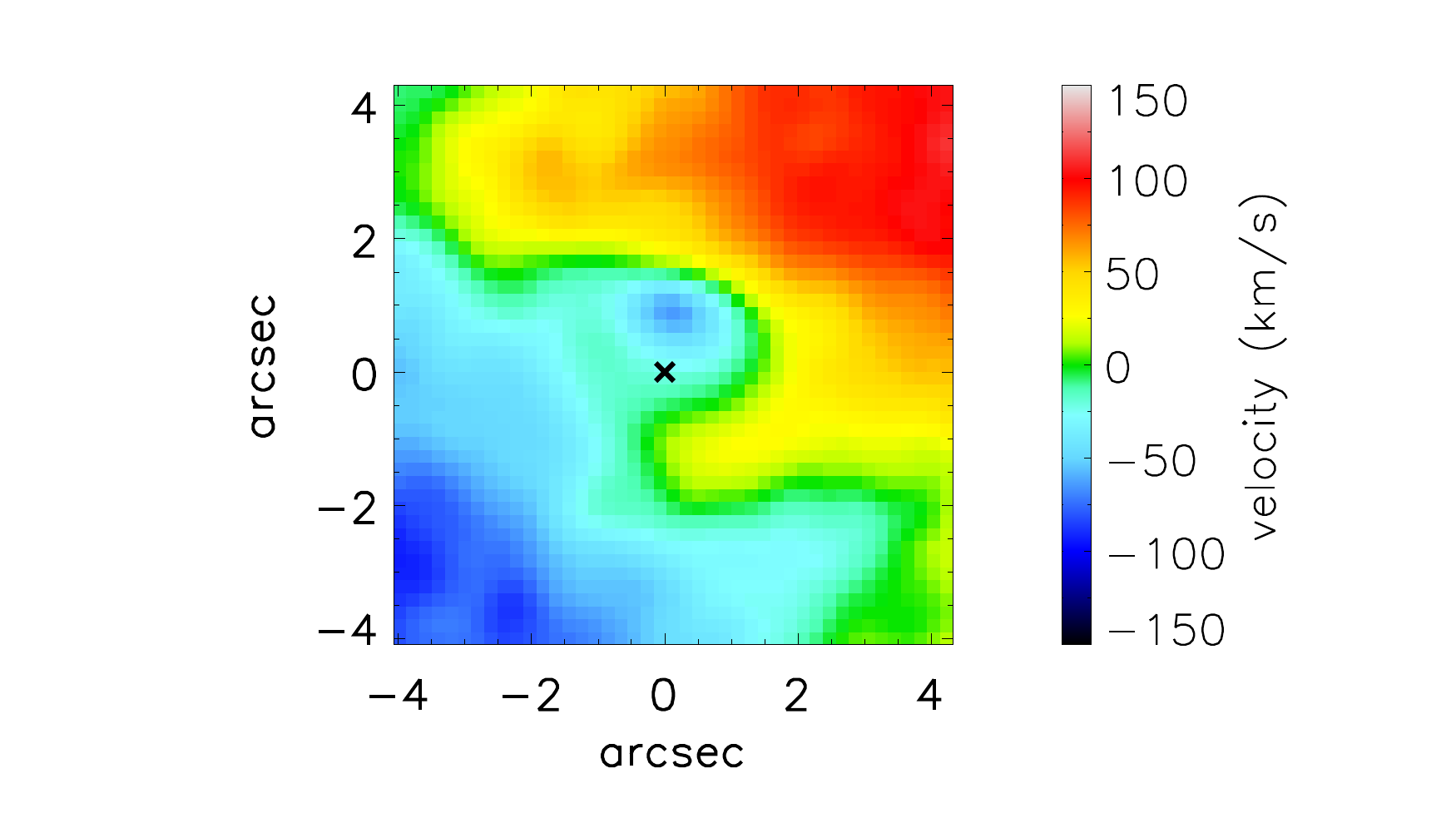}\hspace{-1.5cm}
\includegraphics[width=0.38\textwidth]{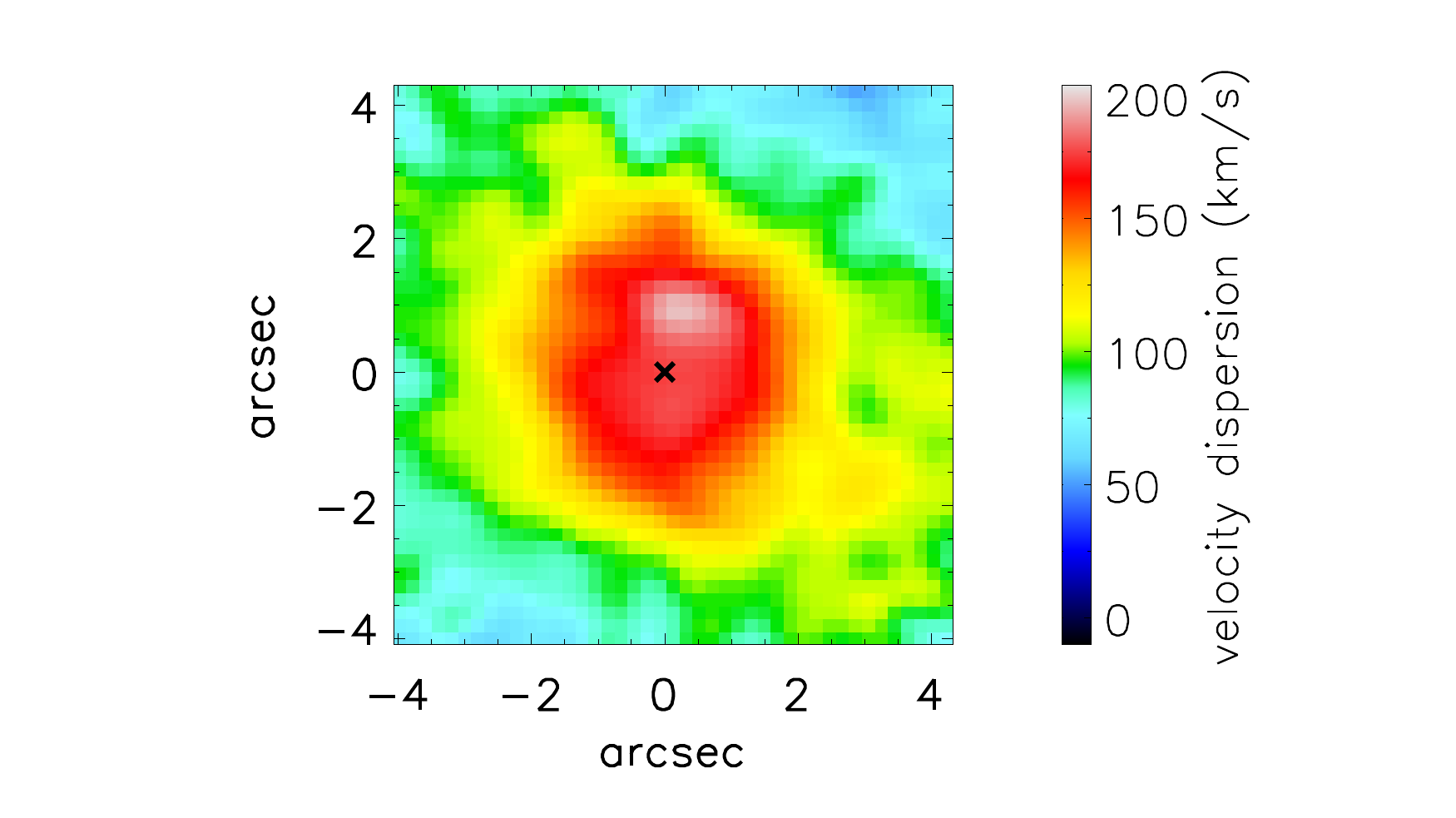}\\[-0.4cm]
\hspace{-1.0cm}
\includegraphics[width=0.2\textwidth]{plots/labels_OI.pdf}\hspace{-2.0cm}
\includegraphics[width=0.38\textwidth]{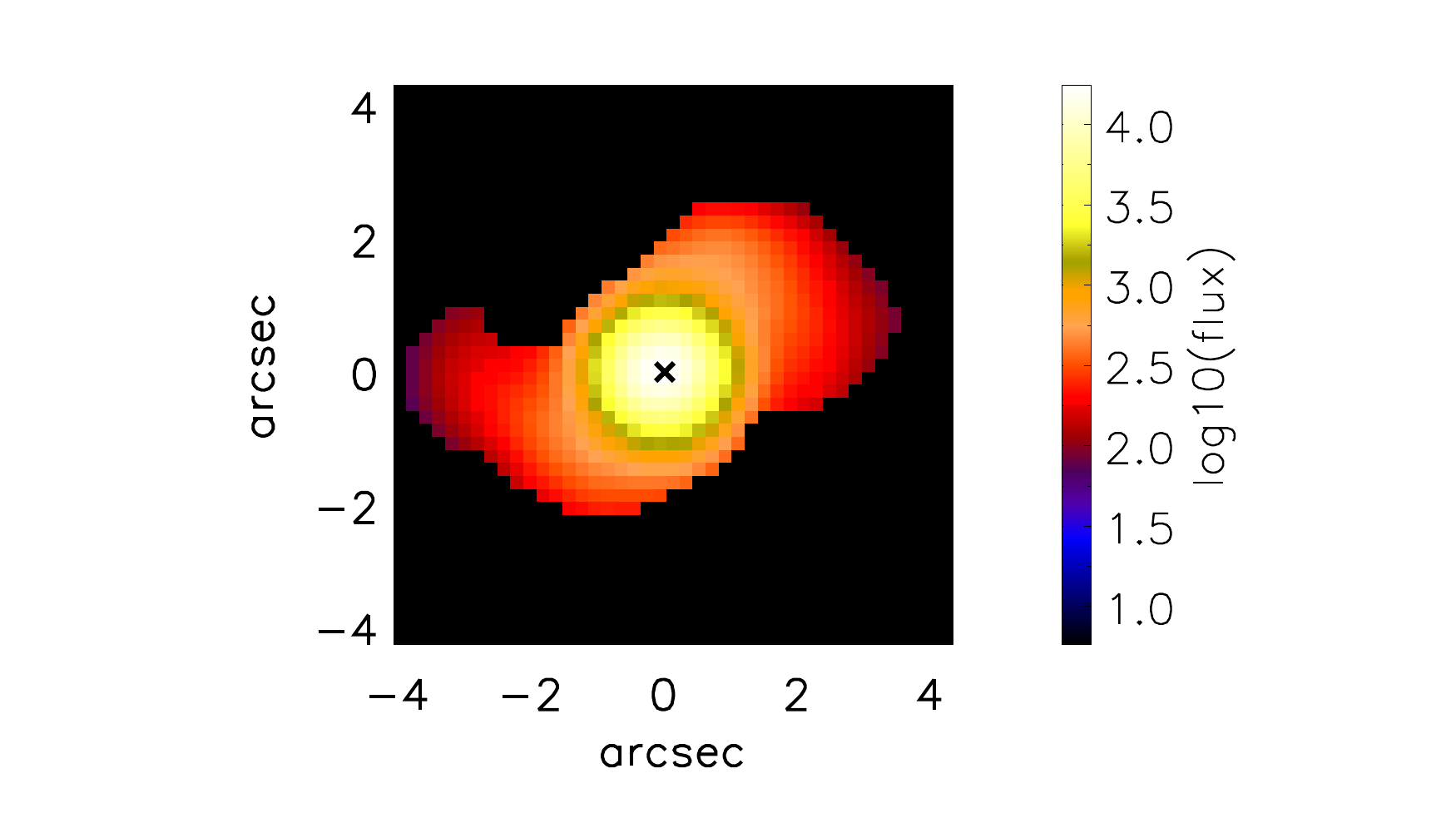}\hspace{-2cm}
\includegraphics[width=0.38\textwidth]{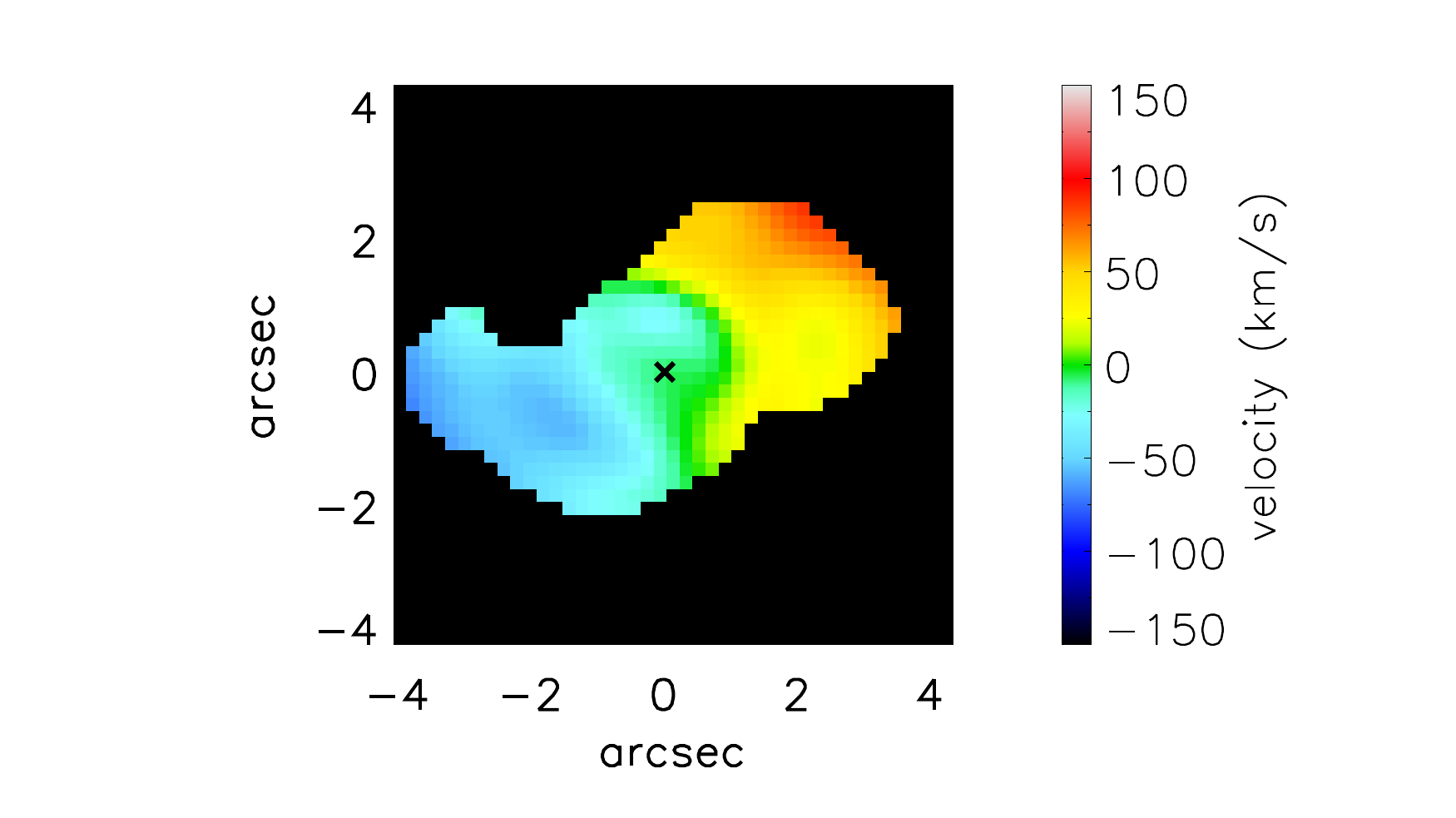}\hspace{-1.5cm}
\includegraphics[width=0.38\textwidth]{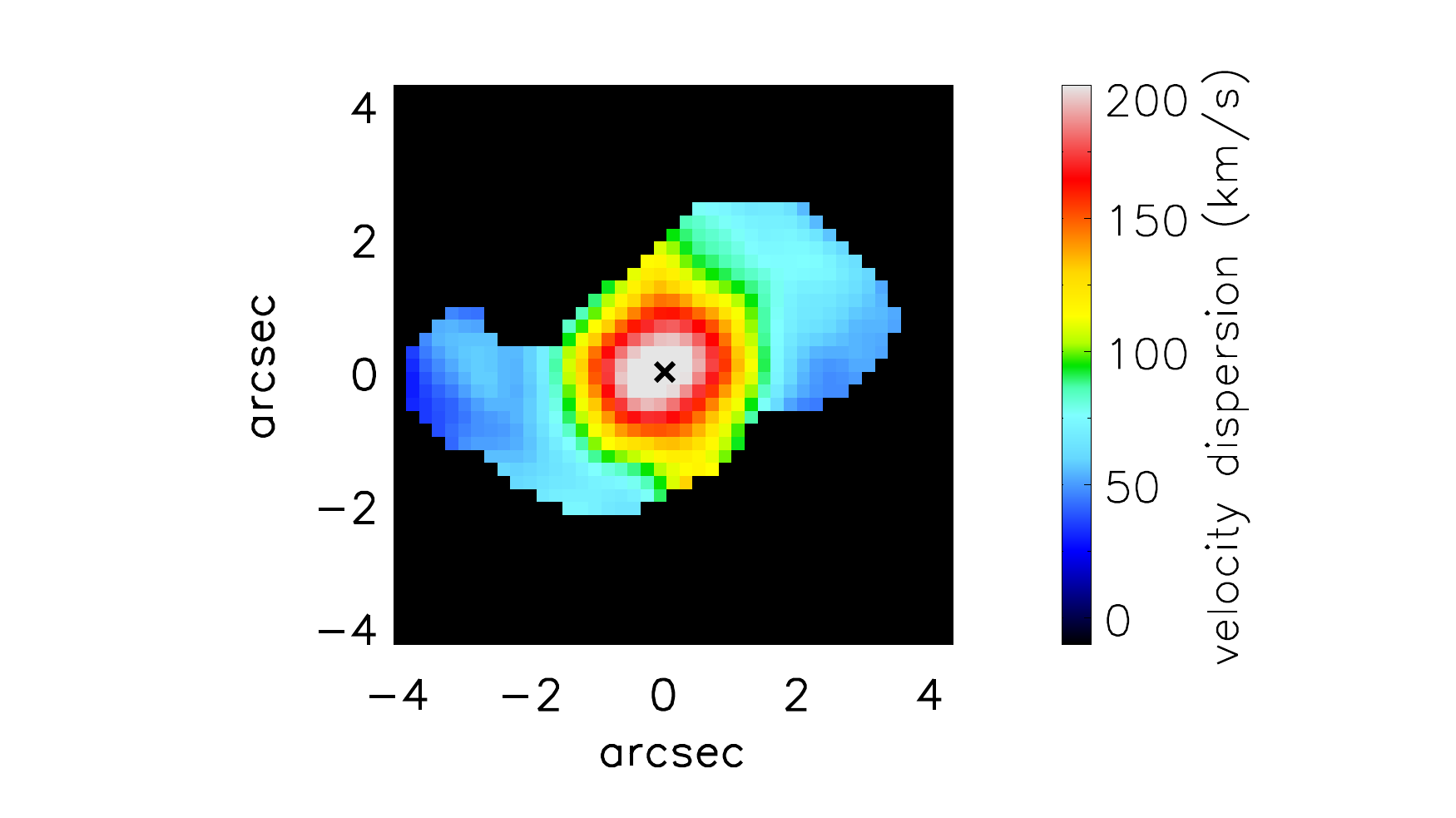}\\[-0.4cm]
\hspace{-1.0cm}
\includegraphics[width=0.2\textwidth]{plots/labels_NII.pdf}\hspace{-2.0cm}
\includegraphics[width=0.38\textwidth]{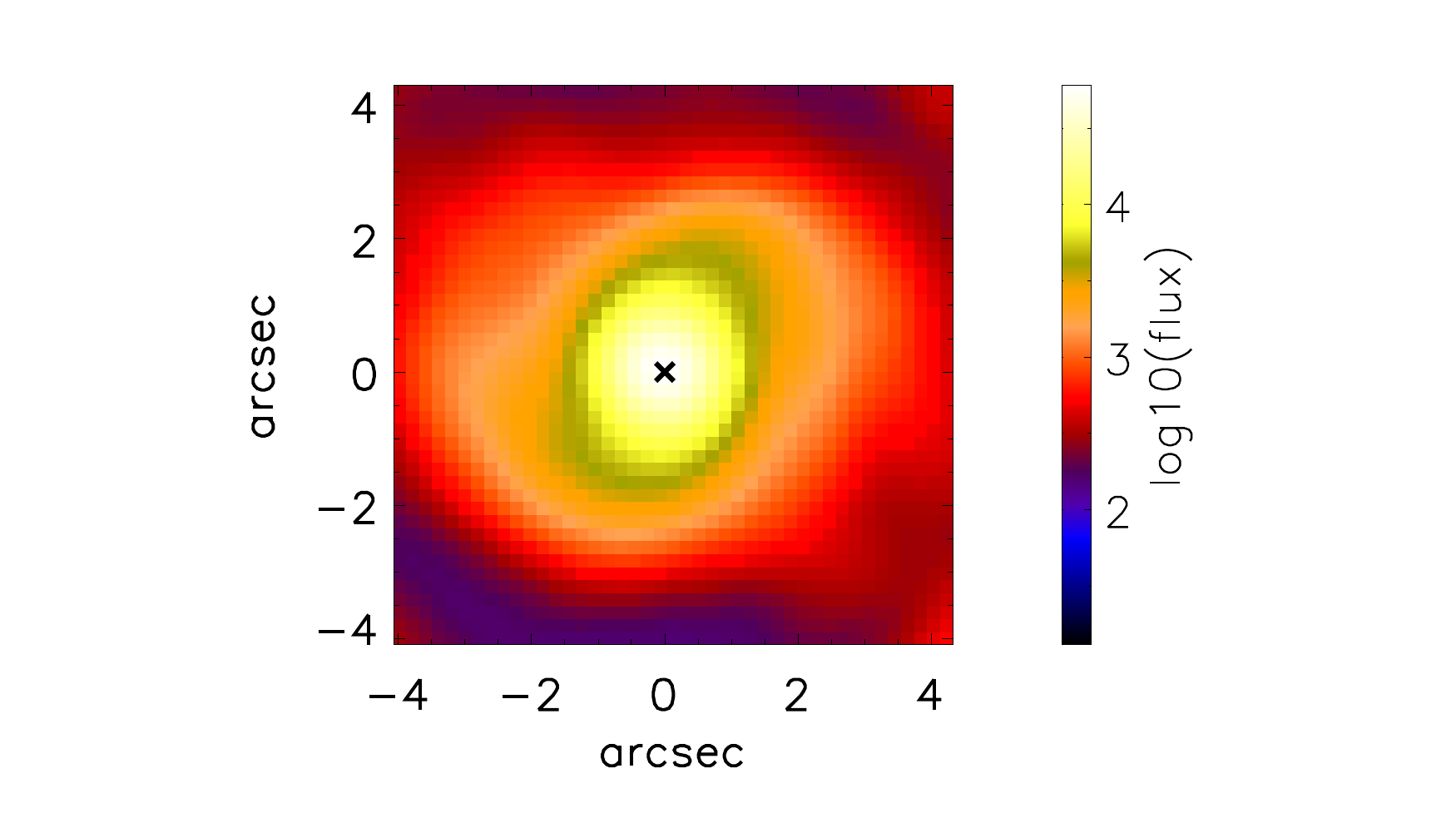}\hspace{-2cm}
\includegraphics[width=0.38\textwidth]{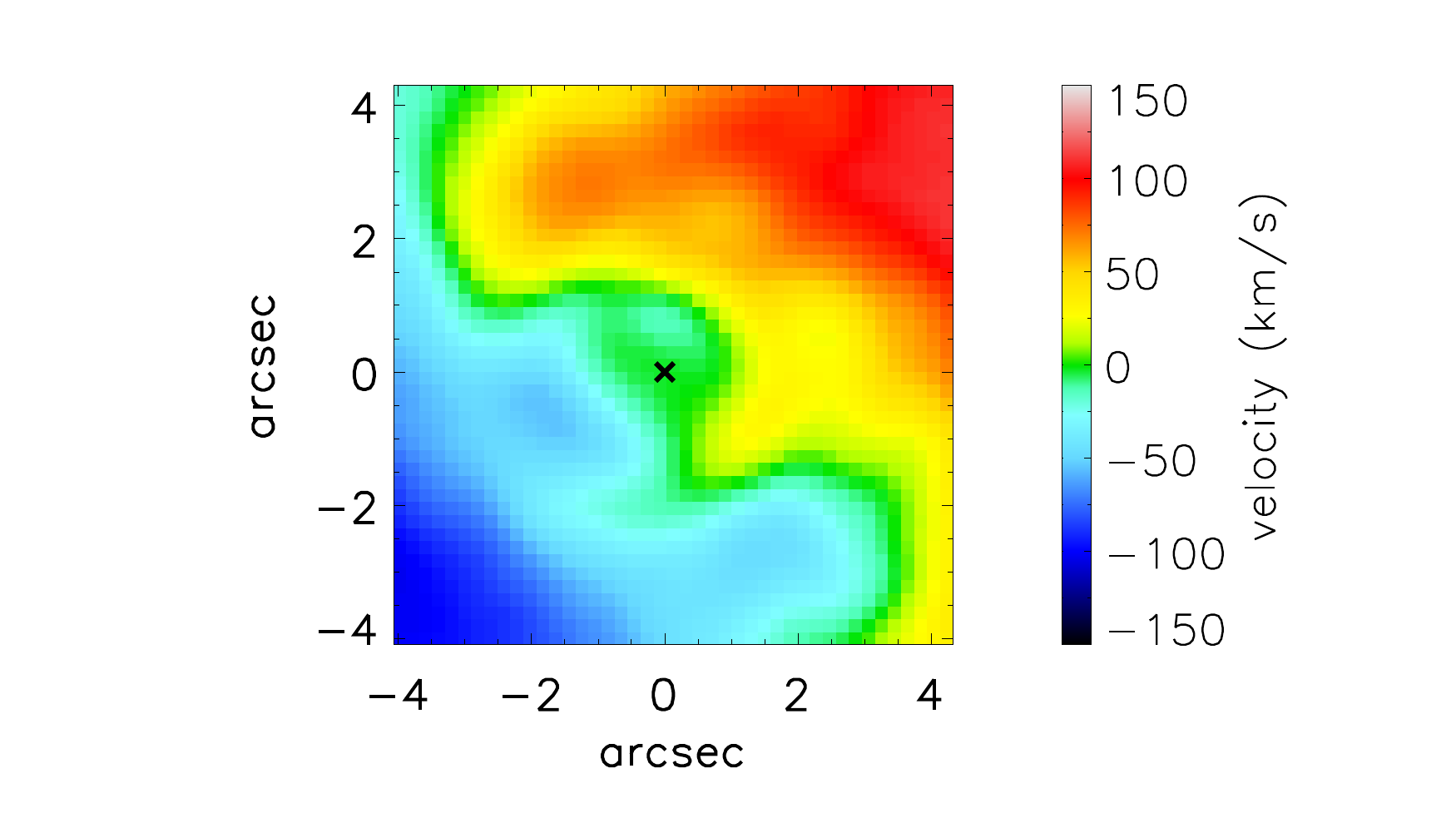}\hspace{-1.5cm}
\includegraphics[width=0.38\textwidth]{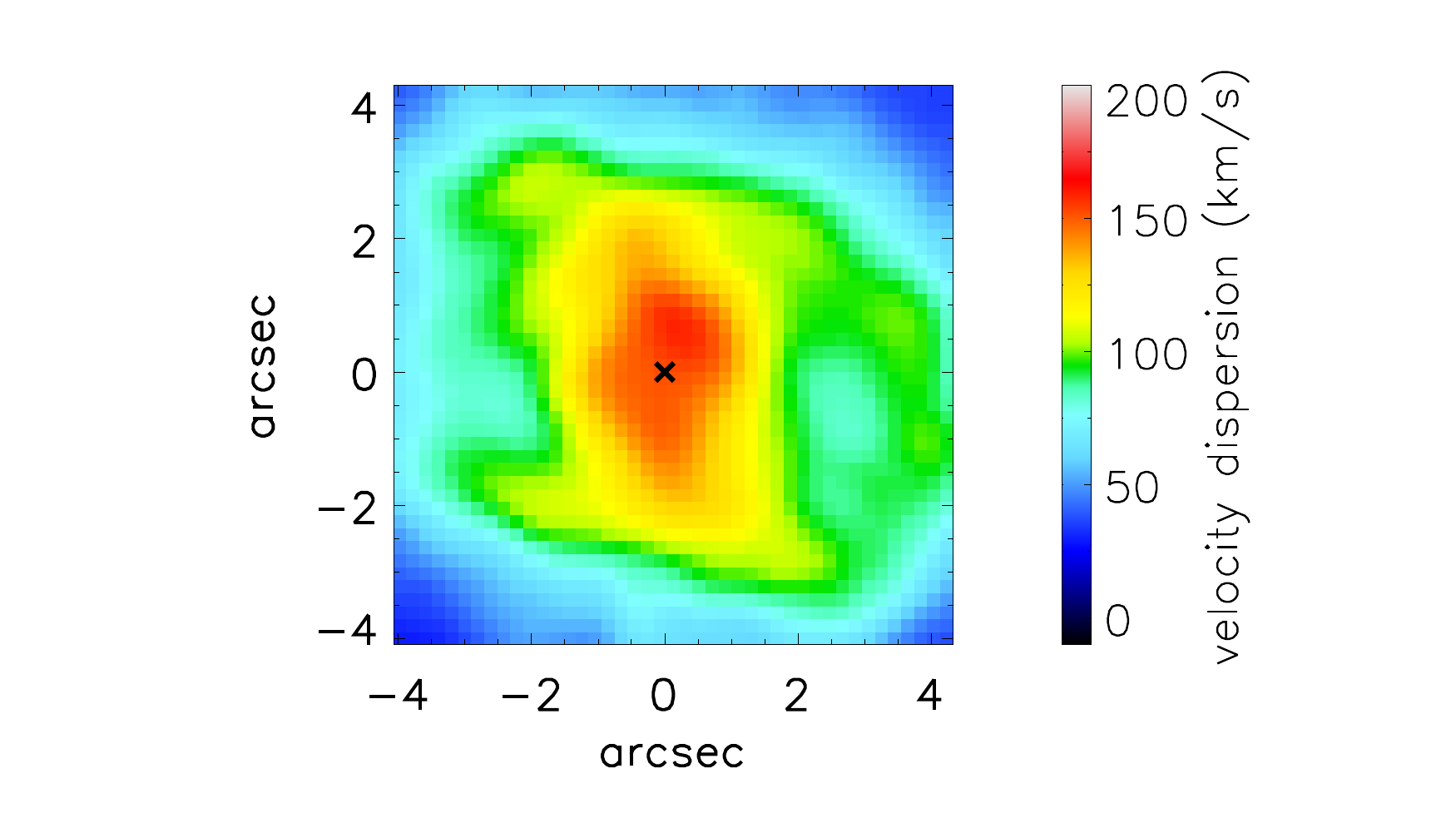}\\[-0.4cm]
\hspace{-1.0cm}
\includegraphics[width=0.2\textwidth]{plots/labels_SII.pdf}\hspace{-2.0cm}
\includegraphics[width=0.38\textwidth]{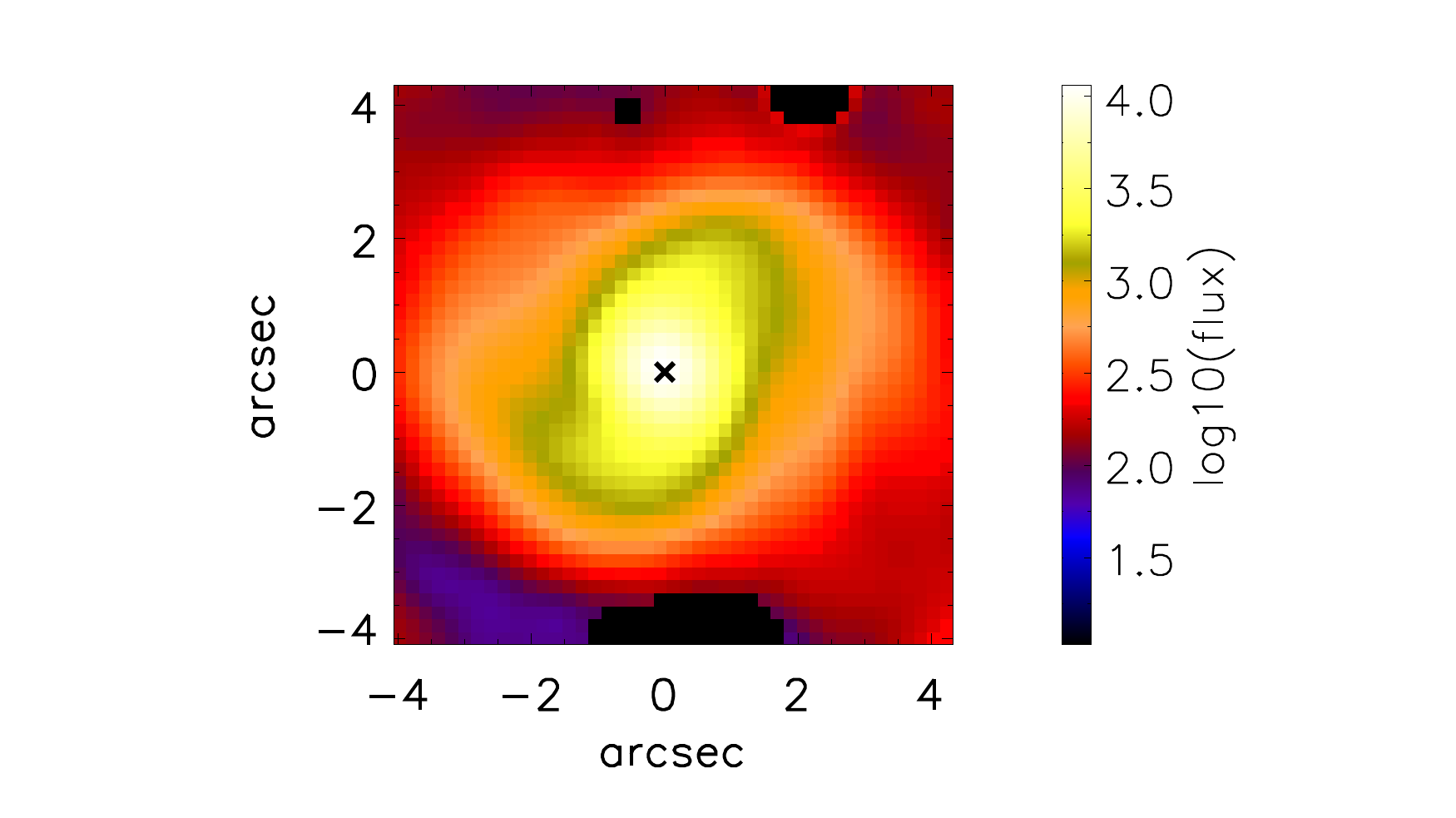}\hspace{-2cm}
\includegraphics[width=0.38\textwidth]{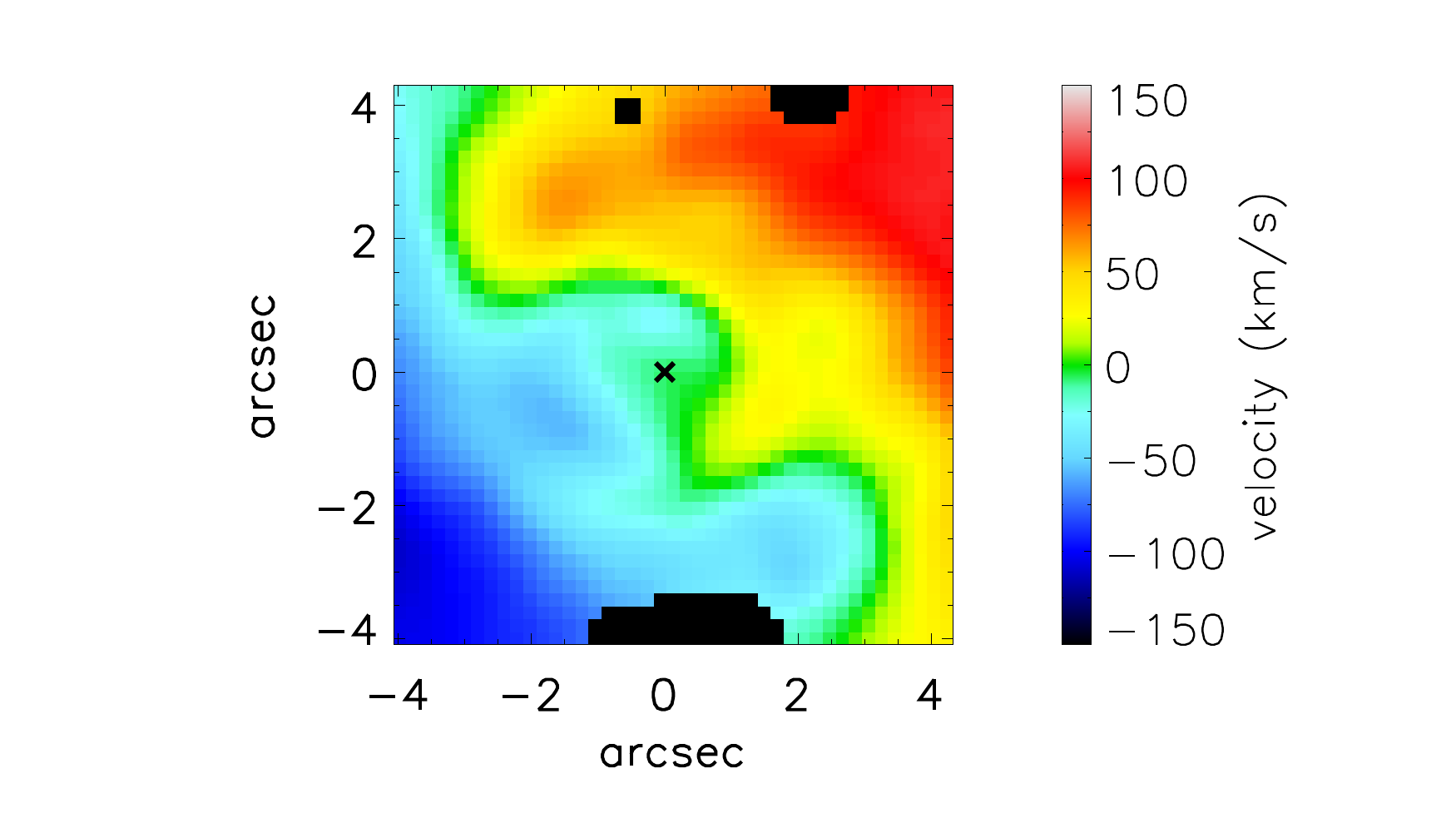}\hspace{-1.5cm}
\includegraphics[width=0.38\textwidth]{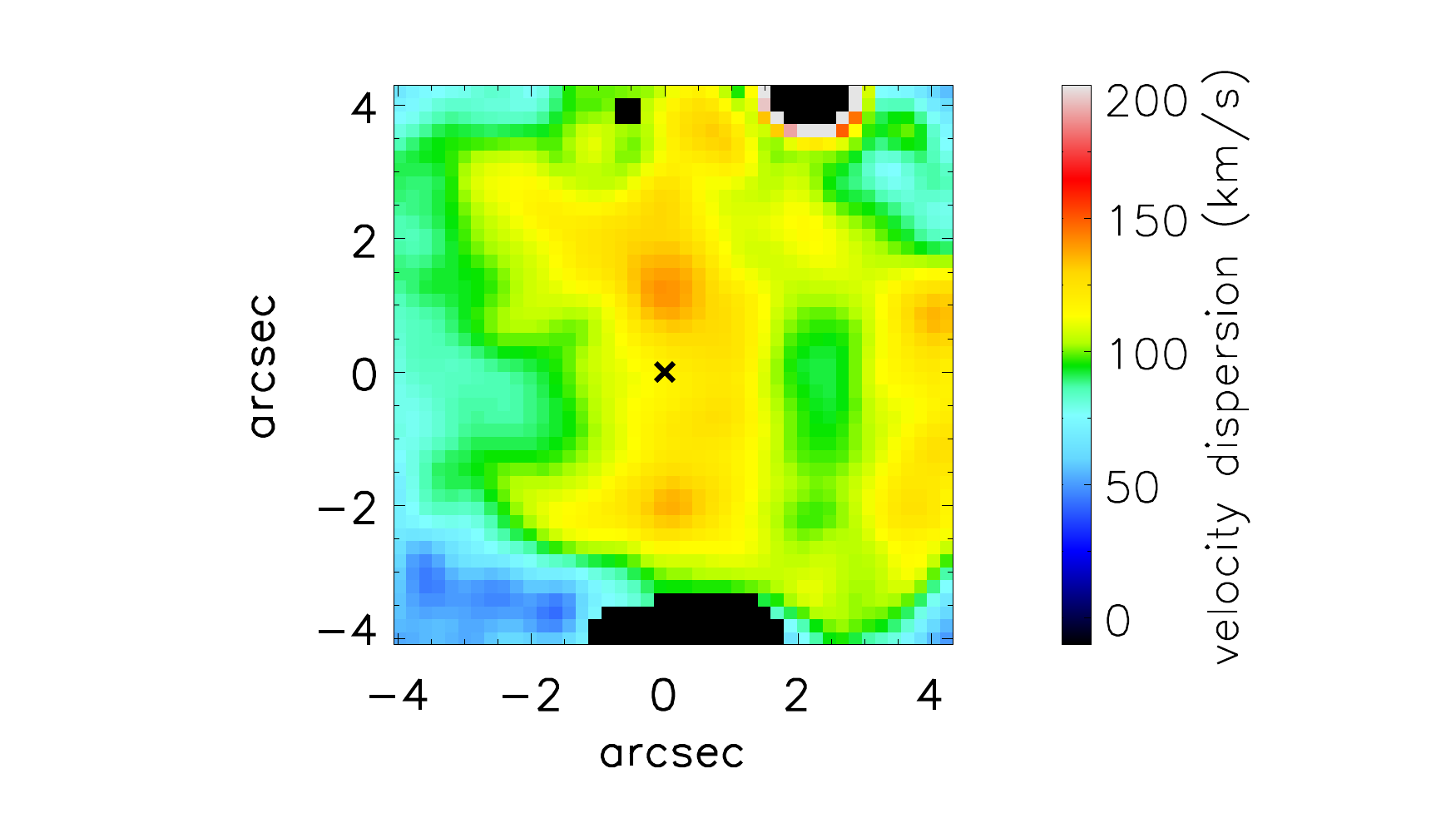}
\caption{Close-up views of the ionised gas maps in Fig.~\ref{gas_dynamics} showing the nuclear region (central $\sim$ 8 arcsec - 4 kpc). Each row corresponds to a different tracer, from top to bottom: H$\alpha$, [O III], [O I], [NII] and [S II]. Only the narrow components of the lines are shown in the panels. Columns from left to right show the line flux (in a logarithmic scale for visualisation purposes), velocity and velocity dispersion maps, respectively. Black cross indicates the AGN position as determined from the H$\alpha$ broad emission line peak flux. North is up, East is left. Regions in black have been masked out due to low signal to noise (see text for details).}
\label{gas_dynamics_zoom}
\end{figure*}

The results on the gas kinematics are shown in Fig.~\ref{gas_dynamics}. Fig~\ref{gas_dynamics_zoom} shows a close-up view of the dynamics in the central 8 arcsec $\times$ 8 arcsec of the galaxy. For visualisation purposes the images were smoothed by a Gaussian kernel matching the spatial resolution of the data. 
In Figs.~\ref{gas_dynamics} and \ref{gas_dynamics_zoom} we only show the emission lines that were detected at a significant signal-to-noise ratio and only their narrow components. [N I] is not shown as it is not significantly detected. H$\beta$ is not shown as its distribution and velocity is similar to H$\alpha$ but detected at a lower S/N. The spatial distribution of the broad components of H$\alpha$, H$\beta$ and He I are not shown in Figs.~\ref{gas_dynamics} and \ref{gas_dynamics_zoom} as they are confined to the nucleus and not resolved in our data.

\begin{figure*}
\begin{center}
\includegraphics[width=0.38\textwidth]{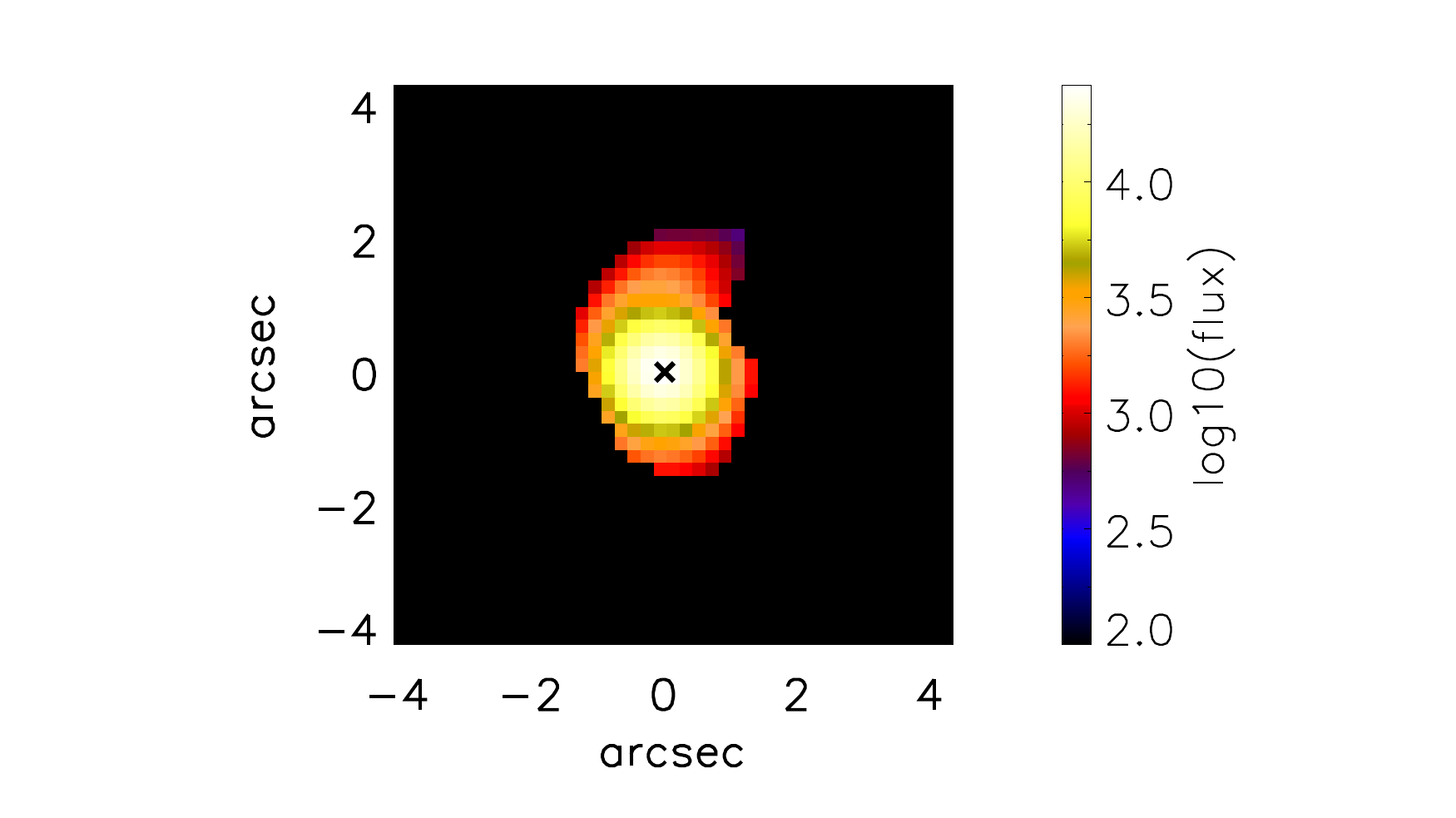}\hspace{-2cm}
\includegraphics[width=0.38\textwidth]{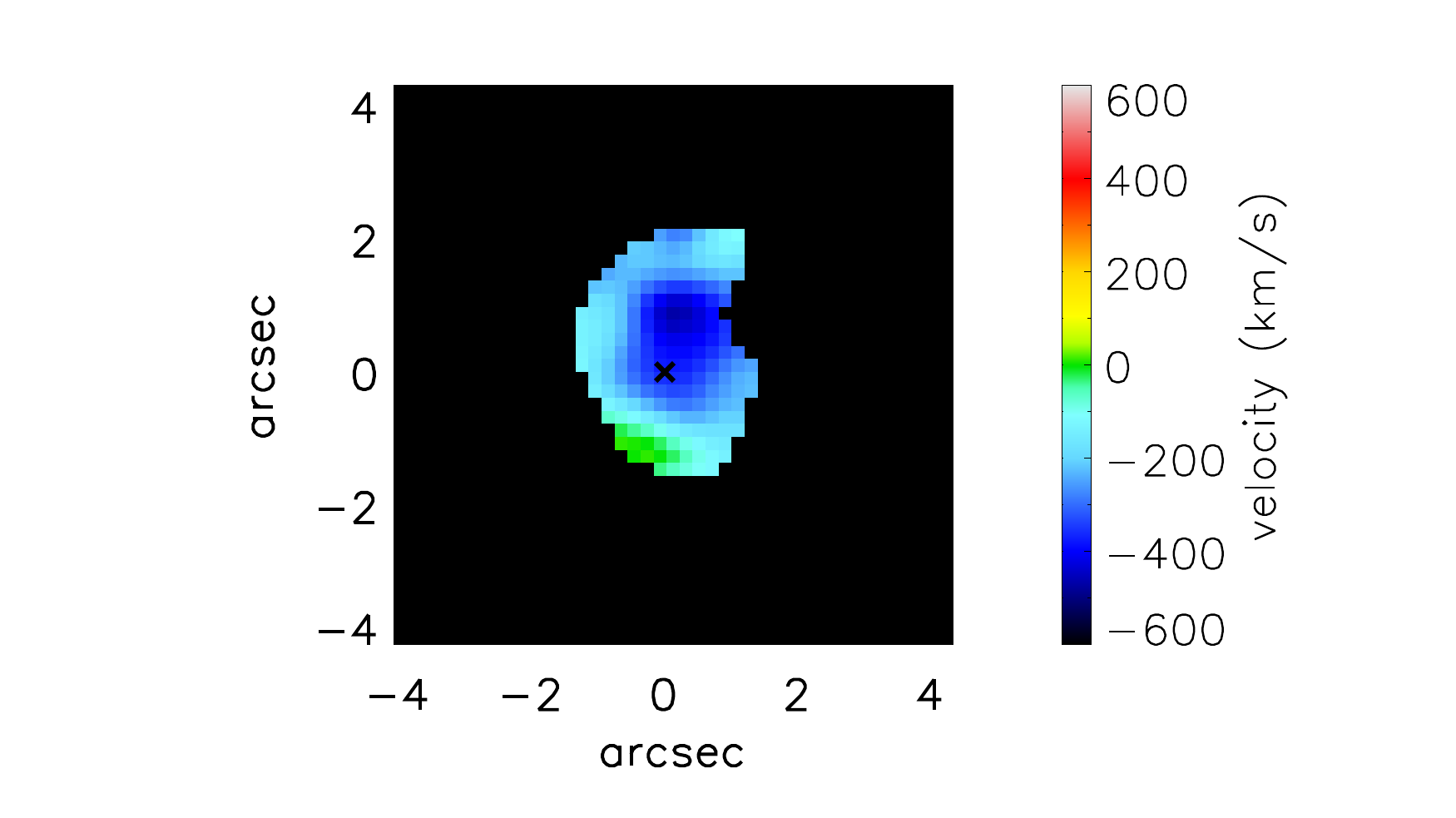}\hspace{-1.5cm}
\includegraphics[width=0.38\textwidth]{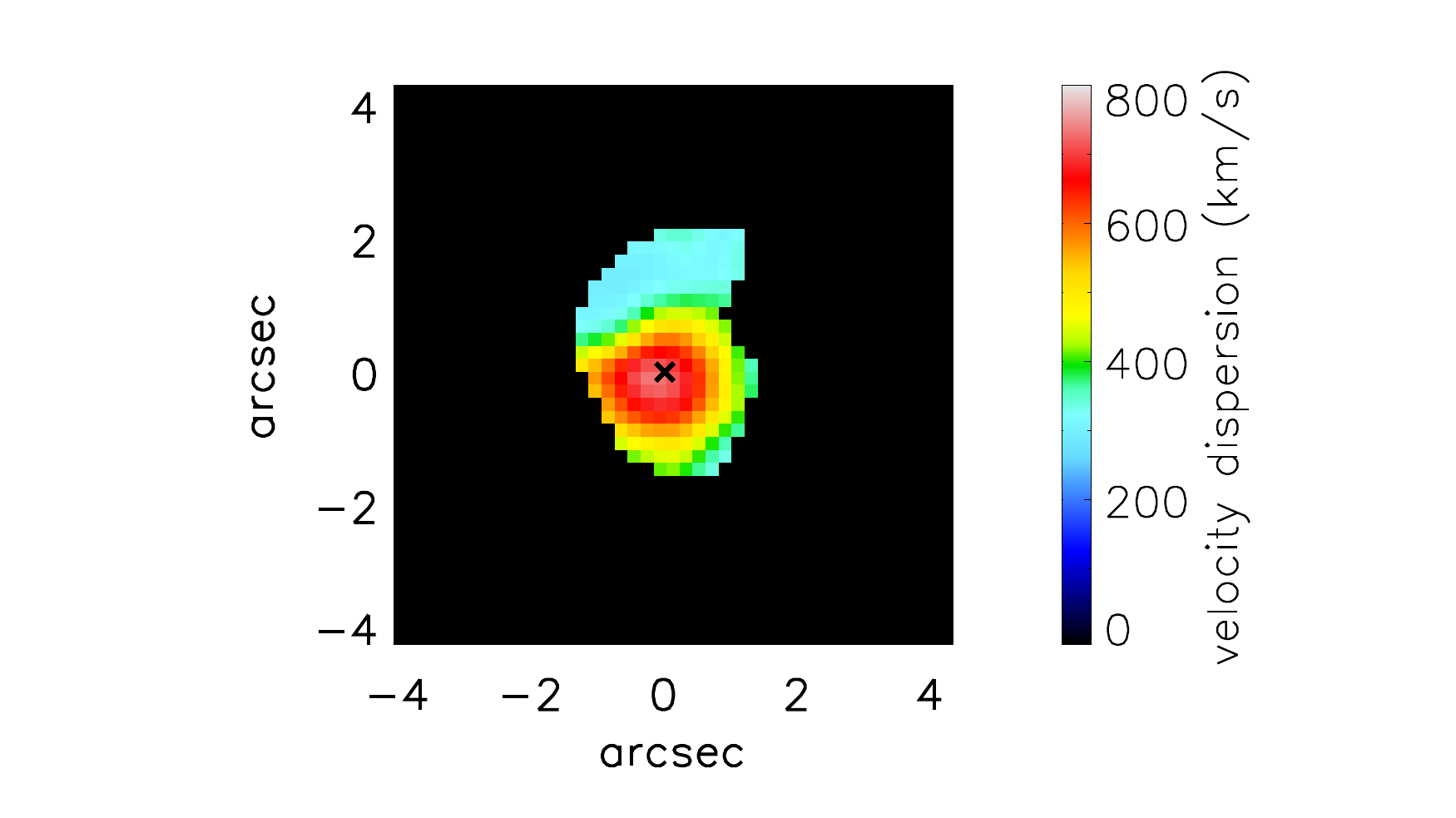}\hspace{-1.5cm}\\[-0.4cm]
\end{center}
\caption{Maps of the broad, blue-shifted component of the [O III] line detected in the nuclear region. Columns from left to right show the line flux, velocity and velocity dispersion maps respectively. Black cross indicates the AGN position as determined from the H$\alpha$ broad emission line peak flux. North is up, East is left.}
\label{gas_dynamics_OIII}
\end{figure*}

We detect an additional broad component in the \mbox{[O III]} line, blue-shifted with respect to the central velocity of the narrow component. This component is also marginally detected in the [N II] line, albeit at a lower S/N. The [O III] line complex is fit using two Gaussian components: a narrow component and a broad blue-shifted component. An example of the fit to the [O III] emission is shown in the top panel of Fig.~\ref{halpha_hbeta}. The distribution, velocity and velocity dispersion of the broad [O III] component is shown in Fig.~\ref{gas_dynamics_OIII}.

\begin{figure*}
\begin{center}
\includegraphics[width=0.9\textwidth]{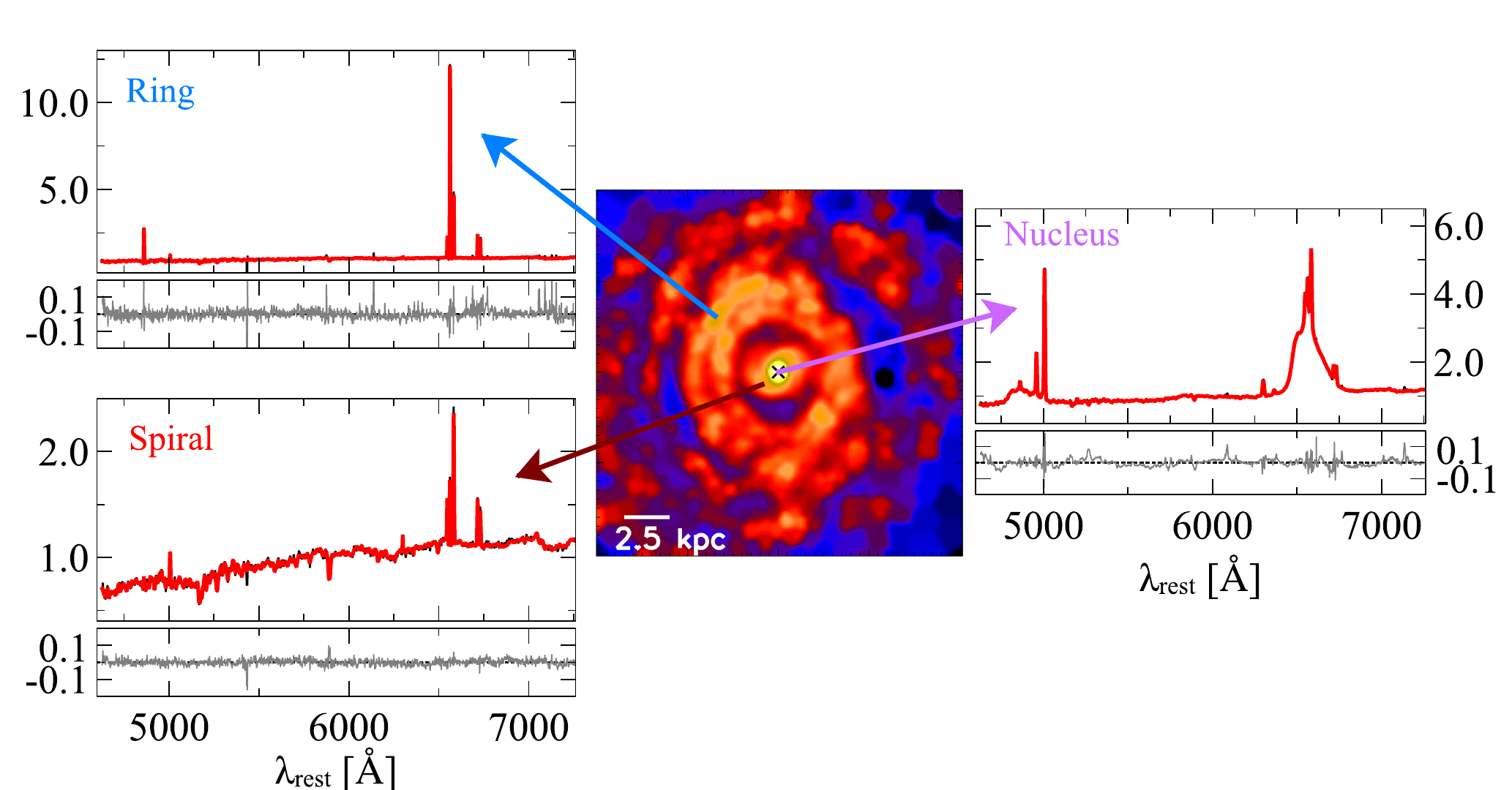}
\end{center}
\caption{Representative \textsc{gandalf} fit results for three different spatial regions. Each panel shows the galaxy spectrum (black), the best fit model (red) and the residuals (grey) for the distinct spatial regions discussed in Section~\ref{sec:ionised_gas}. The central panel shows the narrow H$\alpha$ emission line flux map.}
\label{gandalf_panels}
\end{figure*}

As can be seen in the left column of Fig~\ref{gas_dynamics}, the flux distribution of the ionised gas shows three distinct structures: nuclear emission in the central 2 arcsec (1 kpc), a nuclear spiral structure within a radius $r < 4$ arcsec (2 kpc) (more clearly seen in the close-up view in Fig.~\ref{gas_dynamics_zoom}) and a ring-like structure at $r \sim 9$ arcsec (4.5 kpc) surrounded by larger radii spiral-arms. H$\alpha$, [N II] and [S II] show a similar spatial distribution tracing the large scale structure referred to above. [O III], likely due to its higher ionisation potential, is strongly detected in the nucleus but only weakly detected in the ring. Representative fits for each spatial region are shown in Fig.~\ref{gandalf_panels}. The flux errors are typically lower than 5 per cent in the nucleus, ring of ionised gas emission and nuclear spiral arms. 

In the ring the gas is distributed in clumps emitting with higher intensity, clearly seen in the H$\alpha$ flux map (top row, left column of Fig.~\ref{gas_dynamics}). These likely correspond to individual H II regions that appear as knots of increased line emission. The most popular formation scenario for such a ring is an increased local density of gas and stars. This increase in density is caused by an orbital resonance with a non-axisymmetric perturbation in the gravitational potential, such as that caused by a bar for example (e.g. \citealt{schwarz81}, \citealt{buta86}). Due to evidence pointing to the presence of a bar in Mrk 590, this would be a likely explanation for the formation of the ring we observe.

The two-arm nuclear spiral structure (see left column of Fig.~\ref{gas_dynamics_zoom}) is likely a region of increased gas density and shocks in a gas disc (e.g. \citealt{yuan&kuo97},  \citealt{maciejewski04}). 

The velocity fields of the ionised gas (middle columns of Figs.~\ref{gas_dynamics} and \ref{gas_dynamics_zoom}) have a complex shape, as the gas is more susceptible to dynamical perturbations than the stars. The zero-velocity line shows an S-like shape in the central 10 arcsec, is relatively smooth with small deviations along a P.A. of $\sim$ 30 degrees outside the central 10 arcsec and then has a strong twist towards the outskirts of the field-of-view (r $\sim$ 15 arcsec).
The velocity field is similar to that observed in other local Seyfert galaxies (e.g. NGC 5643 - \citealt{davies14}) with nuclear spiral structures. The twists in the zero-velocity line are likely due to non-circular gas flows in the disc. Departures from circular motion are expected in nuclear spirals (e.g. \citealt{maciejewski04}). Such nuclear spirals can be generated by a non-axisymmetric perturbation in the galactic potential, such as a bar, acting in a rotating disc of gas and stars (\citealt{maciejewski04}, \citealt{binney&tremaine08}).

In terms of the velocity dispersion, the gas shows increased dispersion in the nucleus, reaching values of $\sigma \sim 140 - 180$ km\,s$^{-1}$ in the central 4 arcsec and in general $\sigma > 70$ km\,s$^{-1}$ within the region of the nuclear spiral (r $<$ 4 arcsec) (right colum of Figs.~\ref{gas_dynamics} and \ref{gas_dynamics_zoom}). In almost all the tracers, but more clearly seen in the [S II] and [N II] maps, there appears to be a partial ring or shell of increased velocity dispersion at $\sim$ 4 arcsec from the centre, more clearly seen west of the nucleus. This region broadly coincides with the leading part of the two nuclear spiral arms, assuming that the spiral arms are trailing. It could be associated with shocks that are expected to be present when nuclear gas spirals are observed (e.g. \citealt{maciejewski04}). For r $>$ 4 arcsec the velocity dispersion decreases and shows values of $\sim$ 40 - 50 km\,s$^{-1}$ at the scale of the H$\alpha$ ring. This lower velocity dispersion and enhanced H$\alpha$ emission may be associated with increased star formation in the outer regions. If that is the case, the gas is expected to be in a colder dynamical state with lower velocity dispersion to be able to form stars, as seen for example in the starforming ring of M100 \citep{allard05} and of NGC 7742 \citep{martinsson18}. 

As can be seen from Fig.~\ref{gas_dynamics_OIII} the broad line component of [O III] originates in the nucleus. The emission is blue-shifted, with a mean velocity of V$_{\rm [O III]\,broad}$ $\sim$ -400 km\,s$^{-1}$ with respect to the systemic velocity of the galaxy. The line dispersion is large, with values that reach $\sigma = 700$ km\,s$^{-1}$ in the nucleus. This component is likely associated with an outflow, likely driven by the AGN, due to its location and velocity. It is extended in the N-S direction, similar to what has been observed with \emph{HST} imaging \citep{schmitt03}. The narrow line components presented in Fig.~\ref{gas_dynamics_zoom} also show increased velocity dispersion in the N-S direction, which may be associated with the nuclear outflow propagating to larger radii. In [S II] emission in particular, we see a cone-like structure in the velocity dispersion map, with increased velocity dispersion ($\sim$ 150 km\,s$^{-1}$) at 1.5 arcsec from the nucleus. The velocity dispersion for the other ionised gas tracers also show an elongation along the same North-South axis. This elongation is similar to the [O III] outflowing component geometry, which indicates that the ionised gas bicone is tracing an outflow which could be driven by the AGN (e.g. \citealt{storchi-bergmann07}). Ultra fast outflows at smaller physical scales ($<1$ light-day) have also been detected in Mrk~590 \citep{gupta15}. The outflow we observe in [O III] may be the propagation of the ultra fast outflow to larger radii as the outflow velocity decreases.

A more in depth discussion of the dynamical structures and their role in the fuelling of AGN will be discussed in paper II.

\subsection{Emission line diagnostics}
\label{sec:line_ratios}
The narrow emission lines observed can be excited by different mechanisms such as photoionisation by young stars, photoionisation by the AGN continuum or shock-wave heating \citep{baldwin81}. Integral field spectroscopy data allow us to determine the dominant excitation mechanisms at each spatial position in the galaxy. We use a diagnostic based on prominent emission line flux ratios first proposed by \cite{baldwin81} and modified by \cite{veilleux&osterbrock87}. We use the theoretical regions in line diagnostics diagrams defined by \cite{kewley06} to distinguish between excitation mechanisms. Our line diagnostics include H$\beta$ $\lambda$4861, [O III] $\lambda$5007, H$\alpha$ $\lambda$6563, and the lower ionisation lines [O I] $\lambda$6300, [N II] $\lambda$6583 and [S II] ($\lambda$6716 + $\lambda$6731). %
We determine the emission line fluxes from the \textsc{gandalf} best fit model determined in Section \ref{sec:kinematics}. 

\begin{figure*}
\centering
\includegraphics[width=0.24\textwidth]{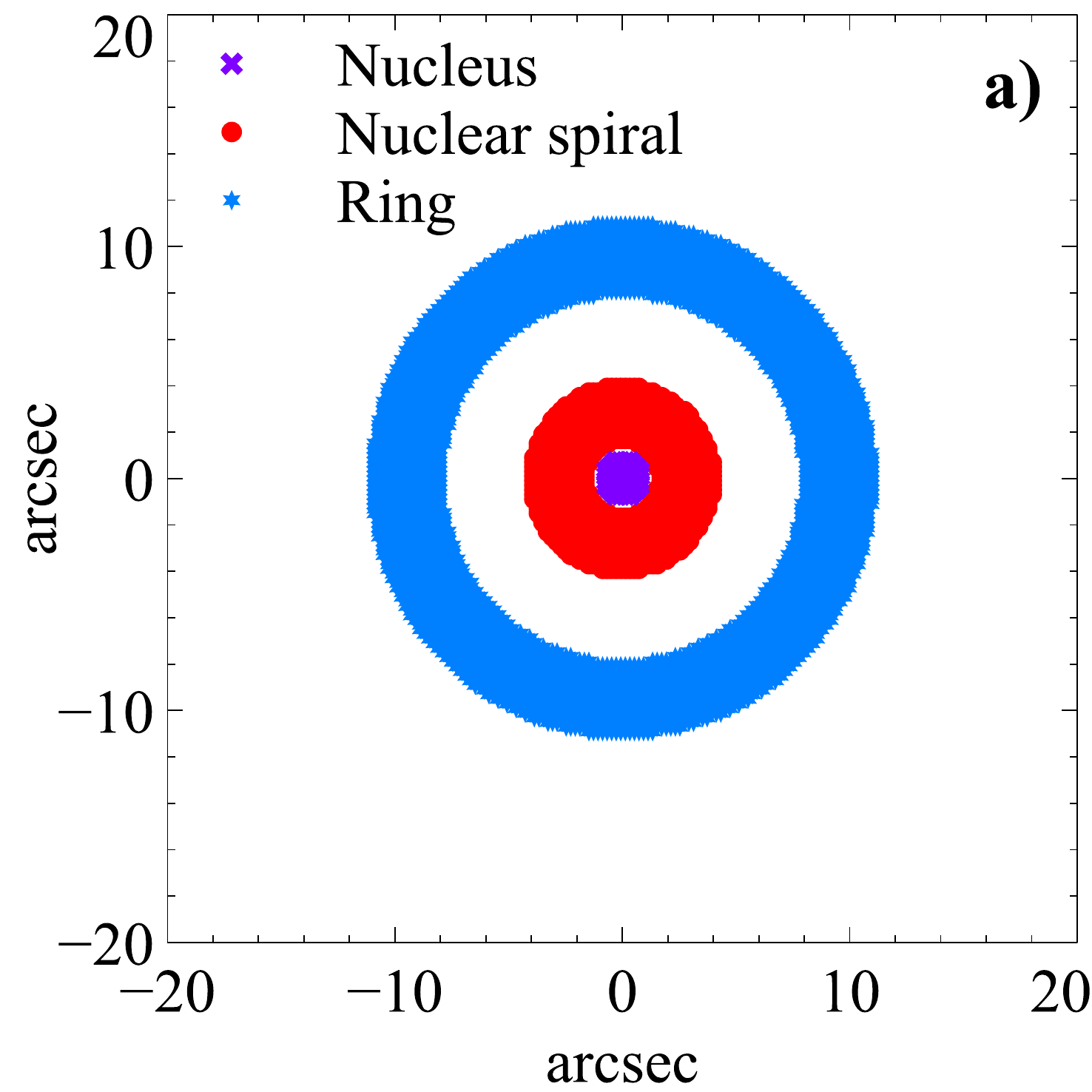}
\includegraphics[width=0.24\textwidth]{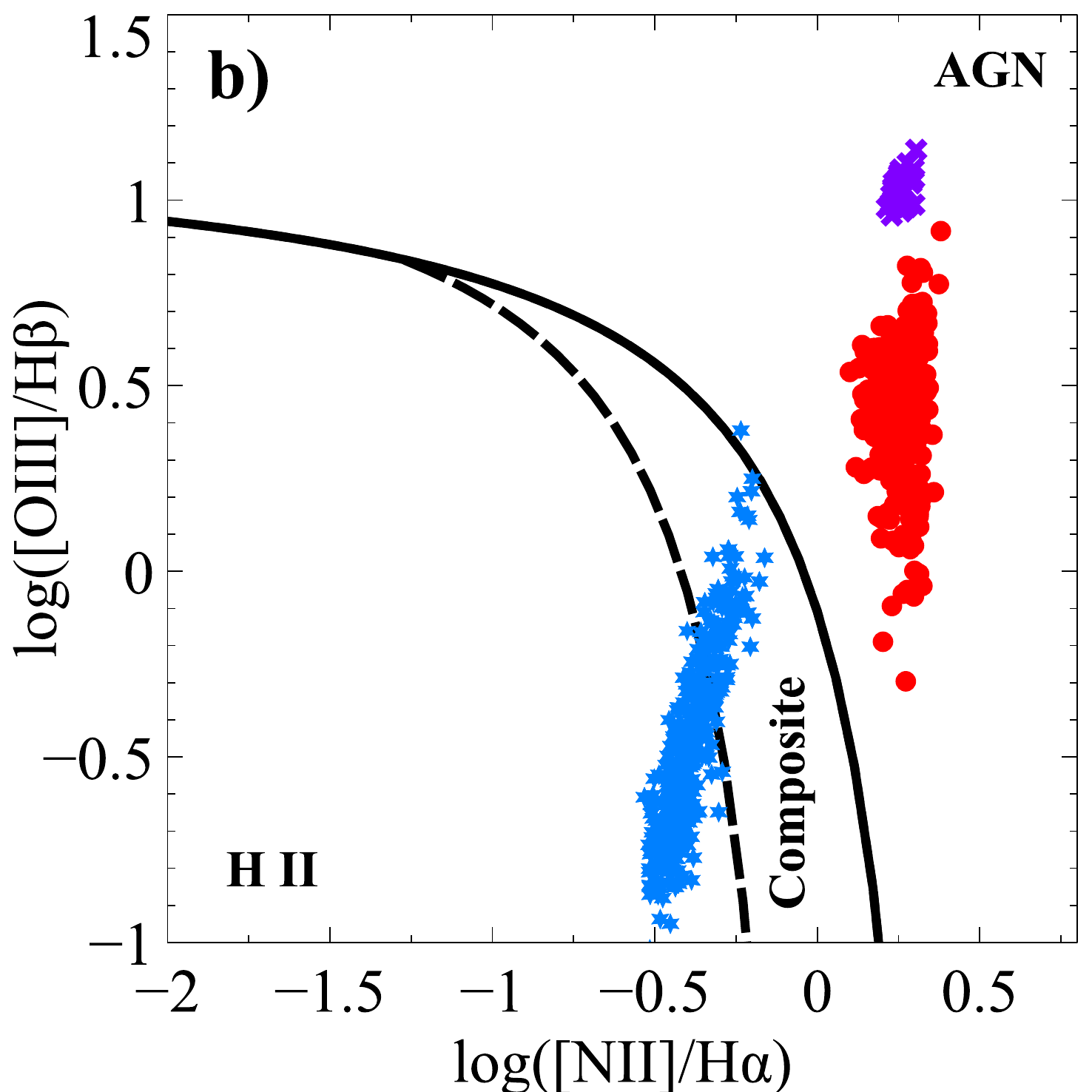}
\includegraphics[width=0.24\textwidth]{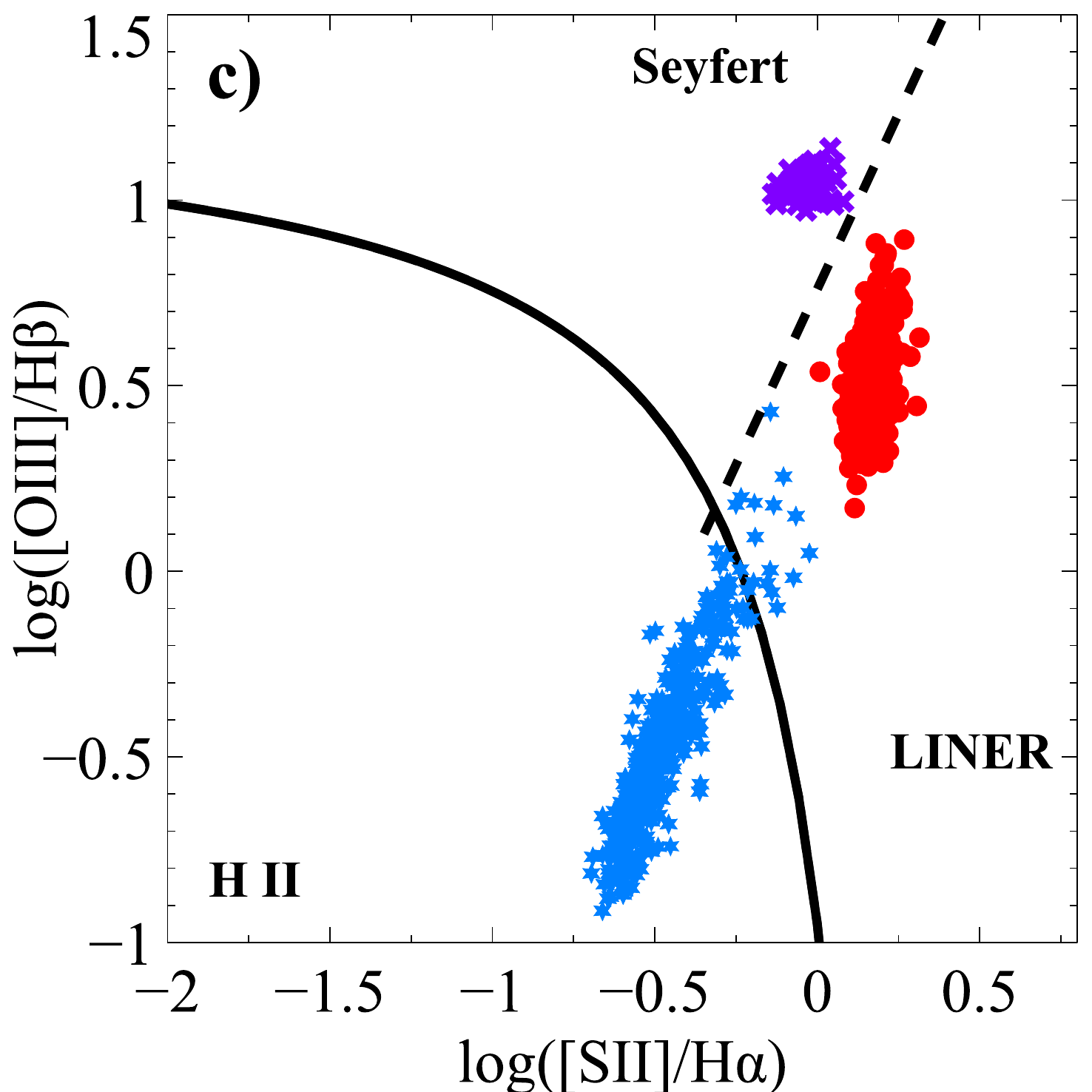}
\includegraphics[width=0.24\textwidth]{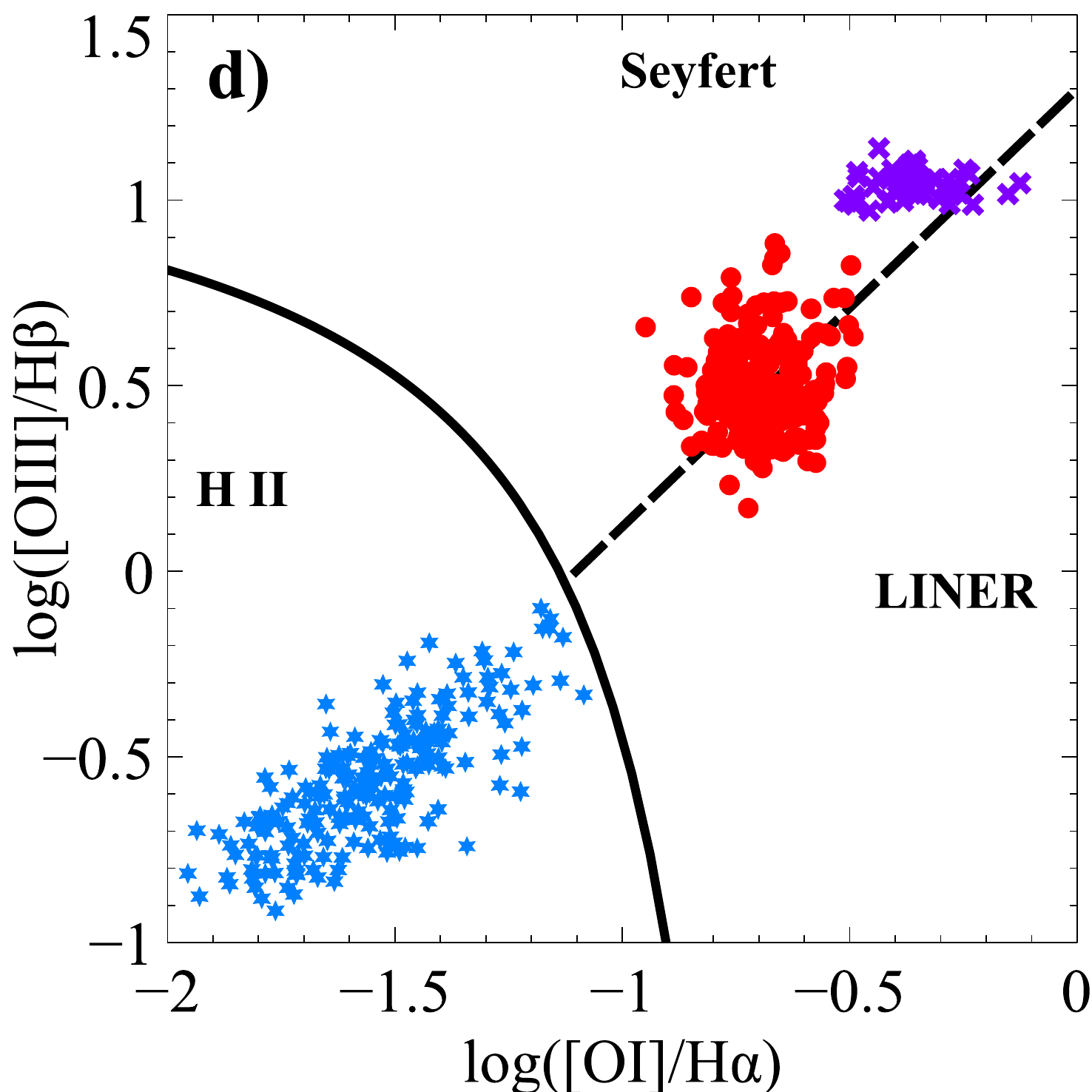}
\caption{Pixel-by-pixel BPT diagrams of the central region of the Mrk~590 galaxy. Panel a) shows the selected regions in the field of view: nucleus (purple crosses), nuclear spiral (red circles) and ring of strong H$\alpha$ emission (blue stars). Panels b), c) and d) show three different line diagnostic diagrams, where the symbols and colours correspond to the regions marked in the left panel. The labeled regions in the line diagnostic panels indicate the dominating excitation mechanism, with the solid and dashed boundary lines from \citealt{kewley06}.}
\label{BPT}
\end{figure*}

To determine the dominating excitation mechanisms in different parts of the galaxy, we isolate three specific spatial regions in the field of view: the nucleus, the nuclear spiral structure and the ring (Fig.~\ref{BPT}\,a). The resulting line ratios for each region with the \cite{kewley06} boundaries overlaid are shown in Fig.~\ref{BPT}\,b), c) and d). 

We can see that in the nucleus the dominant excitation mechanism is the AGN, consistent with excitation by a Seyfert-like continuum. The nuclear spiral appears to be similar to a low-ionisation nuclear emission region (LINER), which can be associated with the AGN and/or with shocks (created by supernova or AGN driven outflows). In the nuclear spiral region we observe an increased velocity dispersion in the gas, which is also consistent with shock-heating. Shocks are expected to increase the [N II]/H$\alpha$ and [S II]/H$\alpha$ ratios and also to broaden the emission lines (e.g. \citealt{davies_rebecca17}). The environment in the ring can be associated with H II regions where the gas is ionised by O and B stars. Between the nuclear spiral and the ring (not shown in Fig.~\ref{BPT}), there is a transition region where the line ratios indicate that the ionisation is likely to have significant contributions of both young stars and AGN. We see a radial variation of the line ratios likely due to a radial change of the relative contribution from AGN and star formation. For example, the [O III]/H$\beta$ ratio (Fig.~\ref{BPT}) is higher in the nucleus and decreases with increasing radius. This is consistent with a decreasing contribution of the AGN to the photoionisation as the distance to the AGN increases. This radial change in line ratios is similar to what has been observed in other integral field spectroscopy studies (e.g. \citealt{scharwachter11}, \citealt{davies_rebecca14}).

\begin{figure}
\includegraphics[width=0.45\textwidth]{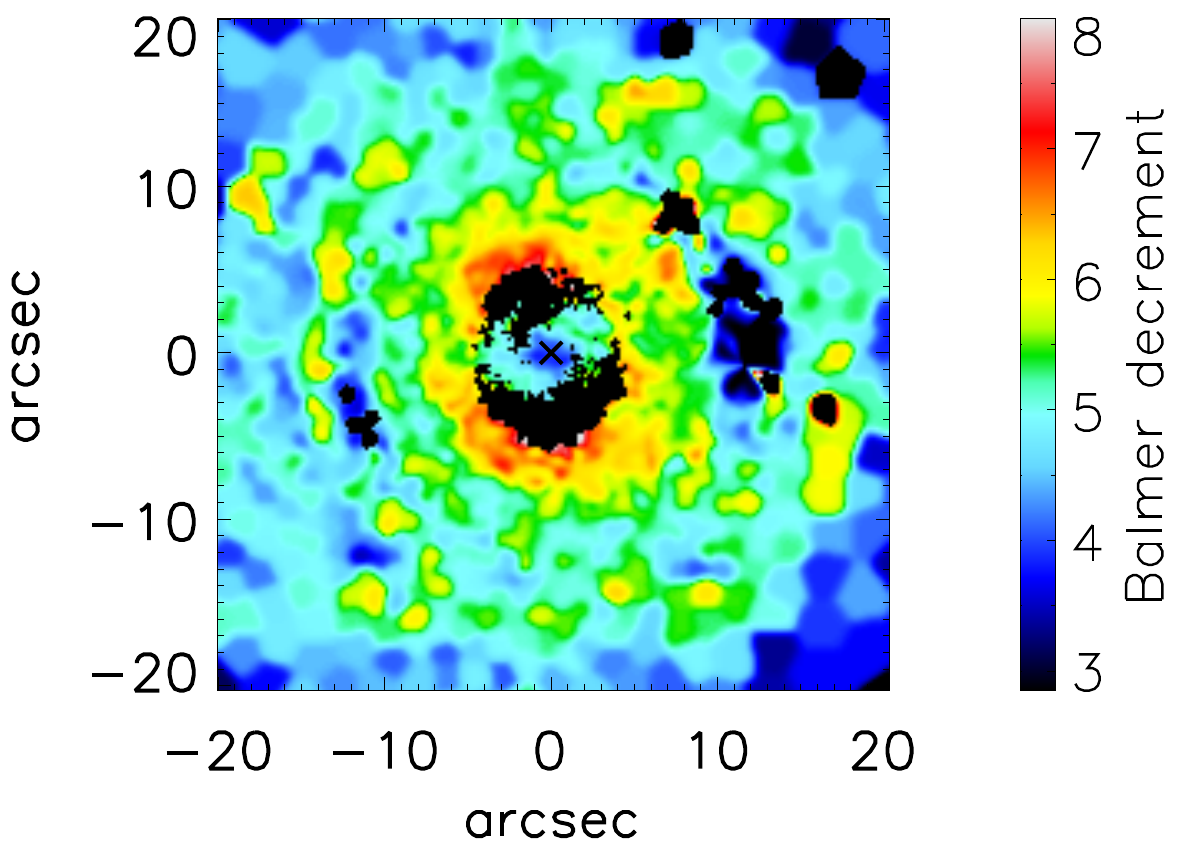}
\caption{H$\alpha$/H$\beta$ the flux ratio between the H$\alpha$ and H$\beta$ narrow emission lines also referred to as the Balmer decrement. Areas with higher ratios are associated with increased dust reddening. Areas shown in black were masked out due to a S/N $<$ 5 in either the H$\alpha$ or H$\beta$ narrow emission lines.}
\label{Balmer}
\end{figure}

The ratio between the hydrogen lines H$\alpha$/H$\beta$, the Balmer decrement, can be used to estimate the dust reddening in the host galaxy. Due to the wavelength dependence of dust extinction, the presence of dust will increase the H$\alpha$/H$\beta$ ratio observed when compared to the expected theoretical value (2.86 for case B recombination; \citealt{osterbrock89}). We calculate the Balmer decrement map from the flux maps of H$\alpha$ and H$\beta$. In Fig.~\ref{Balmer} we show the map of H$\alpha$/H$\beta$ Gaussian smoothed to the spatial resolution of our data. 
This map can be understood as a proxy of the dust distribution in the galaxy. The regions in black have been masked out due to the low S/N of H$\beta$ (S/N $<$ 5) in those regions, using the same S/N threshold criteria as for the analysis in Section~\ref{sec:gas_analysis}. Most of the dust is distributed outside the nuclear spiral region, co-spatial with the H$\alpha$ ring. This is perhaps not surprising since star formation is taking place in this ring. Notably the highest dust reddening (H$\alpha$/H$\beta$ $\sim$ 7.5) occurs in the boundary between the nuclear spiral and the peak flux of the H$\alpha$ ring, suggesting a possible shielding of radiation by the dust. 

\subsection{Molecular gas distribution}
\begin{figure*}
\centering
\includegraphics[width=0.5\textwidth]{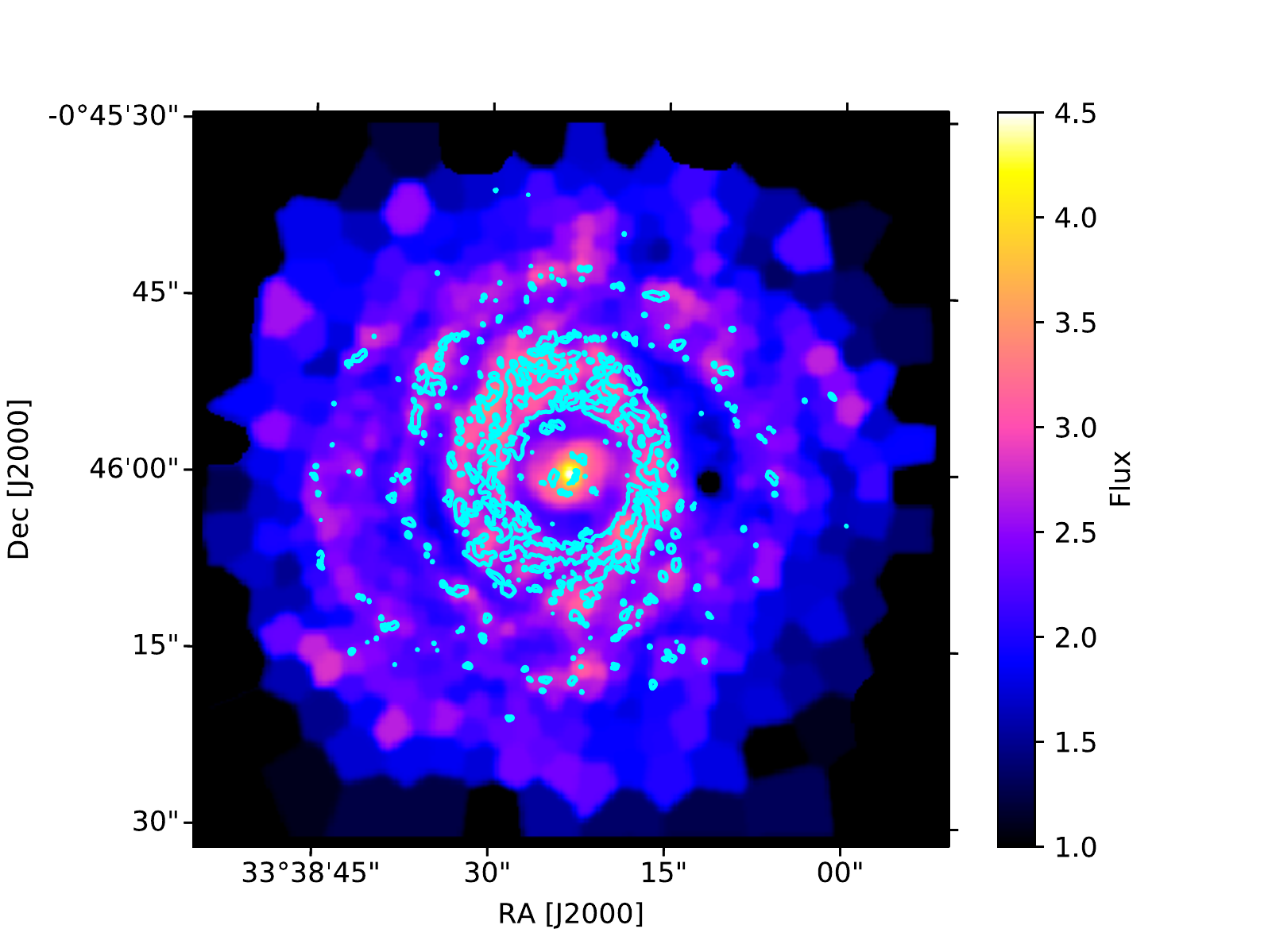} \hspace{-0.8cm}
\includegraphics[width=0.5\textwidth]{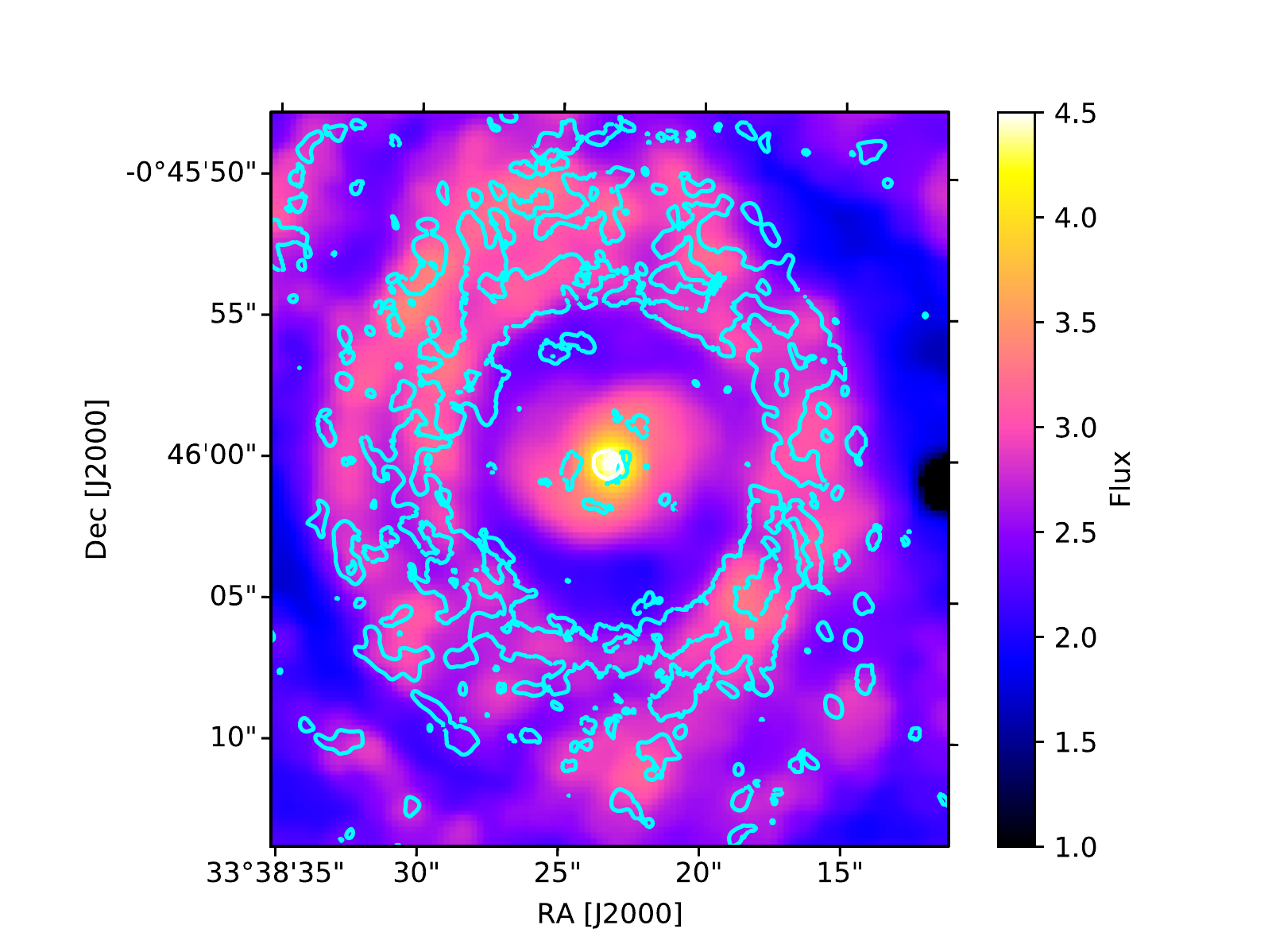}
\caption{ALMA CO(1-0) integrated flux contours in cyan (Koay et al. in prep) overlaid in the H$\alpha$ flux map. The contours are at the level of 20\% of the peak integrated flux (0.097 Jy beam$^{-1}$\,km\,s$^{-1}$). The left panel shows the entire MUSE field of view. The right panel shows a close-up view of the central 25 arcsec$^{2}$ (12 kpc$^{2}$). The white central ellipse shows the 5$\sigma$ contour level of the 344 GHz sub-mm continuum detection obtained with ALMA (\citealt{koay16a}).}
\label{halpha_contour}
\end{figure*}

\begin{figure}
\centering
\includegraphics[width=0.5\textwidth]{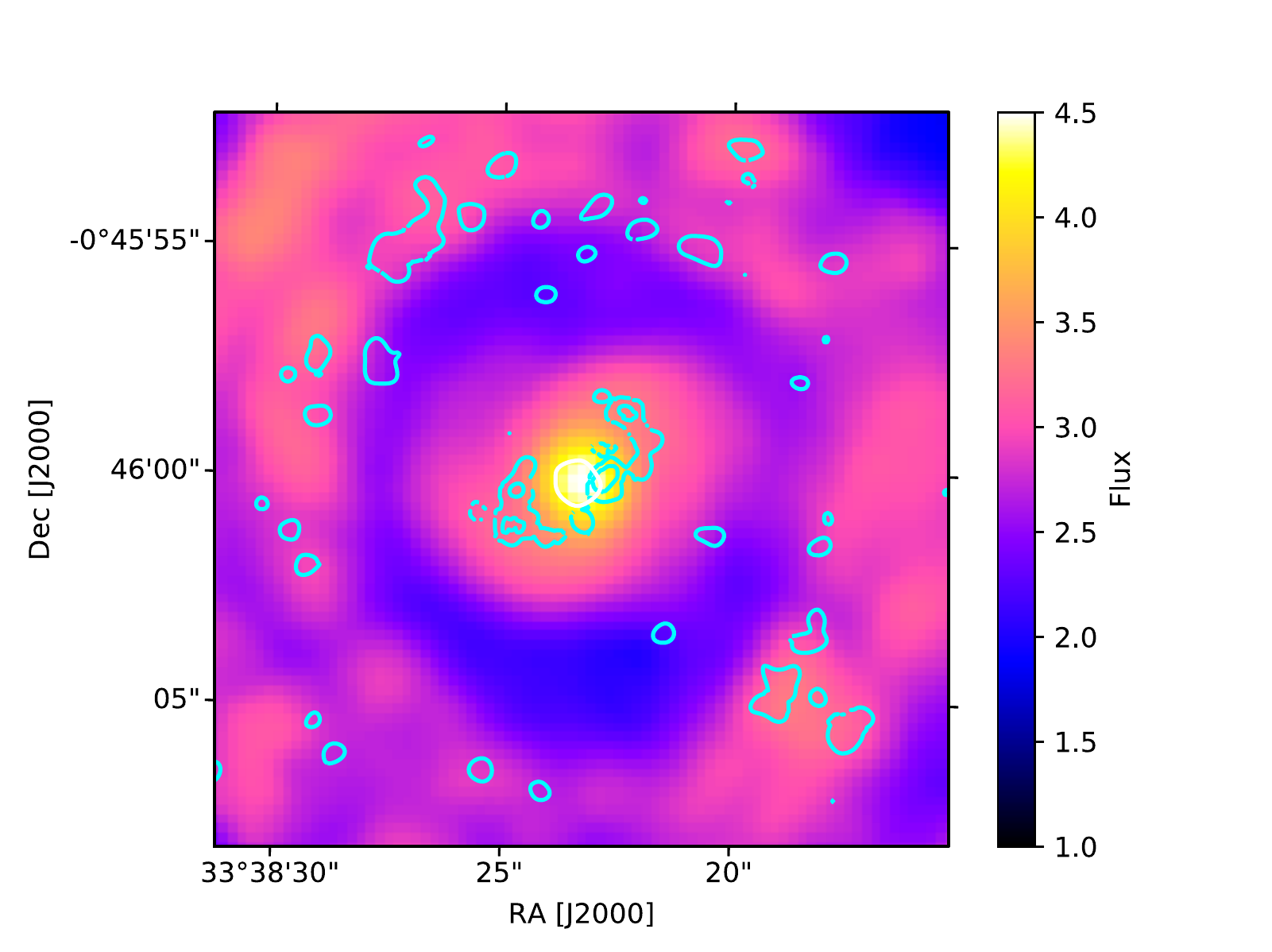}
\caption{ALMA CO(3-2) integrated flux contours in cyan (\citealt{koay16a}) overlaid in the H$\alpha$ flux map. The contours are at the level of 15\% and 50\% of the peak integrated flux (0.27 Jy beam$^{-1}$\,km\,s$^{-1}$). Only a close-up view of the central $\sim$16 arcsec$^{2}$ (8 kpc$^{2}$) is shown, to match the smaller field of view of the ALMA observations. The white central ellipse shows the 5$\sigma$ contour level of the 344 GHz sub-mm continuum detection obtained with ALMA (\citealt{koay16a}).}
\label{halpha_contour_co32}
\end{figure}

In the previous sections we determined the spatial distribution of ionised and warm molecular gas in Mrk~590. Here we compare the spatial distribution of the ionised and warm molecular gas with the cold molecular gas distribution from ALMA observations of CO(3-2) \citep{koay16a} and CO(1-0) (Koay et al. in prep). The CO(3-2) flux maps were presented by \cite{koay16a} and show a ring-like structure of radius $\sim$1 kpc and emission from a clump of gas located $\sim 200$ pc west of the centre. 
The centre is assumed to be the position of the sub-mm continuum emission, where the cm-wavelength radio core is also detected \citep{koay16b}, and no CO line is detected at that position. The CO(1-0) emission broadly follows the CO(3-2) emission, although it is more weakly detected in the nucleus. The detailed analysis of the CO(1-0) maps will be presented by Koay et al. (in prep). In Fig.~\ref{halpha_contour} and Fig.~\ref{halpha_contour_co32} we show contour maps of CO(1-0), CO(3-2) and sub-mm continuum emission overlaid on the H$\alpha$ flux map to compare the cold molecular gas and ionised gas distributions. 

The CO(1-0) appears to trace the inner boundary of the H$\alpha$ ring and the large radii spiral arms to the North of the ring. The strongest H$\alpha$ emission in the ring at $r \sim$ 9 arcsec is external to the ring of CO(1-0) emission (Fig.~\ref{halpha_contour}), which has also been observed in other Seyfert galaxies \citep{storchi-bergmann07}. There is no significant detection of CO(1-0) in the nucleus or the nuclear spiral. The position of the sub-mm continuum is consistent with the AGN position determined from the broad emission lines, considering the size of the optical PSF (FWHM $\sim$ 0.6 arcsec). This suggests that gas is present in the nucleus but due to the local excitation conditions, such as those caused by the presence of the AGN (e.g. \citealt{casasola08}, \citealt{dumas10}), CO(1-0) or CO(3-2) emission is not favoured. 
This hypothesis is strengthened by the warm molecular gas distribution traced by H$_{2}$ 2.12$\micron$ emission. The H$_{2}$ 2.12$\micron$ emission is stronger in the nucleus but we also detect extended emission that seems to match the spatial distribution of the ionised gas. Curiously, the H$_{2}$ 2.12$\micron$ emission is extended towards the West, which could be mapping the clump of CO(3-2) emission located at $\sim$ 200 pc West of the nucleus observed in the ALMA data \citep{koay16a}. The CO(3-2) emission is also not detected in the nucleus \citep{koay16a}. The detection of nuclear H$_{2}$ 2.12 $\micron$ emission indicates that it is a better tracer of molecular gas for the excitation conditions at place in the nucleus of Mrk~590. This has been observed in other Seyfert galaxies, such as for example in NGC 4151, where H$_{2}$ 1-0 S(1) is detected in the nucleus but CO(2-1) is not \citep{dumas10} . 

The Balmer decrement shown in Fig.~\ref{Balmer} which is tracing the extinction by dust, matches the cold molecular gas distribution inferred from the CO emission, indicating that the cold molecular gas and dust are co-spatial, as found in other galaxies of the local Universe (e.g., \citealt{casasola17}). The H$\alpha$ ring of emission peaks at slightly larger radii, suggesting that it could be tracing the outer layer of the molecular gas and dust emission region.

\subsection{Stellar population}
In this section we carry out a spatially resolved study of the stellar age and metallicity of the galaxy. Our goal is to characterise the central stellar population of Mrk~590 to study its past star formation history and the relation between the stellar population and the gas tracers. We use the stellar population synthesis code \textsc{STARLIGHT} (\citealt{cidfernandes05}, \citeyear{cidfernandes09}) to model the observed spectra. Below we summarise the modelling procedure, but for more details on \textsc{starlight} see \cite{cidfernandes04}, \cite{cidfernandes05} and references therein. 
We use the set of theoretical stellar spectral templates from the E-MILES evolutionary population synthesis models \citep{vazdekis16}, covering a range of stellar ages (0.0631 - 17.7828 Gyr) and metallicities (-2.32 $<$ [Fe/H] 
$<$ 0.22). We used the canonical base models obtained with the Padova isochrones \citep{girardi00} and a Chabrier initial mass function \citep{chabrier03}. Since Mrk~590 is an active galaxy, we add an extra template to model the featureless continuum (FC) as in \cite{cidfernandes04} and \cite{riffel09}. The continuum template is of the form of $F_{\nu} \propto \nu^{-1.5}$. We apply STARLIGHT to a binned data cube with an average S/N per bin of 40. Since the AGN continuum emission originates from a spatially unresolved region in the nucleus of the galaxy, it is expected that its contribution is only significant in the central pixels of the data cube, with a spatial extension dictated by the PSF of the observations. Therefore for the bins covering the nucleus of the galaxy (r $<$ 2 arcsec) we use all the stellar templates and the FC template. For the remaining bins we use only the stellar templates as the AGN continuum is not expected to contribute in those regions.

\begin{figure*}
\centering
\includegraphics[width=0.32\textwidth]{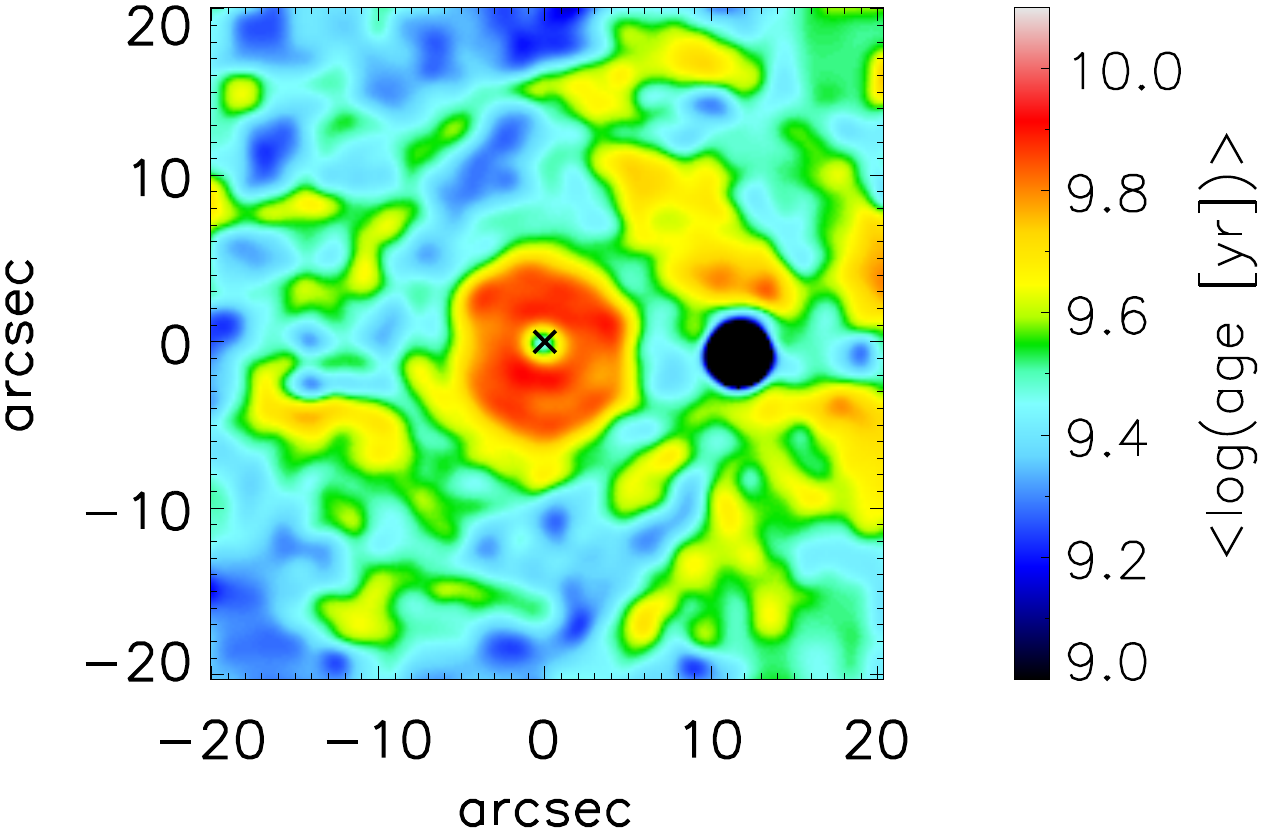}\hspace{0.1cm}
\includegraphics[width=0.32\textwidth]{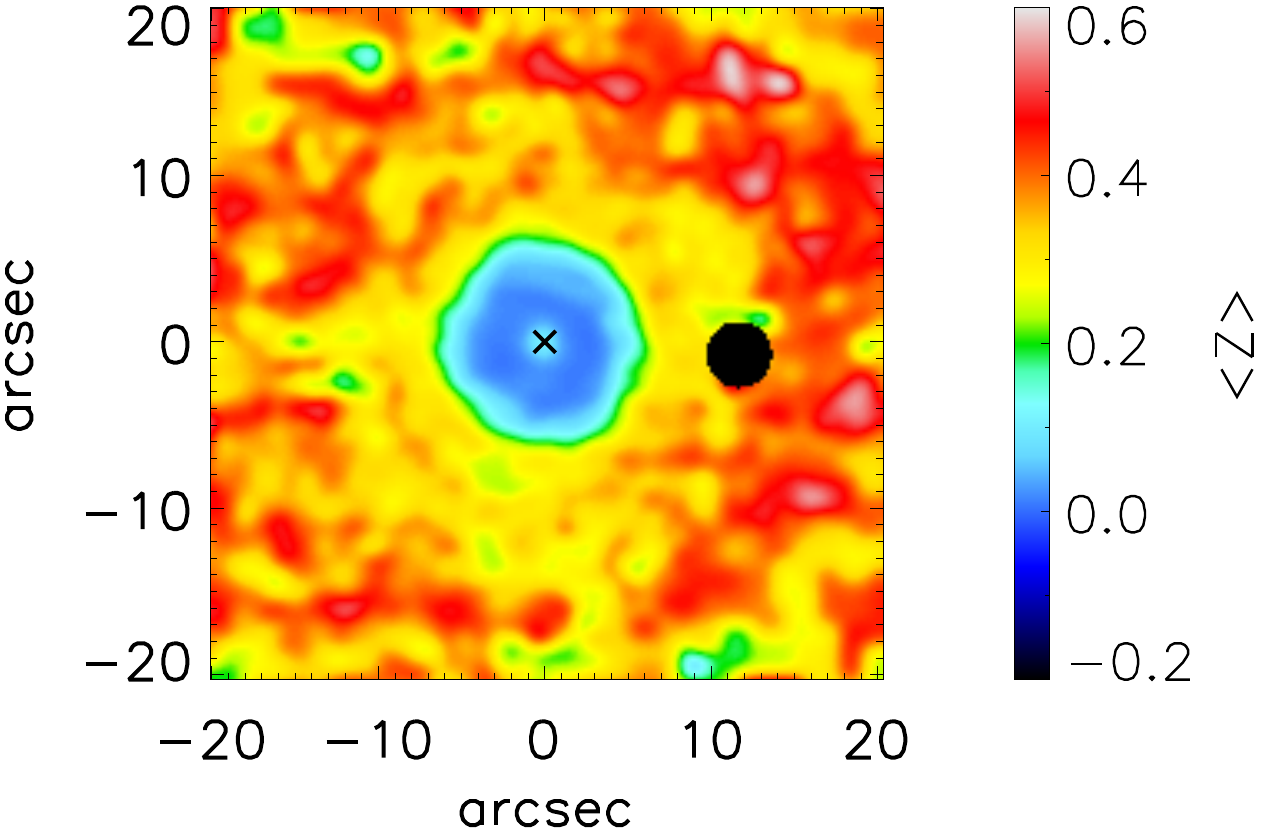}\hspace{0.1cm}
\includegraphics[width=0.32\textwidth]{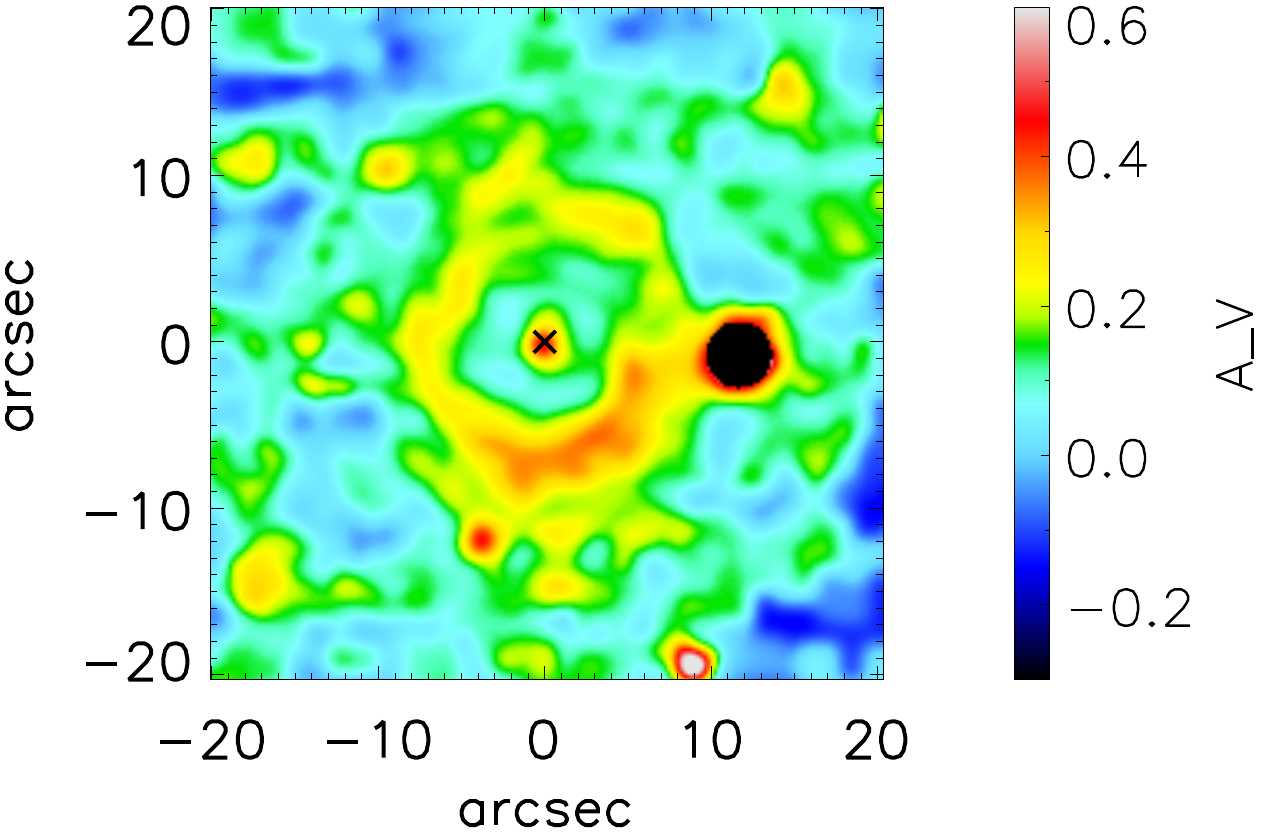}\hspace{0.1cm}
\caption{Stellar population properties determined from stellar population synthesis with \textsc{starlight}. Left: Luminosity weighted mean stellar age. Centre: Luminosity weighted mean metallicity. Right: Spatial distribution of the V-band extinction (A$_{V}$).The black cross indicates the AGN position.}
\label{stellar_pop}
\end{figure*}

\begin{figure*}
\centering
\includegraphics[width=0.33\textwidth]{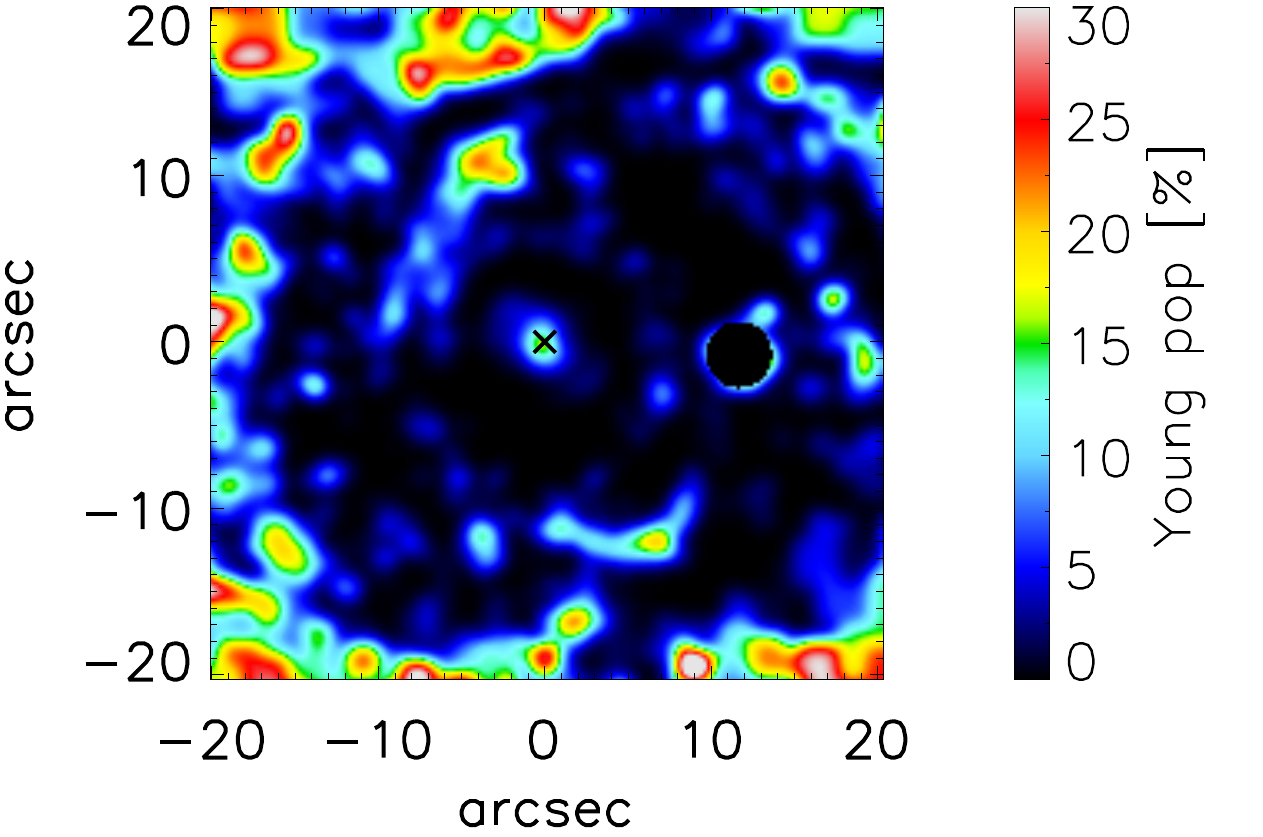}
\includegraphics[width=0.33\textwidth]{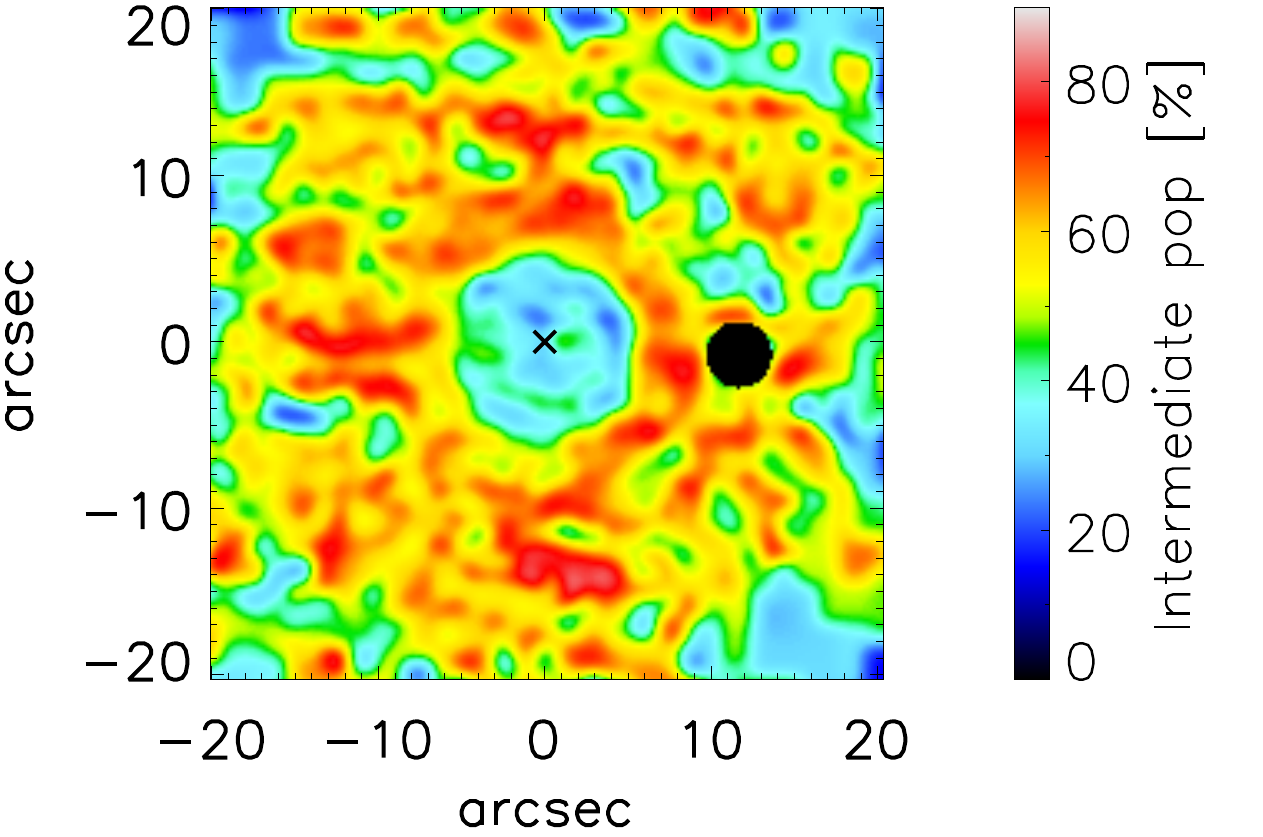}
\includegraphics[width=0.33\textwidth]{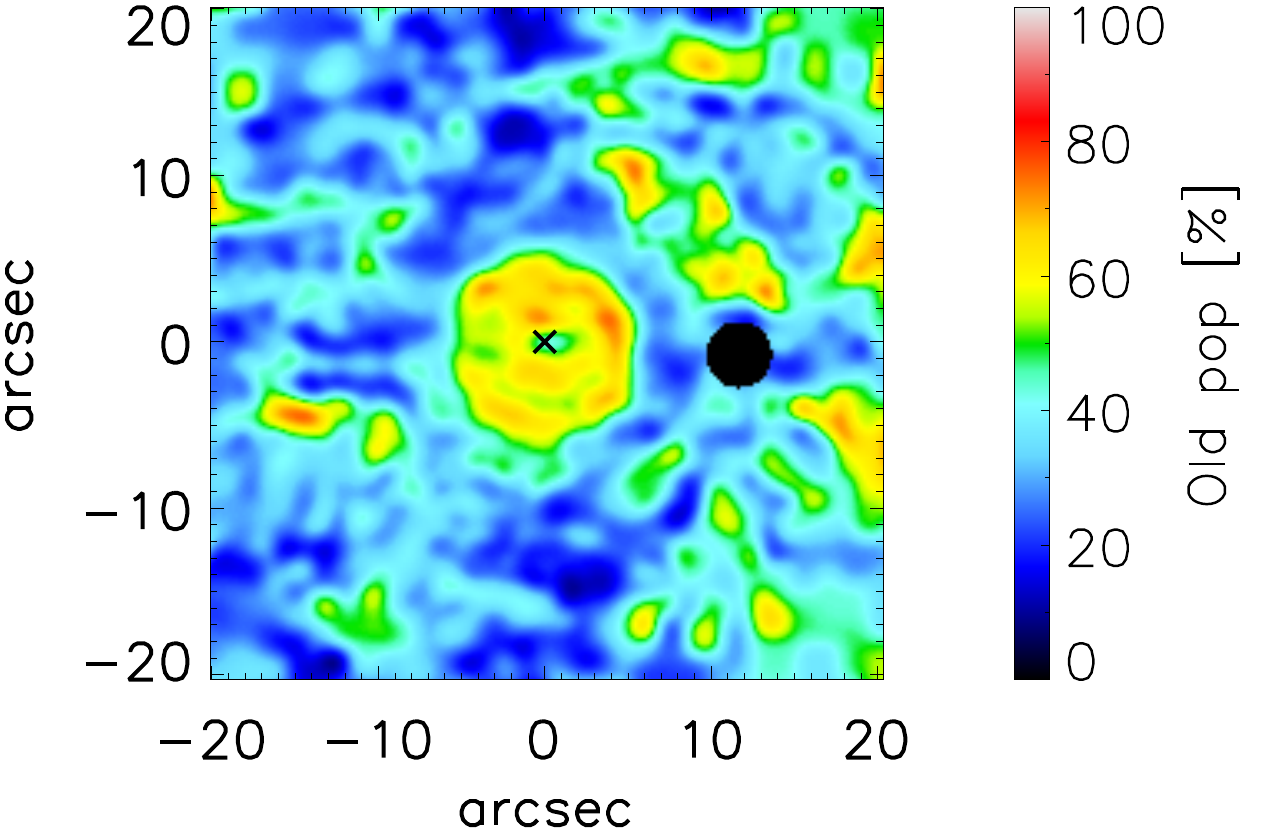}
\caption{Fraction of stars in three stellar age bins. Left: Young age stars ($<$ 1 Gyr). Centre: Intermediate age stars (1 Gyr $<$ age $<$ 5 Gyr). Right: Old age stars ($>$ 5 Gyr).}
\label{stellar_age}
\end{figure*}

The spectrum in each bin is corrected for Galactic reddening using the reddening curve of \cite*{cardelli89}. In \textsc{STARLIGHT} we use a mask to avoid spectral regions that contain emission lines, such as H$\alpha$, H$\beta$, [O III], [N II] and [S II]. For the nuclear regions where an FC template is included, the spectral windows for the masking of H$\alpha$ and H$\beta$ are wider, to account for the AGN broad emission lines. We also flag spectral pixels that show high uncertainties so that they do not affect the \textsc{STARLIGHT} modelling. Our criteria to identify these pixels involve first calculating a 3 sigma-clipped mean of the flux errors in 500 \AA\ wide sections of the spectra, and flag pixels for which the flux errors deviate by 3$\sigma$ or more from the 3 sigma-clipped mean. These flagged pixels are ignored in the fit. We assume a Calzetti dust reddening curve \citep{calzetti00} to model the host galaxy reddening and the code fits the V-band continuum extinction (A$_{V}$) at each spatial bin. The fitted wavelength range is 4644 $-$ 9000 \AA. As output from \textsc{STARLIGHT} we obtain the combination and relative weights of reddened stellar and FC templates that best fit the observed spectrum at each spatial bin. The code also outputs the $\chi^{2}$ of the fit to the observed spectrum. 

From the output of \textsc{starlight} we can also calculate two single parameters to characterise the stellar population of a spatial bin: the luminosity-weighted mean stellar age ($<$log\,t$>$) and the luminosity-weighted mean metallicity \mbox{($<Z>$)}. These two parameters are calculated using:

\begin{equation}
<\log\,t> = \sum^{N}_{j=1} x_{j} \log(t_{j})
\end{equation}

and:

\begin{equation}
<Z> = \sum^{N}_{j=1} x_{j} Z_{j}
\end{equation}

\noindent where $j$ is the stellar template number, out of a total of $N$ templates, x$_{j}$, $t_{j}$ and $Z_{j}$ are the relative weight, age and metallicity of template $j$, respectively. The metallicity is expressed as a mass fraction where the solar metallicity is Z$_{\odot} = 0.02$. The maps with the results for the mean stellar age, mean metallicity and A$_{V}$ are shown in Fig.~\ref{stellar_pop}. The FC contribution is only significant in the PSF region, with a maximum flux contribution of 30 per cent. To evaluate the relative weight of young, intermediate and old stellar populations we bin the stellar ages found by \textsc{starlight} into three age ranges: 
young age population ($<$ 1 Gyr) intermediate age population (1 Gyr $<$ age $<$ 5 Gyr) and old population ($>$ 5 Gyr). The results are shown in Fig.~\ref{stellar_age}. We use Monte Carlo simulations to estimate the uncertainties in the parameters. We do not do this for the entire field-of-view due to the computational resources required, but calculate the uncertainties for specific regions of the field-of-view. We perturb each \textsc{starlight} input spectrum with a Gaussian error with a wavelength-dependent $\sigma$ corresponding to the flux error in each spectral pixel. We run 100 Monte Carlo realisations and estimate the uncertainties in the parameters as the standard deviation of all the realisations. Within the central 10 arcsec the uncertainties in $<$$\log$\,t$>$ are typically 0.05 - 0.1 dex while the uncertainties in $<$Z$>$ are lower than $0.05$ dex and 0.03 mag in A$_{V}$. For the regions outside the central 10 arcsec the uncertainties increase due to the lower signal to noise in the stellar features in each bin. In the ring of H$\alpha$ gas the uncertainties are typically 0.2 - 0.4 dex in $<$$\log$\,t$>$,  lower than 0.2 dex in $<$Z$>$ and 0.2 mag in A$_{V}$. There are additional uncertainties associated with for example the choice of evolutionary synthesis models which we do not evaluate in this work but that have been analysed in detail in \cite{cid-fernandes14}. They find typical uncertainties of the order of 0.15 - 0.25 dex in the mean age and metallicity and 0.1 mag in A$_{\rm V}$ when using different synthesis models.
From the left panel of Fig.~\ref{stellar_pop}, we can see that the average stellar population age is $>5$ Gyr in the central 10 arcsec of the galaxy. At larger radii ($r > 5$ arcsec) there is evidence of younger stellar populations, with mean stellar age of $<5$ Gyr. The H$\alpha$ and CO(1-0) emitting ring at $\sim$9 arcsec radius, in particular, shows a stellar population with an age of 1 - 2 Gyr on average, significantly lower than the one in the central 10 arcsec. The regions of r $\gtrsim$ 5 arcsec with the younger stellar age also show higher metallicity than the older stellar population regions, as expected if they were formed at later times.

The stellar population results and the ionised gas velocity dispersion show that the more recent star formation is occurring in the H$\alpha$ ring. The left panel of Fig.~\ref{stellar_age} shows that there is a significant contribution of stars with age $< 1$ Gyr within the ring. This is consistent with the lower gas velocity dispersion observed in that region providing more favourable conditions to the formation of stars. This is also consistent with the emission line ratios that show excitation by young stars in the ring. 
The accumulation of gas in the ring may be due to dynamical resonances associated with non-axisymmetric perturbations to the potential, such as those produced by a bar. CO(1-0) is detected in that region, indicating that there is still a reservoir of molecular gas to form stars. Recently, a type Ia supernova (ASASSN-18pb) was detected within the host galaxy Mrk~590 \citep{brimacombe18}. The location of this event was measured to be at approximately 4.5 arcsec north and 6.4 arcsec east of the nucleus, which from our data can be identified to be located in the ring of strong H$\alpha$ emission where the relatively younger stellar population is observed and also to coincide with a region of CO(1-0) emission.

As can be seen from the right panel of Fig.~\ref{stellar_age}, within the central 10 arcsec most of the stars have older ages $>$5 Gyr. This is consistent with the higher stellar velocity dispersion observed in this region, indicating that the stars were formed earlier than those in the H$\alpha$ ring. The gas velocity dispersion in the central 10 arcsec, as shown in Section~\ref{sec:ionised_gas}, may be too high for stars to form. And as shown in Section~\ref{sec:line_ratios} the excitation mechanism in this region may have the contribution of shocks, which would cause strong shear and prevent star formation (e.g. \citealt{allard05}). 
There is however a shell-like or ring-like region of slightly younger stellar age within the central 10 arcsec, located at $\sim$ 3 arcsec from the nucleus and shown as a yellow region of age $\sim$ 6 Gyr surrounded by the 10 Gyr old stellar population in the left panel of Fig.~\ref{stellar_pop}. This region coincides with the drop in mean stellar velocity discussed in Section~\ref{sec:stellar_kinematics}. This region may be associated with the dynamical effect of the bar on the nuclear dynamics, which possibly resulted in the accummulation of gas and subsequent localised star formation.

The \textsc{starlight} stellar continuum-measured A$_{V}$ spatial distribution (right panel of Fig.~\ref{stellar_pop}) agrees with the dust distribution inferred from the Balmer decrement. The dust is clearly seen to be distributed within the ring of H$\alpha$ and CO(1-0), consistent with the notion that the ring is a region of more recent star formation. The increased extinction to the south of the nucleus suggests that the South side is the near side of the galaxy (e.g. \citealt{devaucouleurs58}). If this is the case, combined with the gas velocity field information it implies that the nuclear spiral arms are trailing.

\subsection{Stellar and gas dynamics}
A more detailed analysis of the gas dynamics and inflow and outflow structures will be discussed in paper II. Here we do a preliminary summary of our findings regarding the stellar and gas dynamics.

We observe a spiral structure that extends from a radius of 2 kpc down to the nuclear region $r \sim 500$ pc. Nuclear spirals can be caused by the presence of a bar (e.g. \citealt{maciejewski04}). As discussed in Section~\ref{sec:stellar_kinematics}, there is evidence of a bar in Mrk~590, which could be generating the nuclear spiral and also be related, via its orbital resonances, to the accumulation of gas at the position of the gas ring. Nuclear spiral structures have been suggested as a mechanism to transport gas from kiloparsec scales to the central hundred parsecs and have been observed in other Seyferts (e.g. \citealt{fathi06}, \citealt{davies14}). \cite{maciejewski04} shows that in the presence of a supermassive black hole, nuclear spiral shocks are able to drive gas inflow to the nucleus, at the required mass accretion rates to fuel Seyfert-like AGN. At the spatial resolution of our data we are not able to determine the dynamical mechanisms at work within a central distance of r $<$ 500 pc besides measuring an accumulation of gas. This is due to the spatial resolution of our optical data and the low S/N of the near-infrared observations that do not allow us to determine the dynamics of the H$_{2}$ gas. However, we observe that ionised gas and molecular gas are present at scales of tens to hundreds of parsecs. The warm molecular gas in particular is detected down to $r \sim 60$ pc from the black hole.

We observe an outflow of ionised gas in the nucleus of the galaxy, traced by a blueshifted and broad component of the [O III] emission. The outflow is spatially resolved, with a projected total spatial extent of $\sim$1.5 kpc in the North-South direction. The blue-shifted component of the outflow dominates but there are hints of a weaker redshifted component south of the nucleus. These North and South components are consistent with being part of an outflow along a bicone. Considering that the South side is the near side of the galaxy, we are therefore seeing the outflow component to the North being emitted towards the observer, while the redshifted component is being emitted away from the observer and obscured by the gas and dust in the disc of the galaxy and will therefore appear weaker in flux.

The nucleus of Mrk~590 shows complex dynamics due to the effect of non-axisymmetric perturbations to the gravitational potential and respective orbital resonances. Ionised and molecular gas are present in the nucleus, indicating that the AGN is not yet running out of gas to fuel its activity. A nuclear spiral, a dynamical structure that is able to transport gas to the centre of the galaxy, is observed. This indicates that there is a mechanism in Mrk~590 that is able to transport gas from kiloparsec scales or more down to $r \sim 500$ pc. However there is also a nuclear outflow that may be able to remove gas, at least temporarily, from the nucleus of the galaxy. The presence of both inflow and outflow structures within a single galaxy has also been found in other AGN host galaxies in the local Universe both for the ionised and molecular gas (e.g. \citealt{combes13}, \citealt{riffel13},  \citealt{davies14}, \citealt{garcia-burillo14}, \citealt{brum17}).

The physical scales probed by our observations (r $\sim$ 60 pc - 10 kpc) are significantly different from the accretion disc scales, which are necessary to explain AGN type transitions occurring within a matter of years or decades. The dynamical timescale at a radius of r $\sim$ 1 kpc, where the nuclear spiral is observed, is roughly $\sim$ 10 Myr. While this shows that the dynamical features we observe occur on much longer timescales and are not directly responsible for the AGN type transitions of the past few decades, it indicates that there may be a balance between gas inflow and outflow controlling and modulating the amount of gas available to the AGN on longer timescales. This is in line with the occurrence of episodic AGN activity (e.g. \citealt{schawinski15}, \citealt{mathur18}) and the long-term flickering behaviour of AGN, with order of magnitude flux variations on timescales of $\sim 10^{3} - 10^{6}$ yr (e.g. \citealt{hickox14}).

\section{Conclusions}
\label{sec:conclusions}
In this work we analysed the central region of Mrk~590, the host galaxy of a changing-look AGN. We determined the stellar and gas kinematics, the gas spatial distribution and excitation mechanisms and the star formation history of Mrk~590. The AGN in Mrk~590 went through a striking change between 2006 and 2012, showing a dramatic flux decrease and the disappearance of previously strong optical broad emission lines. In this work we found that the optical broad emission lines (H$\alpha$ and H$\beta$) have re-appeared, after being absent for $\sim$ 10 years. The AGN continuum flux however, has not returned to the high values observed in the 1990s when Mrk~590 was classified as a Seyfert 1. While the AGN flux at 5100 \AA\ has increased compared with the 2014 level, it is still $\sim$ 10 times lower than that observed in the 1990s. This indicates that the production of optical broad emission lines is not accompanied by a significant increase in the optical continuum. Although ultraviolet continuum photons are known to be present (see \citealt{mathur18}) and required to explain the optical broad emission lines we observe in this work, optical continuum photons are not being produced at a significant rate. Typical accretion disc emission would produce both ultraviolet from the inner part of the accretion disc and optical continuum photons from larger radii of the disc. This suggests that Mrk~590 may not have a fully built standard accretion disc but rely on other mechanism to produce ultraviolet photons, or that only the inner part of the accretion disc in Mrk~590 is emitting photons that are able to reach us.

The ionised gas in the galaxy shows complex dynamics, with extended emission from the nucleus, a nuclear spiral extending out to a radius of 2 kpc surrounded by a ring of strong emission in ionised gas at a radius of 4.5 kpc. The ring shows a younger stellar population and higher metallicity than in the centre of the galaxy, with knots of increased emission tracing H II regions. A ring of cold molecular gas and dust is also detected and broadly co-spatial but slightly interior to the ring of H$\alpha$ emission, indicating that there is a reservoir of molecular gas to fuel further star formation. We determine that a region of increased H$\alpha$ emission in the ring matches the location of a recently discovered supernova type Ia in Mrk~590.

The nuclear emission is excited by the AGN, while the nuclear spiral excitation has a possible contribution from shocks based on the emission line ratios and increased gas velocity dispersion observed. Nuclear spiral structures are formed by a non-axisymmetry in the galactic potential. In the case of Mrk~590 the nuclear spiral structure is likely due to the presence of a bar in its central arcseconds. The orbital resonances associated with the bar may also be responsible for the accumulation of gas in the nucleus and in the ring. A detailed analysis of the dynamics in Mrk~590 will be the subject of paper II.

We detect the presence of extended warm molecular gas emission in the nucleus, traced by the H$_{2}$ 2.12 $\micron$ emission line. The cold molecular gas traced by the CO(1-0) and CO(3-2) lines is not detected in the nucleus, indicating that H$_{2}$ 2.12 $\micron$ may be a better tracer of molecular gas content and dynamics for the excitation conditions at place in the nucleus of Mrk~590.

We also discovered a nuclear outflow of ionised gas traced by [O III] 5007 \AA\ emission, extending out to $r \sim$ 1 kpc. The blue-shifted component of the outflow dominates the emission, with a radial velocity of V$_{\rm [O III]\,broad}$ $\sim$ -400 km\,s$^{-1}$ with respect to the systemic velocity of the galaxy. 

Ionised and molecular gas have been detected in the nucleus of Mrk~590, indicating that down to the spatial resolution of our near-infrared data ($r \sim 60$ pc from the black hole) there is a reservoir of gas available. A mechanism to drive gas from kiloparsec scales to $r \sim 500$ pc, in the form of a nuclear spiral, is observed in Mrk~590. Nuclear spirals have been shown from simulations and observations in the local Universe, to be able to drive gas to the nucleus at the mass accretion rates needed to power Seyfert-like activity. While the gas inflow is expected to occur in the gas disc and along the spiral arms, a kiloparsec scale nuclear outflow is also detected, which can remove gas from the nucleus of Mrk~590. Although singular due to its changing-look status, Mrk~590 is in this aspect similar to some other Seyfert galaxies in the local Universe that simultaneously show gas inflow and gas outflow structures, suggesting a balance between gas inflow and outflow in their nuclear regions.

\section{Acknowledgements}
The authors thank Rogerio Riffel, Massimo Dotti and Vera Patricio for useful discussions and Raoul Canameras for his help with contour diagrams. SIR, MV and DL acknowledge financial support from the Independent Research Fund Denmark via grant no. DFF-4002-00275 (PI: Vestergaard). Based on observations collected at the European Organisation for Astronomical Research in the Southern Hemisphere under ESO programmes 080.B-0239 and 099.B-0294(A). This paper makes use of the following ALMA data:
ADS/JAO.ALMA\#2013.1.00534.S and ADS/JAO.ALMA\#2015.1.00770.S. ALMA is a partnership of ESO (representing its member states), NSF (USA) and NINS (Japan), together with NRC (Canada), NSC and ASIAA (Taiwan), and KASI (Republic of Korea), in cooperation with the Republic of Chile. The Joint ALMA Observatory is operated by ESO, AUI/NRAO and NAOJ.
GANDALF was developed by the SAURON team and is available from the SAURON website (www.strw.leidenuniv.nl/sauron). See also \cite{sarzi06} for details.
The STARLIGHT project is supported by the Brazilian agencies CNPq, CAPES and FAPESP and by the France-Brazil CAPES/Cofecub program.

\bibliographystyle{mnras}
\bibliography{AGN}
\end{document}